\documentclass[%
 reprint,
 amsmath,amssymb,
 aps,
]{revtex4-2}

\usepackage{graphicx}
\usepackage{dcolumn}
\usepackage{bm}

\usepackage{braket}
\usepackage{subcaption}

\DeclareMathOperator{\Tr}{Tr}
\newcommand{\be}{\begin{equation}}
\newcommand{\ee}{\end{equation}}
\newcommand{\nn}{\nonumber}

\begin{document}

\preprint{APS/123-QED}

\title{Local vs. translationally-invariant \\
slowest operators in quantum Ising spin chains}

\author{Ekaterina Izotova$^{1,2}$}
\email{ekat.izotova@gmail.com}

\affiliation{
$^1$ Skolkovo Institute of Science and Technology,
Bolshoy Boulevard 30, bld. 1, Moscow 121205, Russia \\
$^2$ Moscow Institute of Physics and Technology, Institutsky Lane 9,
Dolgoprudny 141700, Russia 
}


\begin{abstract}
In this paper we study one-dimensional quantum Ising spin chains in external magnetic field close to an integrable point. We concentrate on the dynamics of the slowest operator, that plays a key role at the final period of thermalization. We introduce two independent definitions of the slowest operator: local and translationally-invariant ones. We construct both operators numerically using tensor networks and extensively compare their physical properties. We find that the local operator has a significant overlap with energy flux, it does not correspond to an integral of motion, and, as one goes away from the integrable point, its revivals get suppressed and the rate of delocalization changes from extremely slow to slower than diffusion. The translationally-invariant operator corresponds to an integral of motion; as the system becomes less integrable, at some point this operator changes its nature: from no overlap with any magnetization and fast rate of delocalization, to non-zero overlap with magnetizations $\sigma_{x}$ and $\sigma_{z}$ and slow rate of delocalization.


\end{abstract}

\maketitle


\section{Introduction}

Thermalization is widely studied in the literature. This process occurs in closed many-body quantum systems. In such systems, if we take a finite subsystem, its complement plays a role of a bath and thermalizes it \cite{PhysRevE.97.012140}. This phenomenon has roots in the theory of open quantum systems, where it was proven that a system connected to a thermal bath reaches an equilibrium with a temperature of a bath \cite{benatti2005open, andrianov2020perturbation, andrianov2022franke}. The primary known mechanism of thermalization is Eigenstate Thermalization Hypothesis \cite{srednicki1994chaos, deutsch1991quantum, rigol2008thermalization, Lashkari_2018} (see also the reviews \cite{d2016quantum,gogolin2016equilibration}). But there are systems that do not thermalize, the famous examples are Anderson localization \cite{anderson1958absence}, Many-Body-Localization (see the review \cite{alet2018many}) and also integrable systems, but the latter are not a phase of matter, they rather correspond to the special points in the space of parameters in Hamiltonian.

In this work we are interested in the transition from integrability to non-integrability, as we slowly change the parameters in Hamiltonian. Such close-to-integrable systems are known to have an initial period of prethermalization \cite{lin2017explicit, berges2004prethermalization, mori2018thermalization,PhysRevLett.122.080603,lin2019surviving}, while at the latest period there is a transport of conserved quantities, such as energy or magnetization. At the prethermalization period, the fastest correlations decay, while at this latest period the slowest operator \cite{kim2015slowest, pancotti2018almost} plays a key role.

We consider one-dimensional quantum Ising spin chain in external magnetic field, and its Hamiltonian is

\be H = - \sum_i \sigma_z^{(i)} \sigma_z^{(i+1)} + h \sum_i \sigma_z^{(i)} + g \sum_i \sigma_x^{(i)} \label{Ham}\ee
where $\sigma_{x,y,z}$ are Pauli matrices, $i$ is a site on the chain, $h, g$ are real numbers. We assume periodic boundary conditions, with total number of spin sites being $L$. In some situations we will take a limit $L\rightarrow \infty$.

This system is known to be integrable for
\begin{itemize}
\item $g = 0$ and any real $h$
\item $h = 0$ and any real $g$
\end{itemize}

As we are interested in the vicinity of the integrable point, we consider $3$ cases:
\begin{enumerate}
\item $g \neq 0$ fixed, $h$ near $0$ (non-integrable)
\item $h \neq 0$ fixed, $g$ near $0$ (non-integrable)
\item $h = 0$, various $g$ (integrable)
\end{enumerate}

We do not consider a case $g = 0$ and varous $h$, since it has a trivial integral of motion $\sigma_z$, which we would get in all our later calculations.

In this paper we focus on the latest period of thermalization and, therefore, on the dynamics of the slowest operator. We particularly consider two independent definitions of the slowest operator (local and translationally-invariant ones) and reveal the differences in their physical properties. For doing so, we construct them numerically using tensor network methods \cite{orus2014practical,bridgeman2017hand,biamonte2017tensor,roberts2019tensornetwork,tnorg,tnet}.

The structure of this paper is as follows:
\begin{itemize}
\item \textbf{In section \ref{periods}} we describe periods of thermalization process of a non-integrable system close to an integrable point.
\item \textbf{In section \ref{two_def}} we narrow down to the final period of thermalization and introduce two definitions of the slowest operator.
\item \textbf{In section \ref{tn}} we describe how we construct the two operators using tensor networks.
\item \textbf{In section \ref{sec_entropy}} we calculate entanglement entropy and observe that it is low. Therefore, application of tensor networks is justified.
\item \textbf{In section \ref{sec_hg}} we study the dependence of the slowest operators on the parameters $g$ and $h$. We find that the local operator does not correspond to an integral of motion and has a significant overlap with diffusion mode/energy flux. The translationally-invariant operator, on the opposite, corresponds to an integral of motion; as one goes away from the integrable point, at a specific $h^*$ it changes its nature: from no overlap with any magnetization, to non-zero overlap with magnetizations $\sigma_{x}$ and $\sigma_{z}$.


\item \textbf{In section \ref{sec_N}} we study the delocalization rate of the slowest operator by calculating the scaling with its support size $N$ on the chain. As one goes away from the integrable point, the rate of delocalization of the local slowest operator changes from extremely slow to slower than diffusion. The translationally-invariant operator before the transition delocalizes faster than diffusion, but after - slower than diffusion.



\item \textbf{In section \ref{sec_timeevol}} we study time evolution. In particular, we calculate the dynamics of two-point correlation function and the out-of-time-ordered commutator. We find that, as one goes away from the integrable point, the revivals of the local slowest operator get suppressed. We observe the common features of the time evolution of the two slowest operators.



\item \textbf{In section \ref{sec_conclusion}} we conclude what we have found.

\item \textbf{In section \ref{discussion}} we discuss the remaining questions and possible directions of future research.

\end{itemize}

\section{Periods of thermalization process \label{periods}}

In this section we consider separately thermalization of non-integrable systems and equilibration of integrable systems. Then, combining those two together, we describe a picture of thermalization of non-integrable systems close to an integrable point. We particularly emphasize the resulting two periods of thermalization.

\subsection{Non-integrable systems}

In such systems thermalization can be defined as follows. An average of a local operator $A$ during its evolution reaches thermal average:
\be \bra{\psi(t)} A \ket{\psi(t)} \rightarrow \Tr (\rho_{th} A) \label{nonint}\ee
where $\rho_{th} = \frac{1}{Z_{th}} e^{-\beta H}$. $\beta = \frac{1}{T}$, $T$ is fixed by the condition $\bra{\psi(t)} H \ket{\psi(t)} = \Tr (\rho_{th} H) $. $Z_{th}$ is such that $\Tr \rho_{th} = 1$.

\subsection{Integrable systems}

In such systems there is an extensive amount of integrals of motion $\{Q_i\}$, which commute with Hamiltonian: $[Q_i, H] = 0$ (they also obey: $[Q_i,Q_j]=0$; $Q_0 = H$). The system cannot thermalize, because its dynamics is constrained by these integrals of motion $\{Q_i\}$. Instead, it equilibrates to Generalized Gibbs Ensemble (GGE) \cite{rigol2007relaxation,vidmar2016generalized}:
\be \bra{\psi(t)} A \ket{\psi(t)} \rightarrow \Tr (\rho_{GGE} A) \ee
where $\rho_{GGE} = \frac{1}{Z_{GGE}} e^{\sum_i \mu_i Q_i}$, $Z_{GGE}$ is such that $\Tr \rho_{GGE} = 1$.

\subsection{Non-integrable systems close to an integrable point}

In such systems there is no such set of $\{Q_i\}$, with $[Q_i, H] = 0$. Instead they become $\{O_i\}$, and some of them commute better with $H$, some worse. We emphasize $O_0$, which best commutes with $H$, and call it \textbf{the slowest operator} (first introduced and named in \cite{kim2015slowest, pancotti2018almost}). We define it as a local operator (with support on $N$ consecutive sites), which minimizes the non-negative quantity $\Tr [H,O_0]^\dag [H,O_0] = - \Tr [H,O_0]^2$.

$O_0$ has an important physical meaning. $O_0$ plays a role of an integral of motion for all other operators $A$, because its dynamics is much slower ($\dot{O_0} \sim |[H,O_0]|$). Therefore, the thermalization process can be divided into two periods:

\begin{enumerate}
\item \textbf{Initial period of prethermalization} when all operators equilibrate to $\widetilde{GGE}$:
\be \bra{\psi(t)} A \ket{\psi(t)} \rightarrow \Tr (\rho_{\widetilde{GGE}} A) \label{preth}\ee
where $\rho_{\widetilde{GGE}} = e^{-\beta H + \mu O_0}$.

\item \textbf{Period of final thermalization}:
\be \Tr (\rho_{\widetilde{GGE}} A) \rightarrow \Tr (\rho_{th} A) \label{GGE-th}\ee

\end{enumerate}

The slowest operator $O_0$ (as a part of $\rho_{\widetilde{GGE}}$) plays an important role during the period of final thermalization.

\section{Two definitions of the slowest operator \label{two_def}}

In the rest of the paper we study the slowest operator. It is a local operator $O_0$ with support on $N$ consecutive sites, that minimizes the non-negative quantity $\Tr [H,O_0]^\dag [H,O_0] = - \Tr [H,O_0]^2$.

Since we do not consider faster operators $O_1, O_2, \dots$, from now on we will denote the slowest operator as $O$ instead of $O_0$. We also numerate the spin sites $0 \dots L-1$, as the total number is equal to $L$.

\begin{figure}
\centering
\begin{subfigure}{0.4\textwidth}
    \includegraphics[scale=0.28]{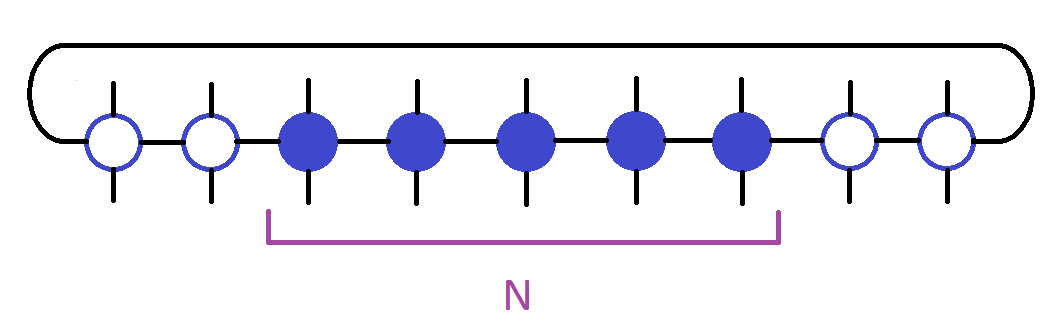}
    \caption{Local ($N=5$, $L=9$)}
    \label{fig:first}
\end{subfigure}
\hfill
\begin{subfigure}{0.4\textwidth}
    \includegraphics[scale=0.22]{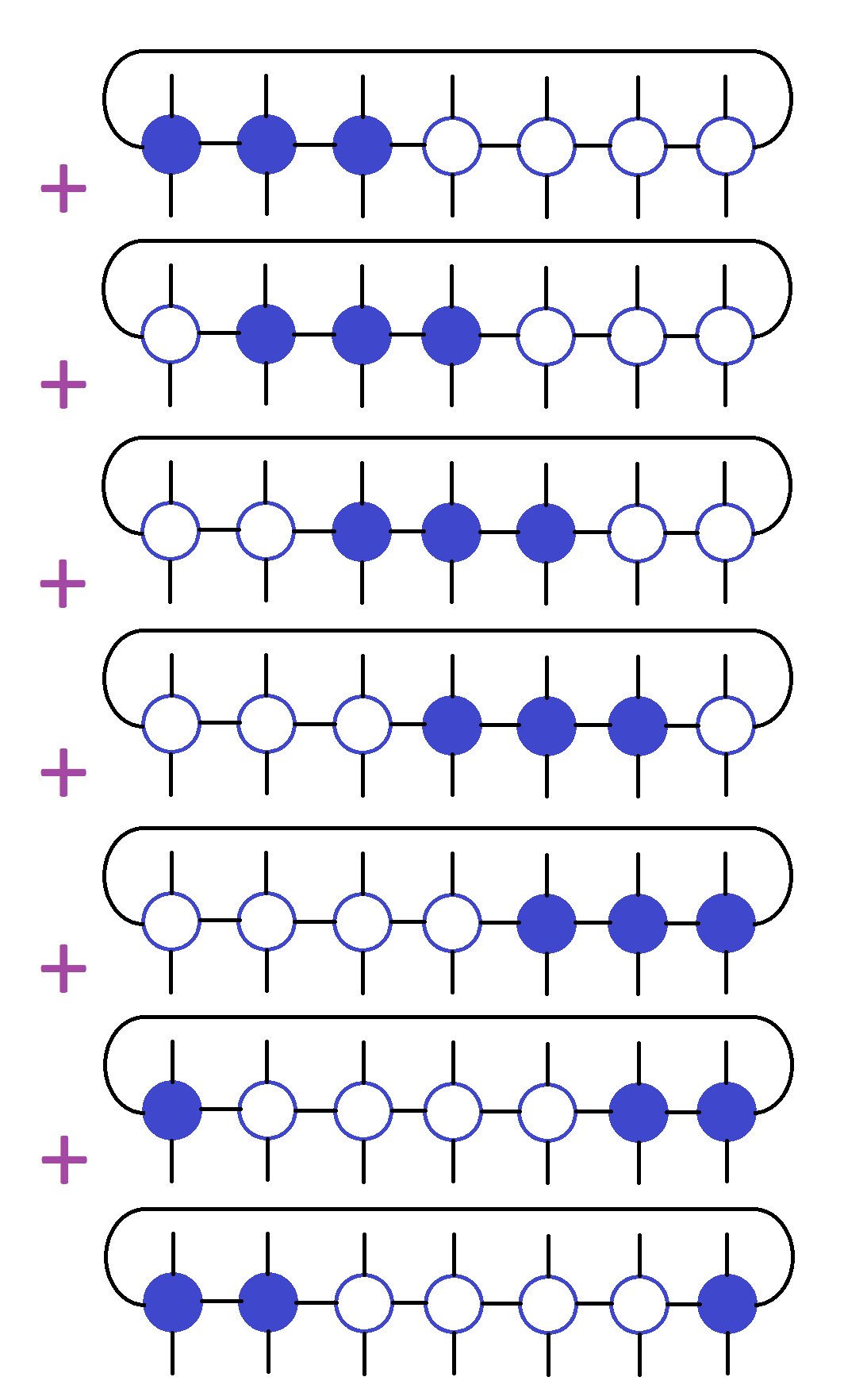}
    \caption{Transaltionally-invariant ($N=3$, $L=7$)}
    \label{fig:second}
\end{subfigure}
        
\caption{(a) Local and (b) translationally-invariant definitions of the slowest operator. Empty circles correspond to identity matrices.}
\label{def}
\end{figure}

Then, there are \textbf{two independent definitions} of the slowest operator $O$ (see Fig. \ref{def}).

\subsection{"Local" definition}
$O$ is a single operator that has support on $N$ consecutive sites.

We also impose other conditions:
\begin{itemize}
\item $\Tr O = 0$ (to exclude identical operator)
\item $O$ is Hermitian
\item $\Tr O^2 = 1$ (normalization)
\end{itemize}

\subsection{"Translationally-invariant" definition}
$O$ is a sum of shifted (by one site) identical operators, each one having support on $N$ consecutive sites: \\
$O = \sum^{L-1}_{i=0} O_i$, where $O_i$ has support on $N$ consecutive sites from $i$ to $i+N-1$

{\small If $i+N-1 > L - 1$, then $O_i$ after the end of the chain (site $L-1$) continues from the beginning (site $0$).}

And we impose other conditons:
\begin{itemize}
\item $\Tr O_i = 0$ (to exclude identical operator)
\item $O_i$ is Hermitian

\item $O_i$ at its first site $i$ is decomposed in a basis $\{ \sigma_x, \sigma_y, \sigma_z \}$ (without $\sigma_0$)

{\small (To be discussed later.) }

\item $\Tr O^2  = 1$ (normalization)


\item $\Tr HO = 0$ 

{\small If we do not impose this condition, the minimization of $- \Tr [H,O]^2$ will give us $O$ equal to Hamiltonian, since the latter is exactly a sum of local terms.}

\end{itemize}

%
%

%

\begin{figure}
\includegraphics[scale=0.21]{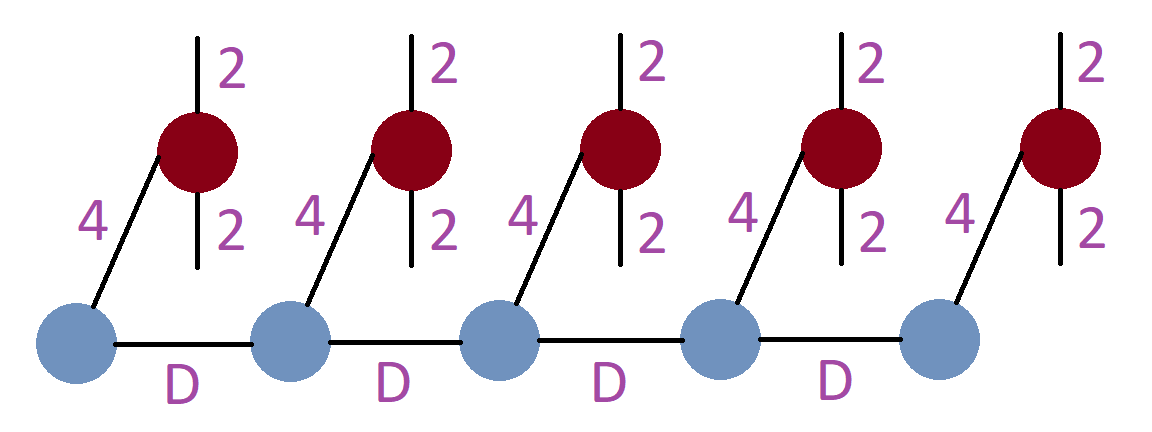}
\caption{Tensor network representation of the slowest operator. The dark red circles correspond to Pauli matrices $\sigma^{(i)}$, blue ones - to the tensor coefficients $A^{(i)}$ in front of them. Numbers correspond to dimensions of the edges. See (\ref{mps_O}).}
\label{mpo_sigma}
\end{figure}

\section{Finding the slowest operator using tensor networks \label{tn}}

The next step is to find the slowest operator $O$ by minimization of $ - \Tr [H,O]^2$. We describe the procedure for local and translationally-invariant slowest operators separately.

\subsection{Local slowest operator}

This optimization problem can be solved by exact diagonalization. For doing so, we need to represent $ - \Tr [H,O]^2$ in a form $\bra{O} \mathcal{H}_{eff} \ket{O}$. Since $O$ has support on $N$ sites, its dimensions are $2^N \times 2^N$. Then, the dimension of $\ket{O}$ is $4^N$, and those of $\mathcal{H}_{eff}$ - $4^N \times 4^N$. $\ket{O}$ can be found by exact diagonalization of $\mathcal{H}_{eff}$. The problem is that it can only be done for small $N$ up to $\sim 8$, because, for bigger $N$, $\mathcal{H}_{eff}$ is a large matrix and exact diagonalization takes too much memory and computational time. That is why we use tensor networks.

\subsubsection{Tensor network representation of $O$}

We represent $O$ in a matrix product state (MPS) form. We use Pauli matrix basis at every site $\{ I, \sigma_x, \sigma_y, \sigma_z \}$ with real coefficients to ensure that $O$ is Hermitian (see Fig. \ref{mpo_sigma}):
\be O = \sum_{\substack{l,m,n,\dots \\ k_0 \dots k_{N-1}}} A^{(0),k_0}_{l} A^{(1),k_1}_{lm} \dots A^{(N-1),k_{N-1}}_n \times\nn\ee
\be \times  \sigma^{(0),i_0,j_0}_{k_0} \otimes \sigma^{(1),i_1,j_1}_{k_1} \dots \otimes \sigma^{(N-1),i_{N-1},j_{N-1}}_{k_{N-1}} \label{mps_O}\ee
where $(i)$ follows a site number, $i = 0 \dots N-1$. $A^{(i)}$ is a real tensor at site $i$. $l,m,n,\dots$ are so-called bond (internal) indices, each bond index can take value from $0$ to $D-1$, where $D$ is a bond dimension. $i_0,\dots,i_{N-1}$ are up physical indices, $j_0,\dots,j_{N-1}$ are down physical indices, and every one of them can be $0$ or $1$, as spin at each site is up or down.  $k_0,\dots, k_{N-1}$ take values from $0$ to $3$, as there are four Pauli matrices. 

In practice, we put $O$ in a canonical form to fulfill the normalization condition $\Tr O^2 = 1$. Then, bond dimension near the boundaries (close to site $0$ or $N-1$) is actually less than $D$. From the beginning of the chain bond dimension is increases as $4, 16, 64, \dots$ until it is cut with fixed $D$. After that it is constant for some time (equal to $D$), until it decreases in the same manner near the second boundary (near site $N-1$).

Tensor network representation is efficient for optimization problems, since MPS has a small number of parameters - $N\times D^2 \times 4$, while a general operator has $4^N$ parameters. On the other hand, MPS is constructed of local tensors and allows one to do optimization site by site - therefore, even more reduce a number of optimization parameters. On the downside, MPS ansatz is only applicable for low entanglement inside $O$ \cite{orus2014practical}. The greater the dimension $D$ is, the higher entanglement one can capture. In the next section, we specifically check that entanglement entropy is low enough, so that we can apply tensor networks.

\begin{figure}
\includegraphics[scale=0.21]{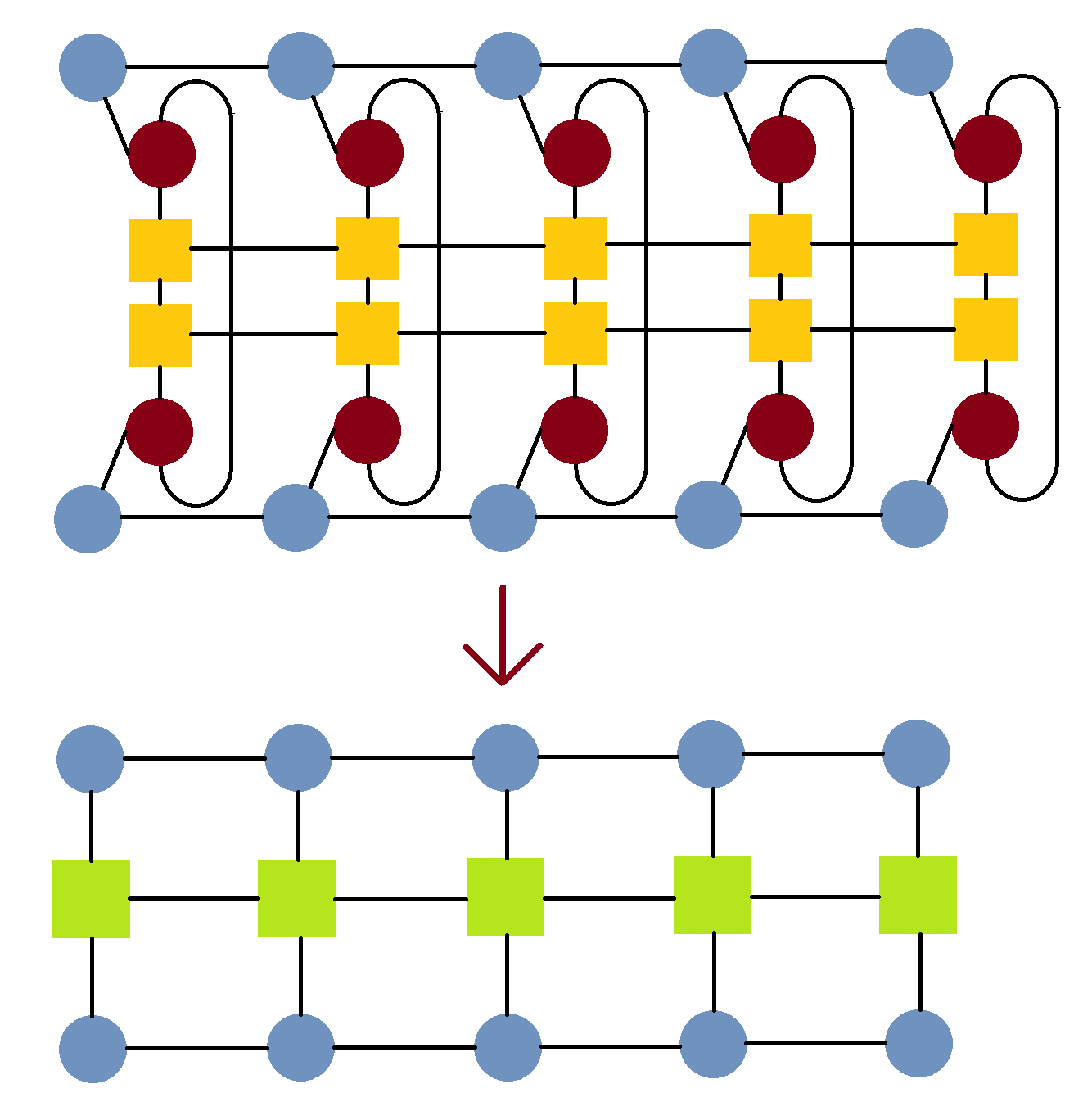}
\caption{Transformation of the term $\Tr (O^\dag H_{loc}^2 O)$ (see (\ref{terms_for_opt})). After combining the resulting tensor networks for all the terms, DMRG algorithm can be applied. Dark red circles correspond to Pauli matrices, blue circles - to coefficients $A^{(i)}$ in front of them, yellow squares - to an MPO of $H_{loc}$.}
\label{tn_1term}
\end{figure}

\subsubsection{Tensor network representation of $-\Tr [H,.]^2$}

To be able to do optimization locally, site by site, we not only have to represent $O$ in an MPS form, but also $-\Tr[H,.]^2$ - in a matrix product operator (MPO) form.

Then, we first decompose $ \Tr ([H,O]^\dag [H,O])$ as follows:
\be  \Tr ([H,O]^\dag [H,O]) = \Tr ([H_{loc},O]^\dag [H_{loc},O]) + \nn\ee
\be + \Tr ([\sigma_z^{(0)},O]^\dag [\sigma_z^{(0)},O]) + \Tr ([\sigma_z^{(N-1)},O]^\dag [\sigma_z^{(N-1)},O]) \label{terms_for_opt0}\ee
where $H_{loc}$ is a part of Hamiltonian (\ref{Ham}) having support on the same $N$ consecutive sites as $O$ does. The last $2$ terms come from the terms in Hamiltonian $\sigma^{-1}_z \sigma^{0}_z$ and $\sigma^{N-1}_z \sigma^{N}_z$ respectively.

In the next step, we decompose the latter expression:
\be  \Tr ([H,O]^\dag [H,O]) = \nn \ee
\be = \Tr (O^\dag H_{loc}^2 O) + \Tr (O^\dag O H_{loc}^2) - 2 \Tr (O^\dag H_{loc} O H_{loc}) + \nn\ee
\be + 2 - 2 \Tr (O^\dag \sigma_z^{(0)} O \sigma_z^{(0)}) + 2 -2  \Tr (O^\dag \sigma_z^{(N-1)} O \sigma_z^{(N-1)}) \label{terms_for_opt}\ee

We also add a term $|\Tr O|^2$ with some positive factor into the optimization problem, to account for the trace condition: $\Tr O = 0$.

\begin{figure}
\centering
\begin{subfigure}{0.49\textwidth}
    \includegraphics[scale=0.16]{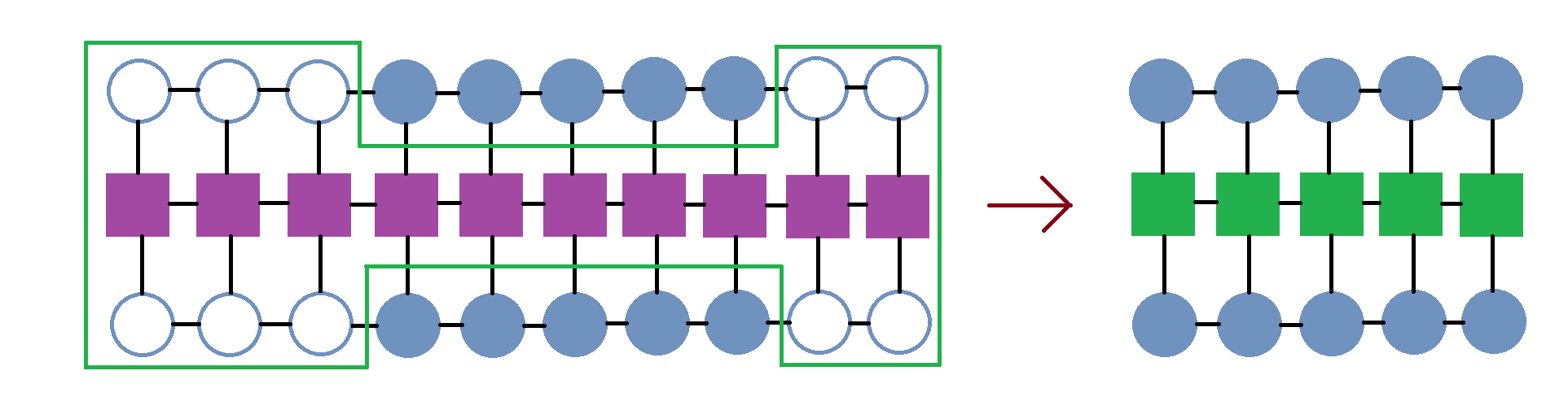}
    \caption{Local}
\end{subfigure}
\hfill
\begin{subfigure}{0.49\textwidth}
    \includegraphics[scale=0.16]{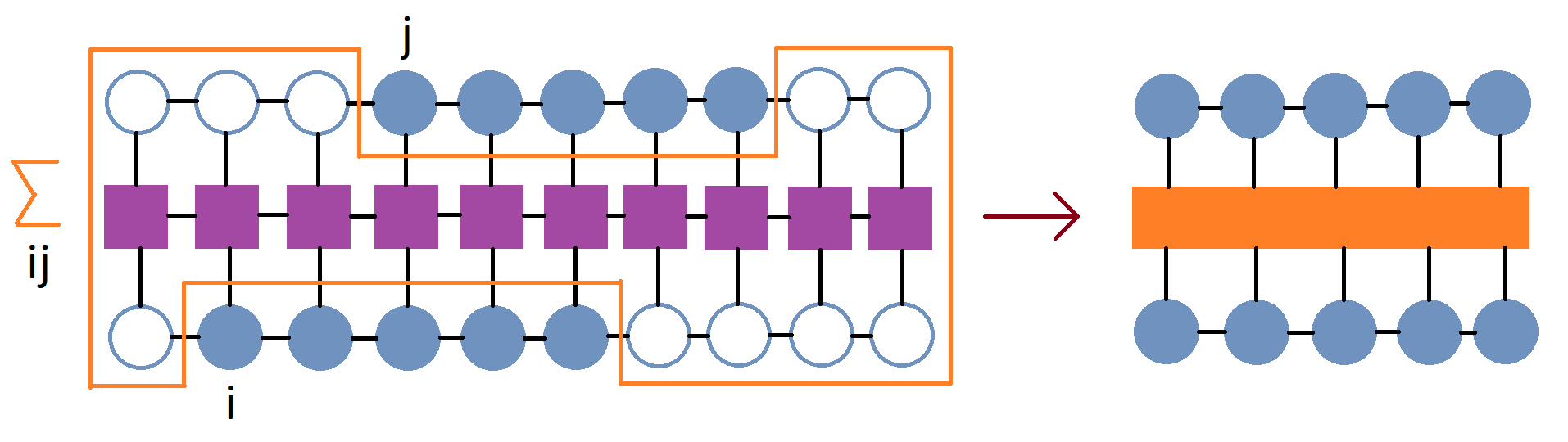}
    \caption{Translationally-invariant}
\end{subfigure}
        
\caption{Reduction of the tensor network, corresponding to $-\Tr [H,O]^2$. DMRG algorithm can be applied to the network on the right. The violet squares correspond to $-\Tr [H,.]^2$ merged with Pauli matrices of $O$. Thus, DMRG algorithm finds coefficients $A^{(i)}$, from which one can construct $O$, using (\ref{mps_O}).}
\label{reduction}
\end{figure}

Each one of the terms can be represented as a tensor network. We illustrate how to do it for the first term (see Fig. \ref{tn_1term}). Then, these tensor networks can be combined into one or calculated separately to reduce computational time. The overall transformation is sketched in Fig. \ref{reduction} (a).

\begin{itemize}
\item {\small
$H_{loc}$ can be put in an MPO form with bond dimension $3$:
\be vL = \begin{pmatrix}
0 & 0 & 1 
\end{pmatrix}, M^{(i)} = 
\begin{pmatrix}
I^{(i)} & 0 & 0\\
\sigma^{(i)}_z & 0 & 0 \\
h \sigma^{(i)}_z + g \sigma^{(i)}_x & - \sigma^{(i)}_z & I^{(i)}
\end{pmatrix}, vR = \begin{pmatrix}
1 \\
0 \\
0
\end{pmatrix} \ee
with $M^{(i)}$ - a matrix at site $i$, $vL$ and $vR$ - adjoining left and right vectors (that can be merged with $M^{(0)}$ and $ M^{(N-1)}$ respectively), such that\\
$H_{loc} = vL \times M^{(0)} \times \dots \times M^{(N-1)} \times vR$.}
\end{itemize}

Everywhere below we effectively calculate $-\Tr [H,O]^2$ for a limit $L \rightarrow \infty$, because $-\Tr [H,O]^2$ has the same value for any $L\geq N+2$ (see (\ref{terms_for_opt0}), (\ref{terms_for_opt})).

\subsubsection{Finding $O$ using DMRG algorithm}

After we have obtained a tensor network in a form depicted in  Fig. \ref{reduction} (a) on the right, we can apply Density Matrix Renormalization Group (DMRG) algorithm \cite{white1992density}. We go site-by-site from left to right and back and do optimization of $A^{(i)}$ at each step. We gradually increase bond dimension: $8, 16, 32, \dots$ until $1024$. We check the convergence at a fixed bond dimension by a relative error of $10^{-7}$; the program terminates if a relative difference between the final answers at a current bond dimension and a previous one is less than $0.5 \%$.

\subsection{Translationally-invariant slowest operator}

\subsubsection{Tensor network representation of $O$}

The tensor network representation of a particular $O_i$ of the sum $O = \sum_i O_i$ is almost the same as (\ref{mps_O}) (see also Fig. \ref{mpo_sigma}). The difference is that the index $k_0$ takes values $1,2,3$ instead of $0,1,2,3$, because, according to the definition, there can only be $\{ \sigma_x, \sigma_y, \sigma_z \}$ (no $\sigma_0$) as basis elements at the first site.

We impose this condition for the following reason. The minimization of $- \Tr [H, \sum_i O_i]^2$ with the condition $\Tr (\sum_i O_i)^2 = 1$ is a generalized eigenvalue problem and much harder to deal with. With the above condition for the first site, $\Tr (\sum_i O_i)^2$ becomes trivial: it is a sum of identical terms $\sum_i \Tr O_i^2$ (the terms $\Tr O_i O_j, i \neq j$ die out, because we take a trace either at the first site of $O_i$, or at the first site of $O_j$). In other words, it is equivalent to the standard normalization condition $\Tr O_i^2 = 1$, and no generalized eigenvalue problem arises.

On the other hand, this condition is just a gauge fixing: in a spin chain with big enough bond dimension $D$, one can represent the same operator using different sets of $N$-site basis elements, we just choose one of these representations.


\subsubsection{Tensor network representation of $-\Tr [H,.]^2$}

For a translationally-invariant operator, we need to minimize $- \Tr [H,\sum_i  O^{(i)}]^2$ with identical operators $\{ O_i \}$, $O_i$ has support on consecutive sites $i \dots i+N-1$. In this case we also use tensor networks for finding $O_i$, but the key difference is that we envelop the summation into the effective operator (see Fig. \ref{reduction} (b)). As obvious from Fig. \ref{reduction} (b), this effective operator cannot be decomposed into local parts, and, therefore, be represented in an MPO form. It complicates the implementation of the DMRG algorithm, since it leads to the contraction of all the tensors of the tensor network at every local step of the algorithm.

Taking into account the definition of this operator, we include the conditions $|\Tr HO|^2$ and $|\Tr O_i|^2$ with some positive factors into an optimization problem.

Everywhere below we fix $L=2N+3$ and claim that we effectively calculate $\Tr \left( [H,\sum_i  O^{(i)}]^\dag [H,\sum_j  O^{(j)}] \right)$ for $L\rightarrow\infty$. The reason is that $[H,O^{(i)}]$ is an operator with support $N+2$. Therefore, only $O^{(i)}$ and $O^{(j)}$ that are close to each other contribute. All other terms die out.


\subsubsection{Finding $O$ using DMRG algorithm}

The algorithm is similar to the local operator case, but just takes much more computational time. We gradually increase bond dimension: $64, 128, \dots$ until $1024$. We check the convergence at a fixed bond dimension by a relative error of $10^{-4}$; the program terminates if a relative difference between the final answers at a current bond dimension and a previous one is less than $0.5 \%$.


\section{Entanglement entropy of the slowest operator \label{sec_entropy}}

In this section we justify using tensor network ansatz for finding the slowest operator. In particular, we prove that even with not very big bond dimension $D$, we still find the exact slowest operator numerically.

It is known in the literature that entanglement entropy $S$ of a quantum state $\ket{\psi}$, represented in a tensor network form, is bounded by $\log D$, where $D$ is a bond dimension \cite{orus2014practical}. In other words, the greater $D$ one consideres, the higher entanglement entropy one can cover.

\begin{itemize}
\item {\small
The quick explanation is as follows. Suppose $\ket{\psi}$ has support on sites $0 \dots N-1$. One can do a bipartition to the left (sites $0 \dots i-1$) and right (sites $i \dots N-1$) parts. Then, the left/right reduced density matrix is defined as $\rho = \Tr_{R(L)} \ket{\psi} \bra{\psi}$. And entanglement entropy reads: $S = - \Tr \rho \log \rho$ (left and right reduced density matrices give the same answer for $S$).  Since the edge between sites $i-1$ and $i$ has dimension $D$, the reduced density matrix $\rho$ (left or right) has size $D\times D$. Then, entanglement entropy $S$ is maximized by the identity matrix, which is in our case: $\rho = \frac{1}{D} I$. For such a matrix, the entropy is $S = \log D$. The statement is proven.}
\end{itemize}

In our problem, we have the slowest operator in a matrix product state form (see (\ref{mps_O}), Fig. \ref{mpo_sigma}). If we prove that, as we increase bond dimension $D$, entanglement entropy converges to the small enough value, then we can claim that the slowest operator corresponds to the exact slowest operator. (Entanglement entropy is used here as a technical tool, no real physical meaning is implied.)

For doing so, we calculate entanglement entropy for the final value of $D$, used in our calculations. We compare it with the maximum entanglement entropy for a given bipartition.

\begin{itemize}
\item {\small
The dimension of the general vector, having support on $N$ consecutive sites, is $4^N$. The bipartition divides it as $4^{i}\times 4^{N-i}$. Then, the size of the left reduced density matrix is $4^i \times 4^i$, and that of the right one is $4^{N-i} \times 4^{N-i}$. Therefore, the maximum entropy is $\log \min (4^{i},4^{N-i})$.}
\end{itemize}



\begin{figure}
\includegraphics[scale=0.21]{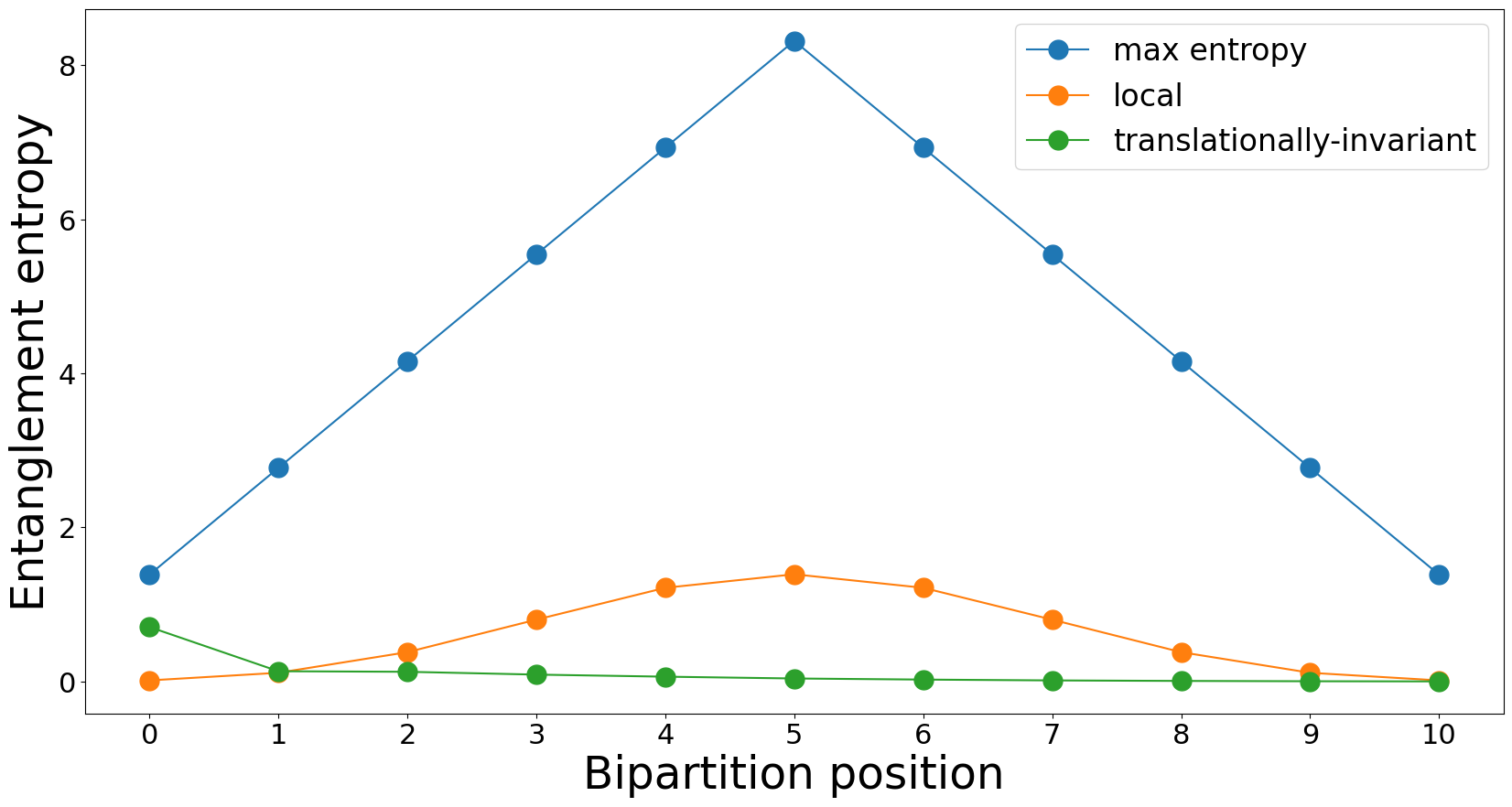}
\caption{\label{entropy} Entanglement entropy as a function of bipartition position for local and translationally-invariant slowest operators ($g=1.05, h=0.1, N=12$). Entanglement is low and allows one to use tensor network representation.}
\end{figure}

The result is depicted in Fig. \ref{entropy}. We observe that entanglement entropy is indeed much smaller than the maximum value, both for local and translationally-invariant definitions of the slowest operator. Therefore, we claim that the slowest operator we find does correspond to the exact slowest operator.

We note that entanglement entropy for the translationally-invariant slowest operator is not left-right symmetric. It is a consequence of the "gauge fixing", that allows only $(\sigma_1,\sigma_2,\sigma_3)$ Pauli matrices as basis elements at the first site, while at all other sites all four of $(I,\sigma_1,\sigma_2,\sigma_3)$ can have non-zero contributions.

\section{Dependence of the slowest operator on the parameters in Hamiltonian \label{sec_hg}}

Here we find how the physical properties of the slowest operator depend on the parameters $g$ and $h$ in Hamiltonian (see (\ref{Ham})). We find the differences between local and translationally-invariant slowest operators. 

In this section we calculate the translationally-invariant slowest operator using exact diagonalization of $-\Tr [H,.]^2$ (the orange operator in Fig. \ref{reduction} (b)). (We do not need large support sizes $N$ and limit ourselves to $N=5,6$. Exact diagonalization is suitable for this task.) In all other sections we use tensor networks and DMRG algorithm.

\subsection{The physical quantities}

We calculate the following quantities.

\subsubsection{$-\Tr [H,O]^2$ as a function of $g$ and $h$}

If $-\Tr [H,O]^2$ tends to $0$, as we approach an integrable point, then $O$ corresponds to an integral of motion of the integrable system.


\subsubsection{Overlap between the slowest operator and probe operators as a function of $g, h$}

We calculate the quantity $\Tr \left( OP \right)$, where $P$ is the probe operator. In this way we find the physical meaning of the slowest operator.

The probe operators are different for local and translationally-invariant slowest operators. All probe operators for the local slowest operator have support on $N$ consecutive sites. For translationally-invariant ones, the probe operators have support on the full chain, i.e. have support $L$.

Those probe operators are:

\begin{enumerate}

\item \textbf{Diffusion mode}

\begin{itemize}
\item For local slowest operator:

local terms of Hamiltonian (see (\ref{Ham})) multiplied by cosine, to form a "bell" shape \cite{kim2015slowest}:

\be E^{(0)} = \sum_{i=0}^{N-2} \cos \left(-\frac{\pi}{2} + \frac{i+\frac{1}{2}}{N}\pi \right) (- \sigma^{(i)}_z \sigma^{(i+1)}_z) + \nn\ee

\be + \sum_{i=0}^{N-1} \cos \left( -\frac{\pi}{2} + \frac{i}{N}\pi \right) (h \sigma^{(i)}_z + g \sigma^{(i)}_x ) \ee

\item For translationally-invariant slowest operator:

\be \sum_{i=0}^{L-1} E^{(i)}\ee
where  $E^{(i)}$ has support on sites $i \dots i+N-1$.

\end{itemize}

\item \textbf{Energy flux}

\begin{itemize}
\item For local slowest operator:

local Hamiltonian terms that belong to the interval of $N$ consecutive sites and an extra boundary term:

\be \sum_{i=0}^{N-2} (- \sigma^{(i)}_z \sigma^{(i+1)}_z) + \nn\ee
\be + \sum_{i=0}^{N-1} (h \sigma^{(i)}_z + g \sigma^{(i)}_x) + (- \sigma^{(N-1)}_z \sigma^{(0)}_z) \ee

\item For translationally-invariant slowest operator:

Hamiltonian (see (\ref{Ham}))

\end{itemize}

\item \textbf{Magnetization}

\begin{itemize}
\item For local slowest operator:

\be M^{(0)}_{x,y,z} = \sum_{i=0}^{N-1} \sigma_{x,y,z}^{(i)}\ee

\item For translationally-invariant slowest operator:

\be \sum_{i=0}^{L-1} M^{(i)}_{x,y,z}\ee
where  $M^{(i)}_{x,y,z}$ has support on sites $i \dots i+N-1$.

\end{itemize}

{\small We often denote magnetization as magnetization1, magnetization2 or magnetization3. They correspond to $\sigma_x$, $\sigma_y$ and $\sigma_z$ magnetizations respectively.}

\end{enumerate}

\begin{figure*}
\centering
\begin{subfigure}{0.47\textwidth}
    \includegraphics[scale=0.22]{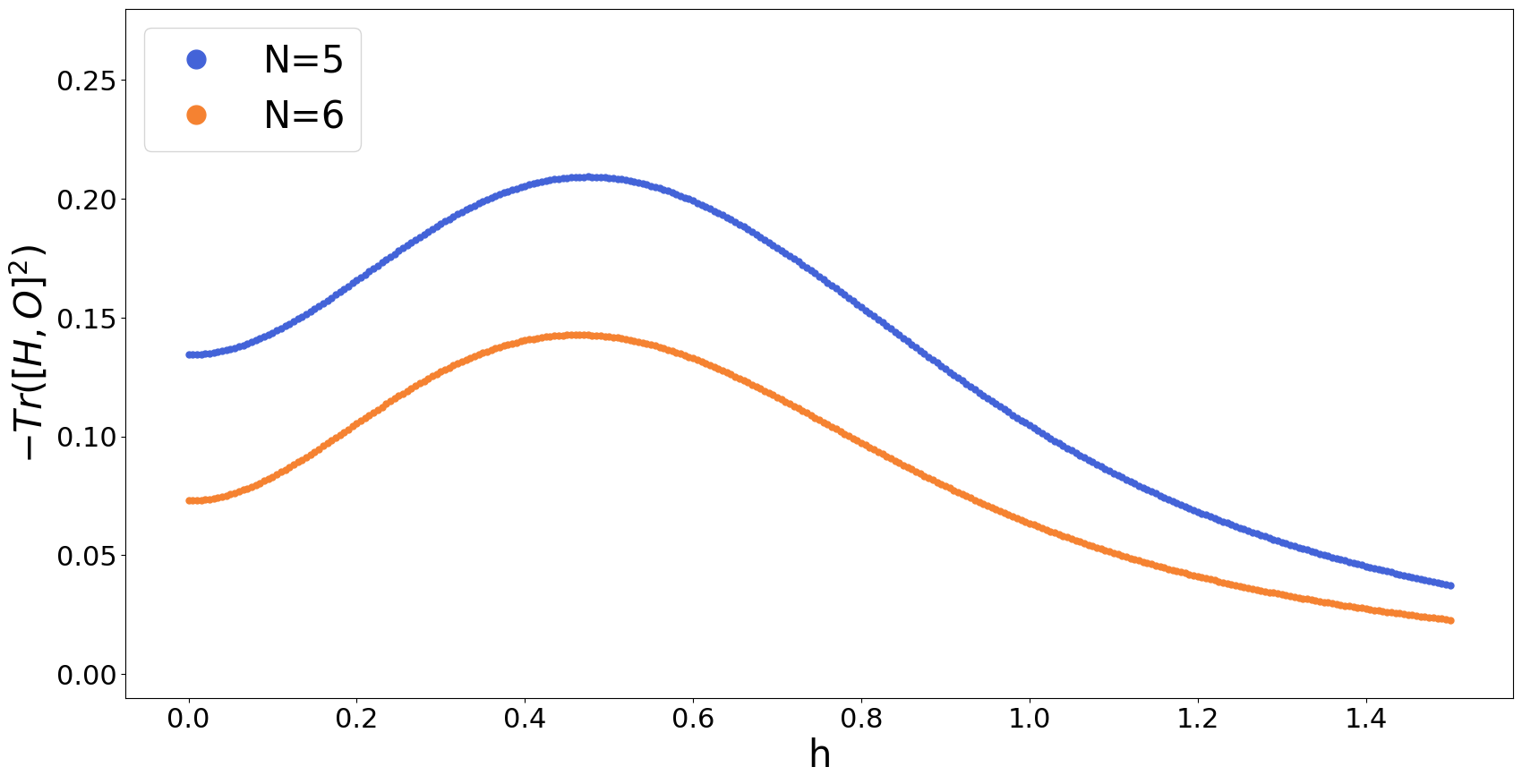}
    \caption{Local ($g=1.05$, various $h$)}
\end{subfigure}
\hfill
\begin{subfigure}{0.47\textwidth}
    \includegraphics[scale=0.22]{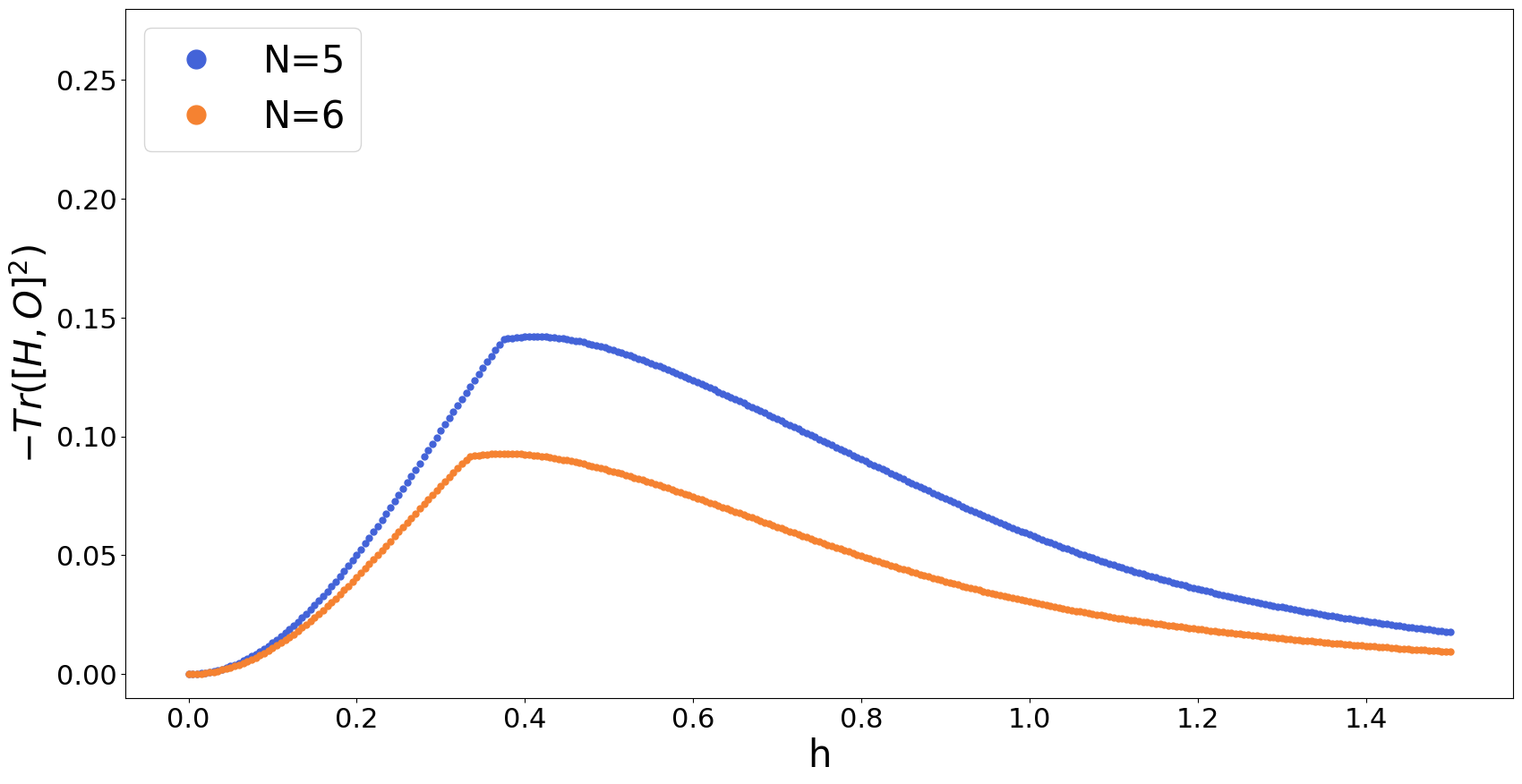}
    \caption{Translationally-invariant ($g=1.05$, various $h$)}
\end{subfigure}

\hfil

\begin{subfigure}{0.47\textwidth}
    \includegraphics[scale=0.22]{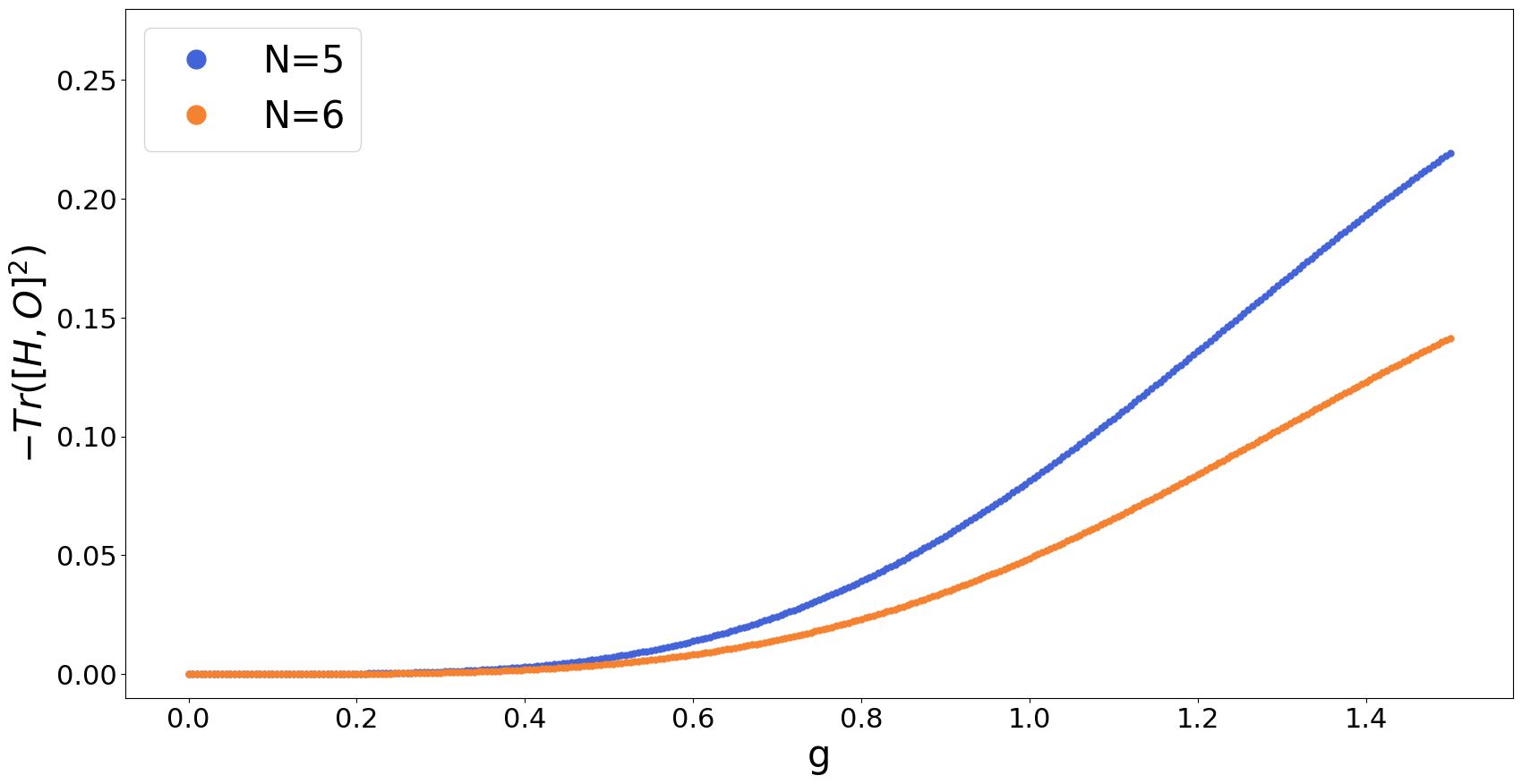}
    \caption{Local ($h=1.05$, various $g$)}
\end{subfigure}
\hfill
\begin{subfigure}{0.47\textwidth}
    \includegraphics[scale=0.22]{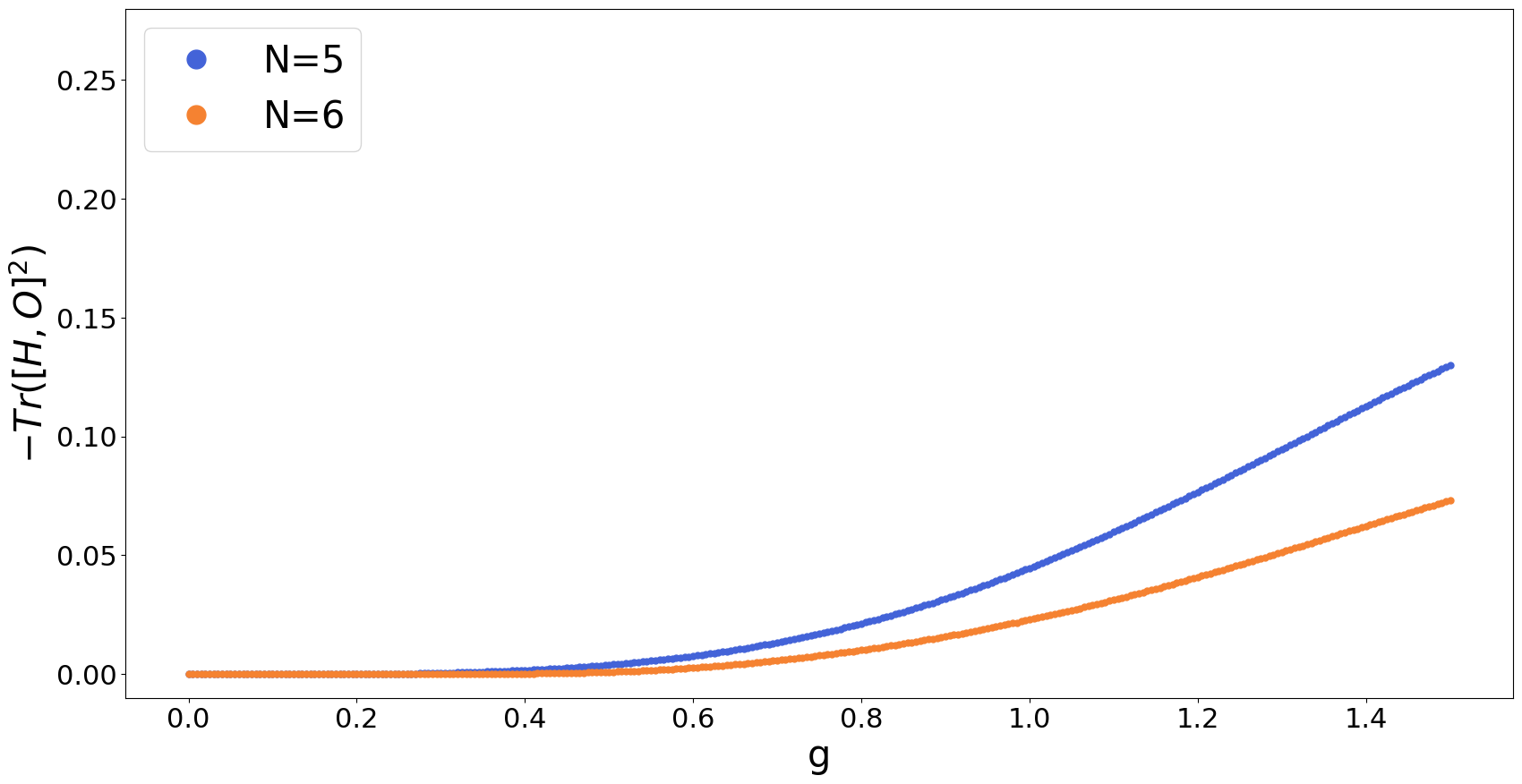}
    \caption{Translationally-invariant ($h=1.05$, various $g$)}
\end{subfigure}

\hfil

\begin{subfigure}{0.47\textwidth}
    \includegraphics[scale=0.22]{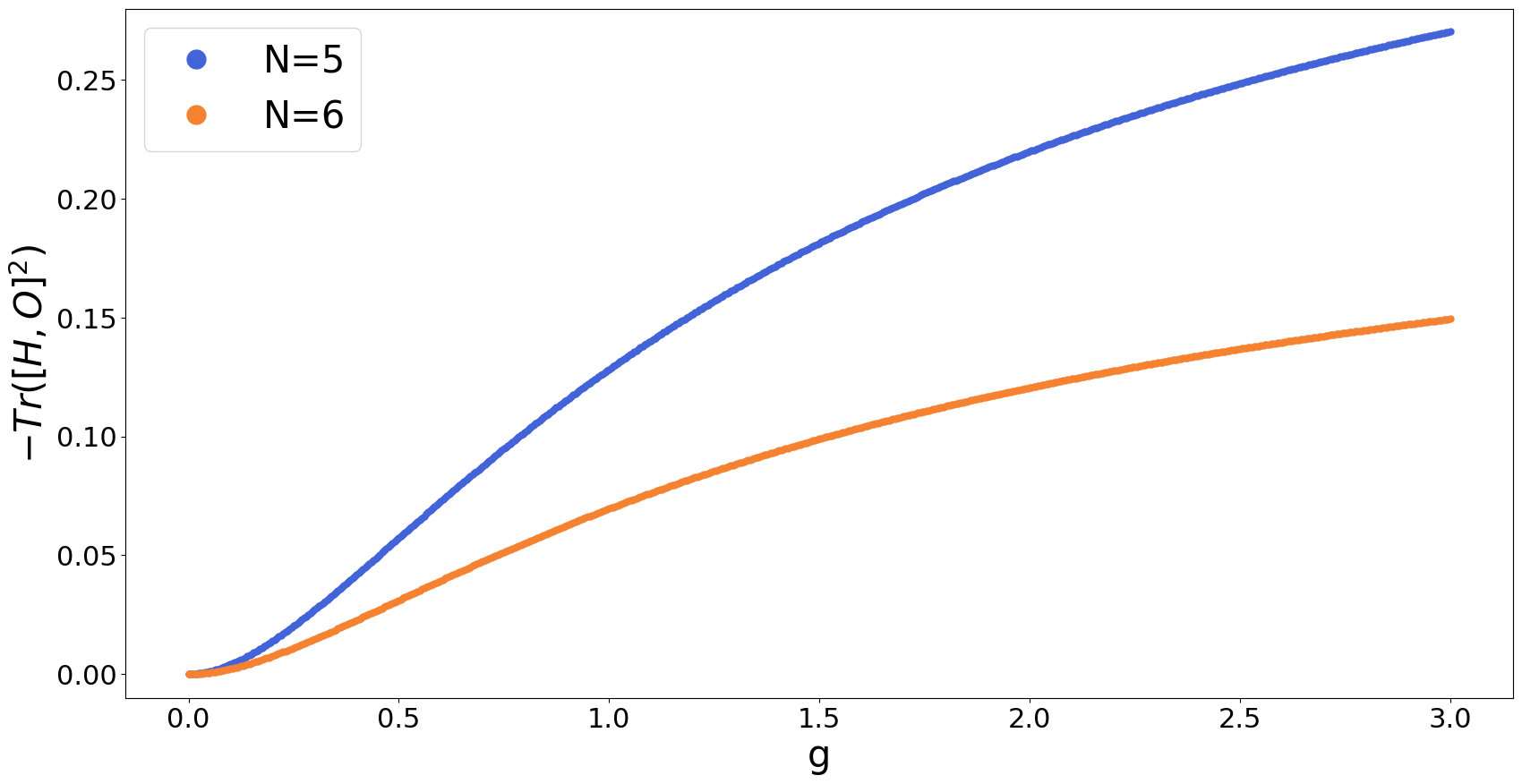}
    \caption{Local ($h=0$, various $g$)}
\end{subfigure}
\hfill
\begin{subfigure}{0.47\textwidth}
    \includegraphics[scale=0.22]{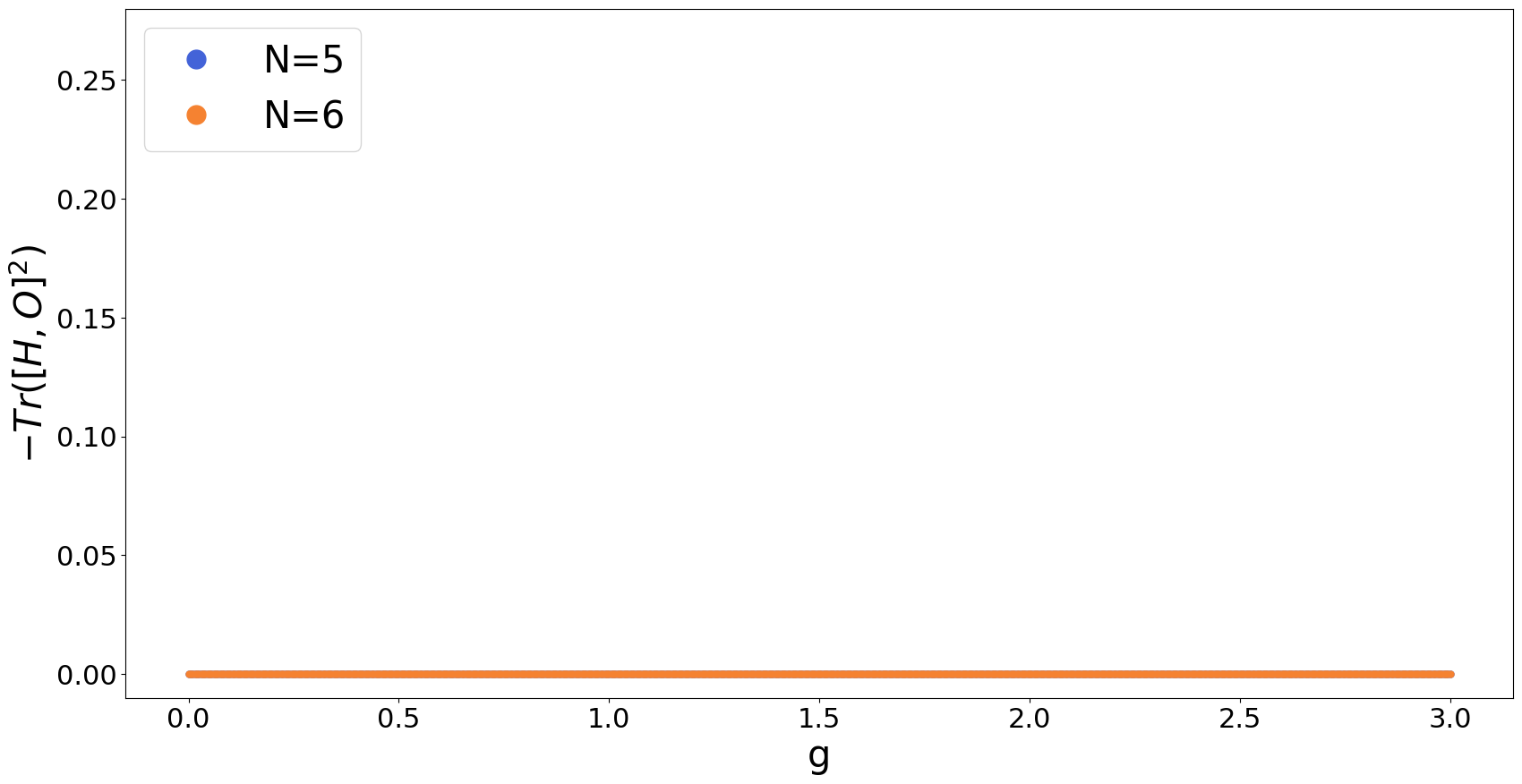}
    \caption{Translationally-invariant ($h=0$, various $g$)}
\end{subfigure}
        
\caption{Scaling of  $-\Tr [H,O]^2$ with $g$ and $h$ for local and translationally-invariant slowest operators (${N=5,6}$).}



\label{scaling_h_g}
\end{figure*}

\begin{figure*}
\centering
\begin{subfigure}{0.47\textwidth}
    \includegraphics[scale=0.22]{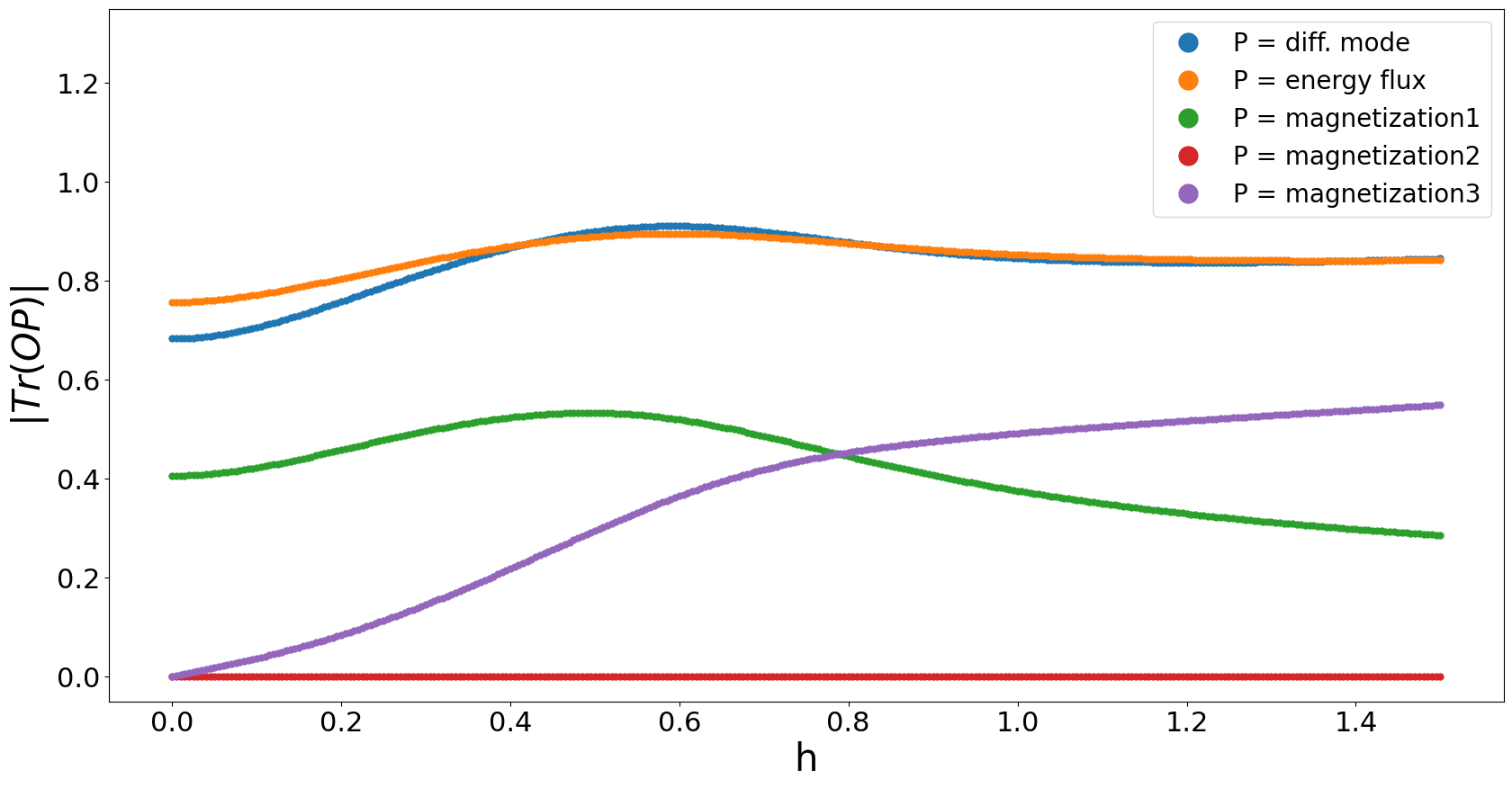}
    \caption{Local ($g=1.05$, various $h$)}
\end{subfigure}
\hfill
\begin{subfigure}{0.47\textwidth}
    \includegraphics[scale=0.22]{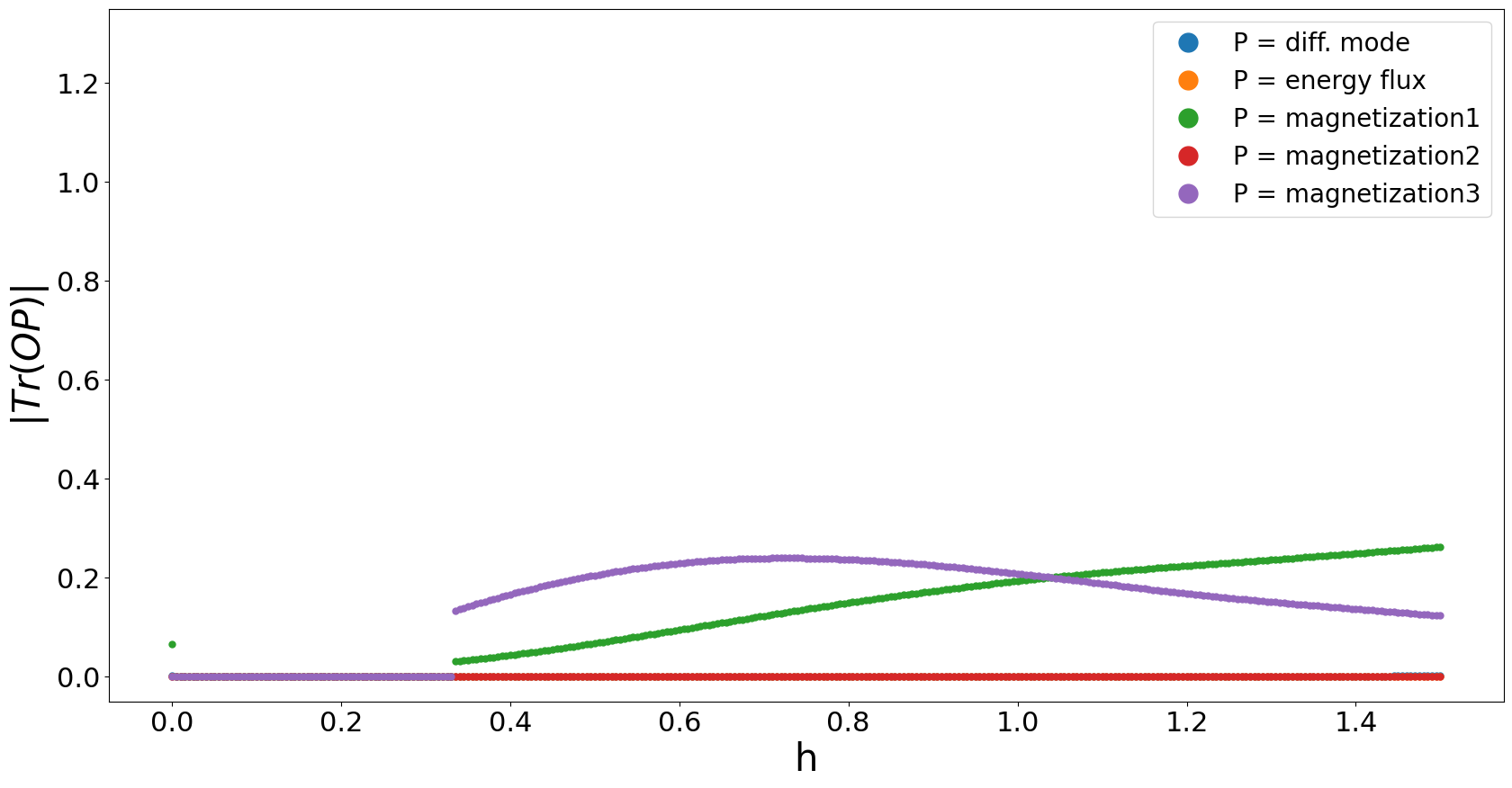}
    \caption{Translationally-invariant ($g=1.05$, various $h$)}
\end{subfigure}
\hfill
\begin{subfigure}{0.47\textwidth}
    \includegraphics[scale=0.22]{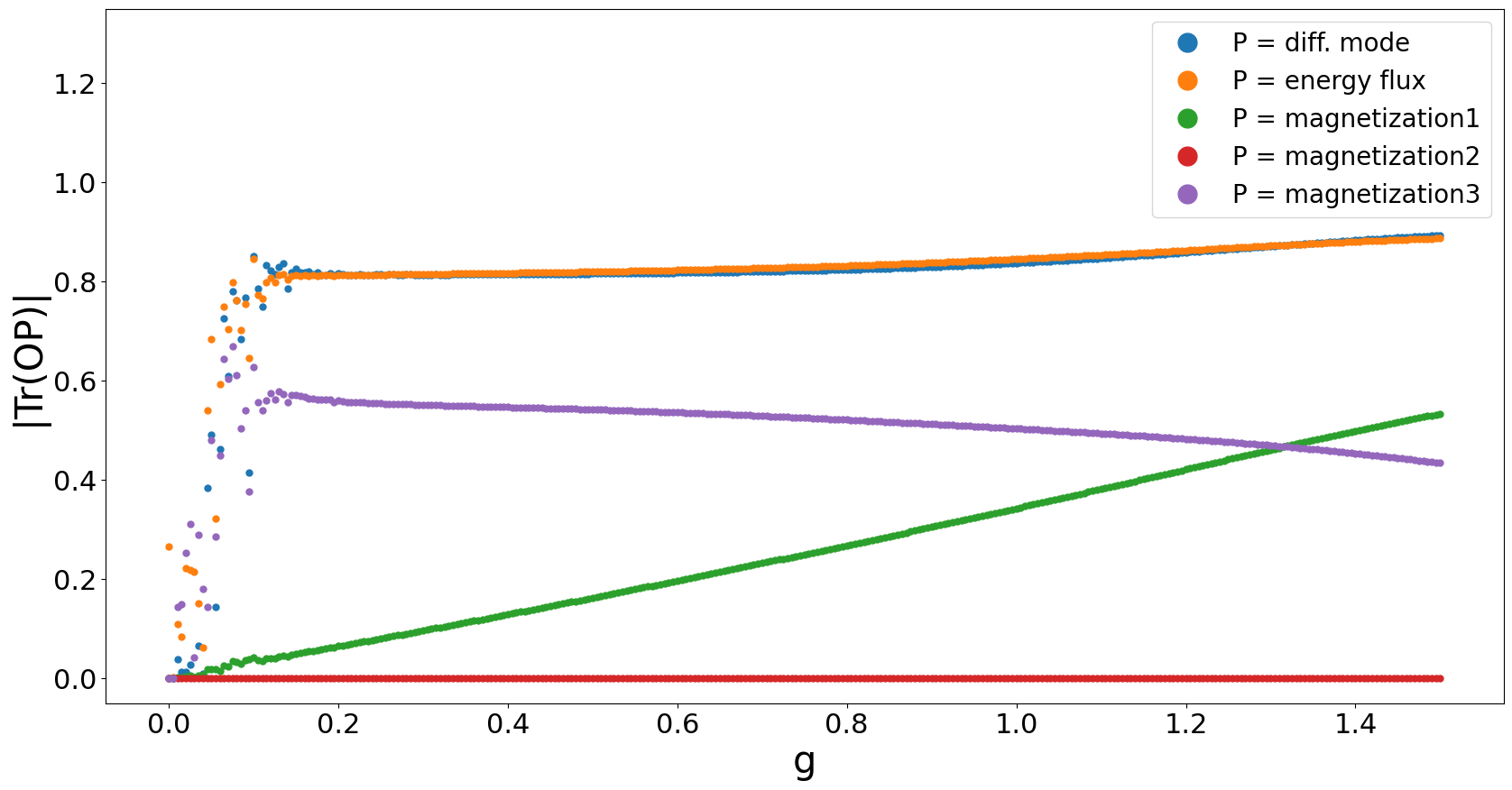}
    \caption{Local ($h=1.05$, various $g$)}
\end{subfigure}
\hfill
\begin{subfigure}{0.47\textwidth}
    \includegraphics[scale=0.22]{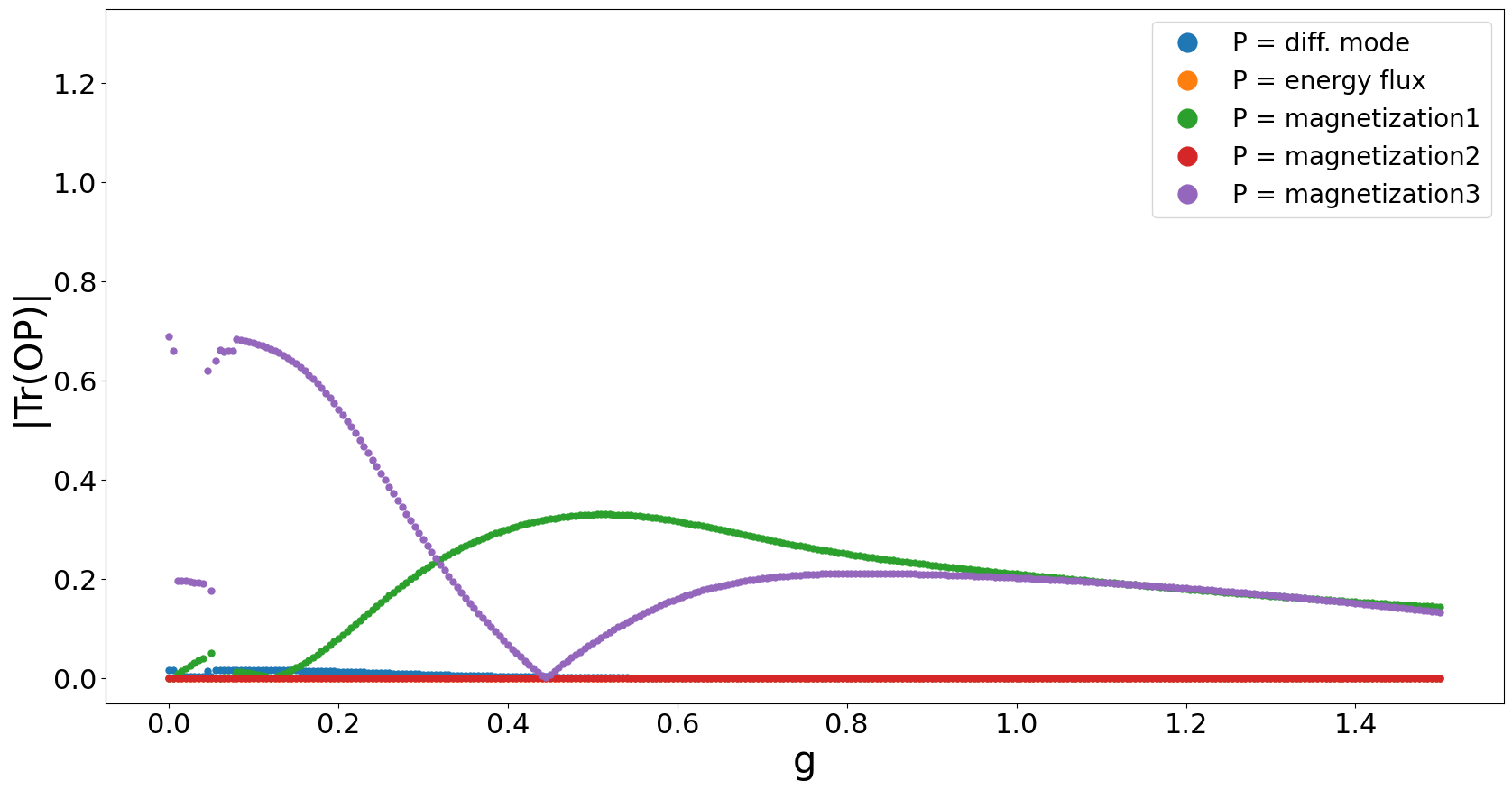}
    \caption{Translationally-invariant ($h=1.05$, various $g$)}
\end{subfigure}
\hfill
\begin{subfigure}{0.47\textwidth}
    \includegraphics[scale=0.22]{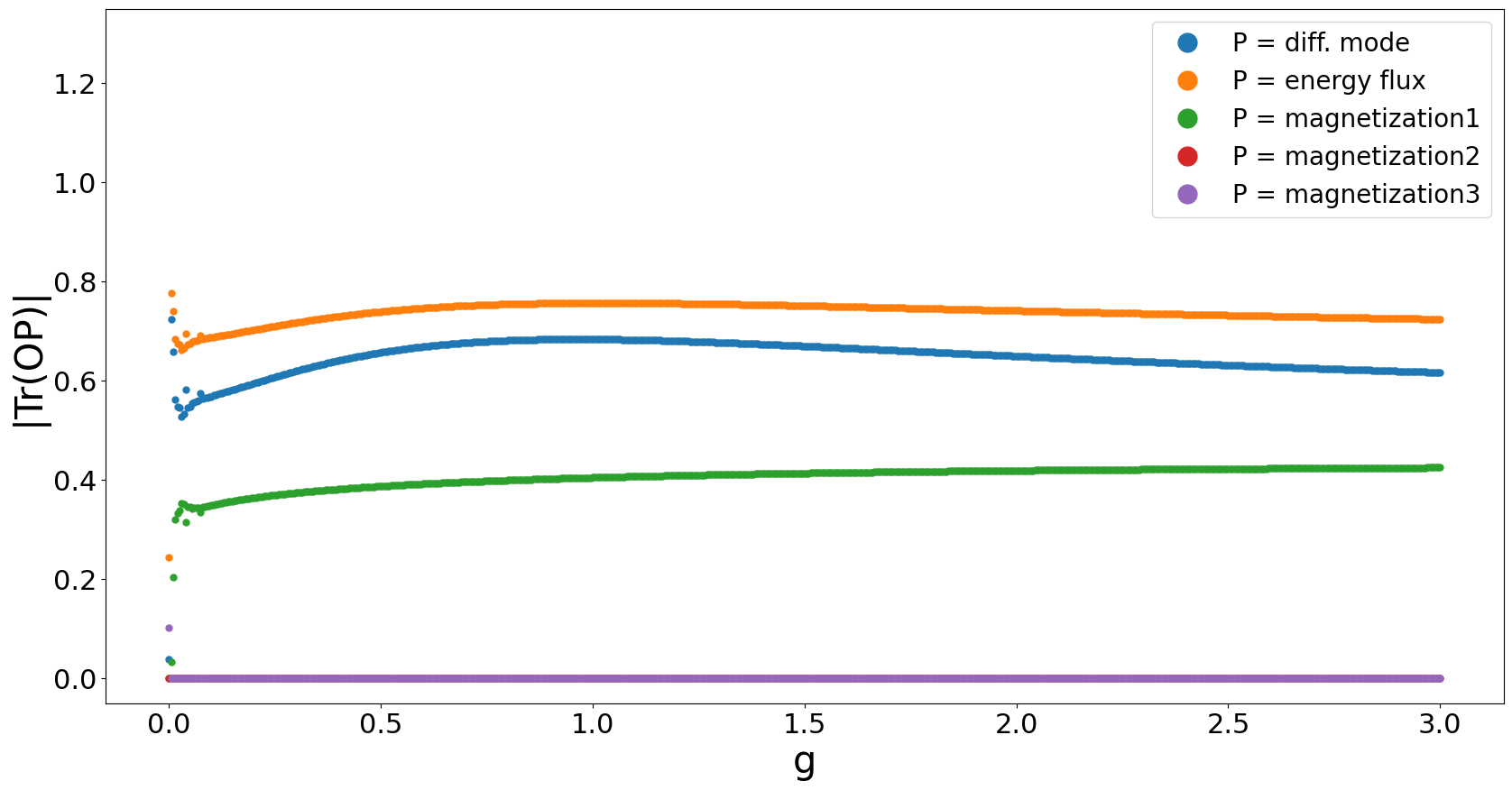}
    \caption{Local ($h=0$, various $g$)}
\end{subfigure}
\hfill
\begin{subfigure}{0.47\textwidth}
    \includegraphics[scale=0.22]{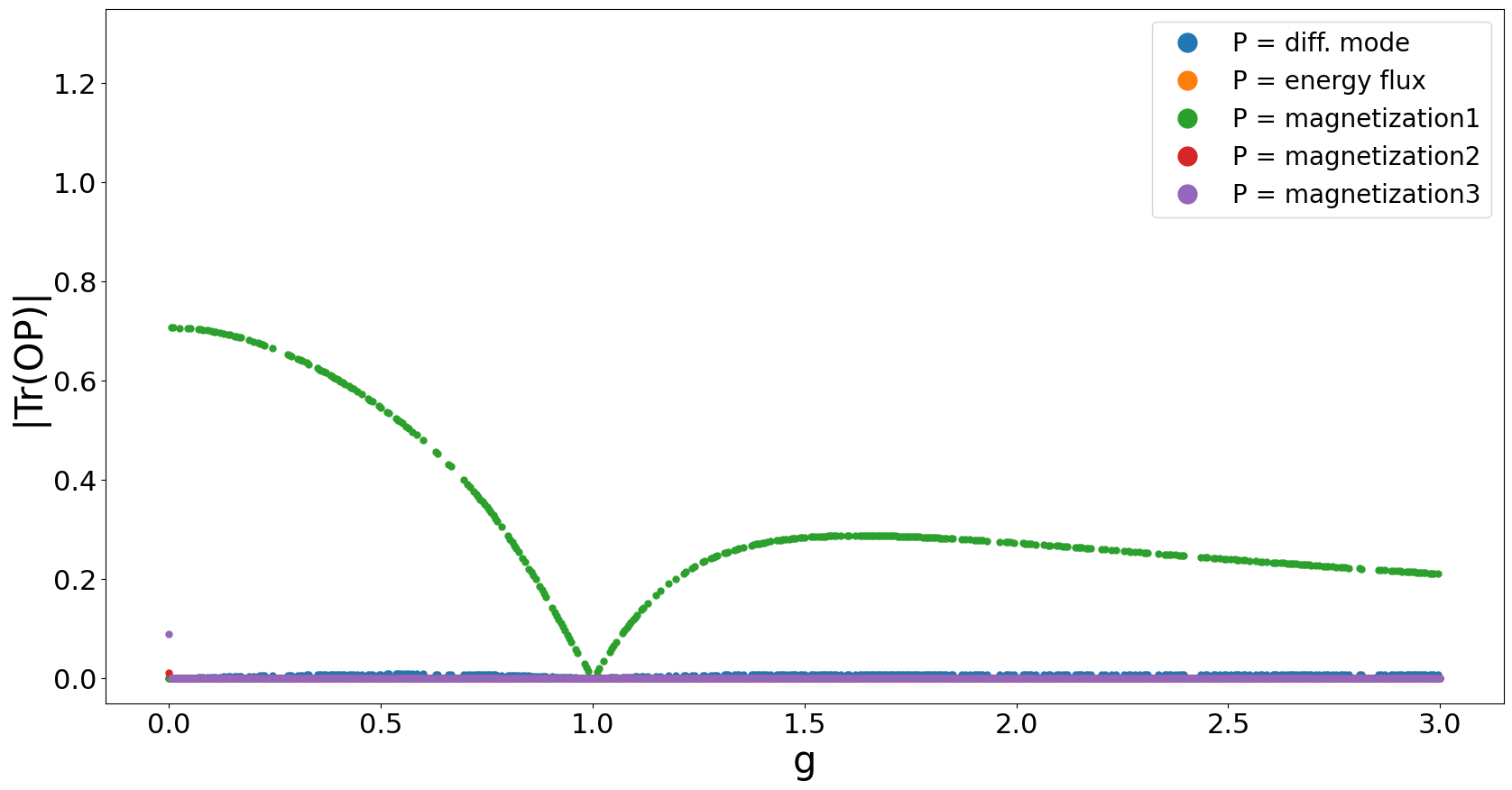}
    \caption{Translationally-invariant ($h=0$, various $g$)}
\end{subfigure}

\caption{Overlap $\Tr (OP)$ of the slowest operator $O$ and a probe operator $P$ as a function of parameters $g$ and $h$. We take $P$ as diffusion mode, energy flux, magnetization1,2,3 (all operators are defined in the text). $N=6$.}
\label{overlap}
\end{figure*}

We plot $-\Tr [H,O]^2$ as a function of $g$ and $h$ in Fig. \ref{scaling_h_g}, and overlap between the slowest operator and probe operators as a function of $g, h$ - in Fig. \ref{overlap}. The subplots on the left - (a,c,e) - correspond to the local slowest operator, while subplots on the right - (b,d,f) - to the translationally-invariant one. The subplots (a) and (b) are concerned with the non-integrable case of fixed $g=1.05$ and various $h$, such that $h=0$ corresponds to the integrable limit. Similarly, subplots (c) and (d) correspond to the non-integrable case of fixed $h=1.05$ and various $g$, while $g=0$ provides the integrable limit. The two bottom subplots (e) and (f) correspond to the integrable case of $h=0$.

\subsection{The results}

Here we make conclusions from Figs. \ref{scaling_h_g} and \ref{overlap}.

\subsubsection{The translationally-invariant slowest operator corresponds to an integral of motion, but the local slowest operator does not (as $h \rightarrow 0$).}

The translationally-invariant operator becomes an integral of motion, as $h\rightarrow 0$ or $g\rightarrow 0$, since the quantity $-\Tr[H,O]^2$ approaches $0$ (see Fig. \ref{scaling_h_g} (b), (d)).

The local slowest operator, on the opposite, does not correspond to an integral of motion, as $h\rightarrow 0$ (see Fig. \ref{scaling_h_g} (a)). (But it does, as $g\rightarrow 0$, see Fig. \ref{scaling_h_g} (c).)


\subsubsection{The curve $-\Tr[H,O]^2$ as a function of $g$ ($h$) has a shape of a deformed bell.}

The curve $-\Tr[H,O]^2$ clearly decreases for big $h$ in Figs. \ref{scaling_h_g} (a,b). But we expect a similar behavior in Figs. \ref{scaling_h_g} (c,d,e). The reason is that, as we go to large $g$ or $h$, the corresponding term in Hamiltonian (\ref{Ham}) dominates over the others, and $\sigma_x$ or $\sigma_z$ starts to play a role of the local slowest operator, and $\sum_i \sigma^{(i)}_x$ or $\sum_i \sigma^{(i)}_z$ - of the translationally-invariant one.

\subsubsection{The translationally-invariant operator changes its nature at a specific $h$, when $g$ is fixed.}

It can be seen in Fig. \ref{scaling_h_g} (b) and Fig. \ref{overlap} (b). The value of $h$ depends on the support size $N$. (For $N=6$, $h^* \sim 0.33$ in both of the graphs.)

Before the transition, the slowest operator does not look like any probe operator we propose.

\subsubsection{For an integrable system of $h=0$, there are translationally-invariant integrals of motion, but no local ones.}


It can be seen in Fig. \ref{scaling_h_g} (f), and then Fig. \ref{scaling_h_g} (e) and (a).


\subsubsection{The local slowest operator looks very much like diffusion mode/energy flux.}

It can be seen in Fig. \ref{overlap} (a,c,e) (orange and blue curves). It is also indicated by the clear correlation between the contribution of magnetization and the corresponding terms in Hamiltonian. The greater $g$ is, the bigger is the contribution of magnetization1 (green curve), the greater $h$ is - the bigger is the contribution of magnetization3 (purple curve).

On the other hand, overlap with magnetization2 is $0$, and it corresponds to the absence of $\sigma_y$ in Hamiltonian.


\subsubsection{The translationally-invariant slowest operator looks most like magnetization1 or magnetization3, but their contribution is not big.}

It can be seen in Fig. \ref{overlap} (b,d,f).

We also note that the overlap with diffusion mode or energy flux is $0$, since $\Tr (H O) = 0$ by definition. Because of the latter, one would expect the "anticorrelation" with respect to the contributions of $g$ and $h$ into Hamiltonian. But it is not the case. This behavior is observed for big $h$ in Fig. \ref{overlap} (b) and big $g$ in Fig. \ref{overlap} (d), but it is clearly violated for small $g$ in Fig. \ref{overlap} (d).

\subsubsection{The translationally-invariant operator is slower than the local one.}

It is clearly seen in Figs.  \ref{scaling_h_g} (a-f), since the curves for the translationally-invariant operator are lower than those for the local one.

\subsubsection{There is a translationally-invariant integral of motion that looks like magnetization1 to a great extent (for $h=0$).}

It can be seen in Fig. \ref{overlap} (f). There are several translationally-invariant integrals of motion, and the algorithm finds one of them. We observe the one corresponding to some overlap with magnetization1. We clearly see the special point at $g=1$. This point corresponds to the known phase transition from ordered ($g<1$) to disordered ($g>1$) phase \cite{chakrabarti2008quantum} (the transition happens when the coefficient in front of $-ZZ$ becomes the same as the coefficient in $gX$).

\section{Dependence of the slowest operator on support size $N$ \label{sec_N}}

The quantity $-\Tr[H,O]^2$ defines the rate of dynamics of the slowest operator $O$. But, if we wish to estimate how the operator $O$ expands over the chain, we need to calculate the dependence of $-\Tr[H,O]^2$ on the support size $N$ of the operator $O$. To understand this, one has to decompose $\Tr O(t) O(0)$ around $t=0$:
\be \Tr O(t) O(0) = 1 - \left( -\Tr([H,O]^2) \right) \frac{t^2}{2} + \dots \label{lambdacorrfuncconnection}\ee

We see that $-\Tr[H,O]^2$ plays a role of $\tau^{-2}$, where $\tau$ is the characteristic time scale of the expansion of $O$ over the chain (at least, for early times). Therefore, one can estimate the rate of expansion by calculating the dependence: $-\Tr[H,O]^2 \sim \frac{1}{N^k}$. The bigger $k$ is - the bigger is the time scale $\tau$ of the expansion of $O$ over the chain, i.e. the slower is the expansion.

In particular, we aim to find, if the rate of expansion corresponds to diffusion, or it is ballistic, or other.

\subsection{The physical quantities}

We calculate the following quantities.

\subsubsection{ $\log -\Tr([H,O]^2$ as a function of $\log N$}

The slope of this graph is equal to $-k$. The less the value $(-k)$ is - the slower is the expansion of $O$ over the chain.

\begin{figure*}
\centering
\begin{subfigure}{0.47\textwidth}
    \includegraphics[scale=0.22]{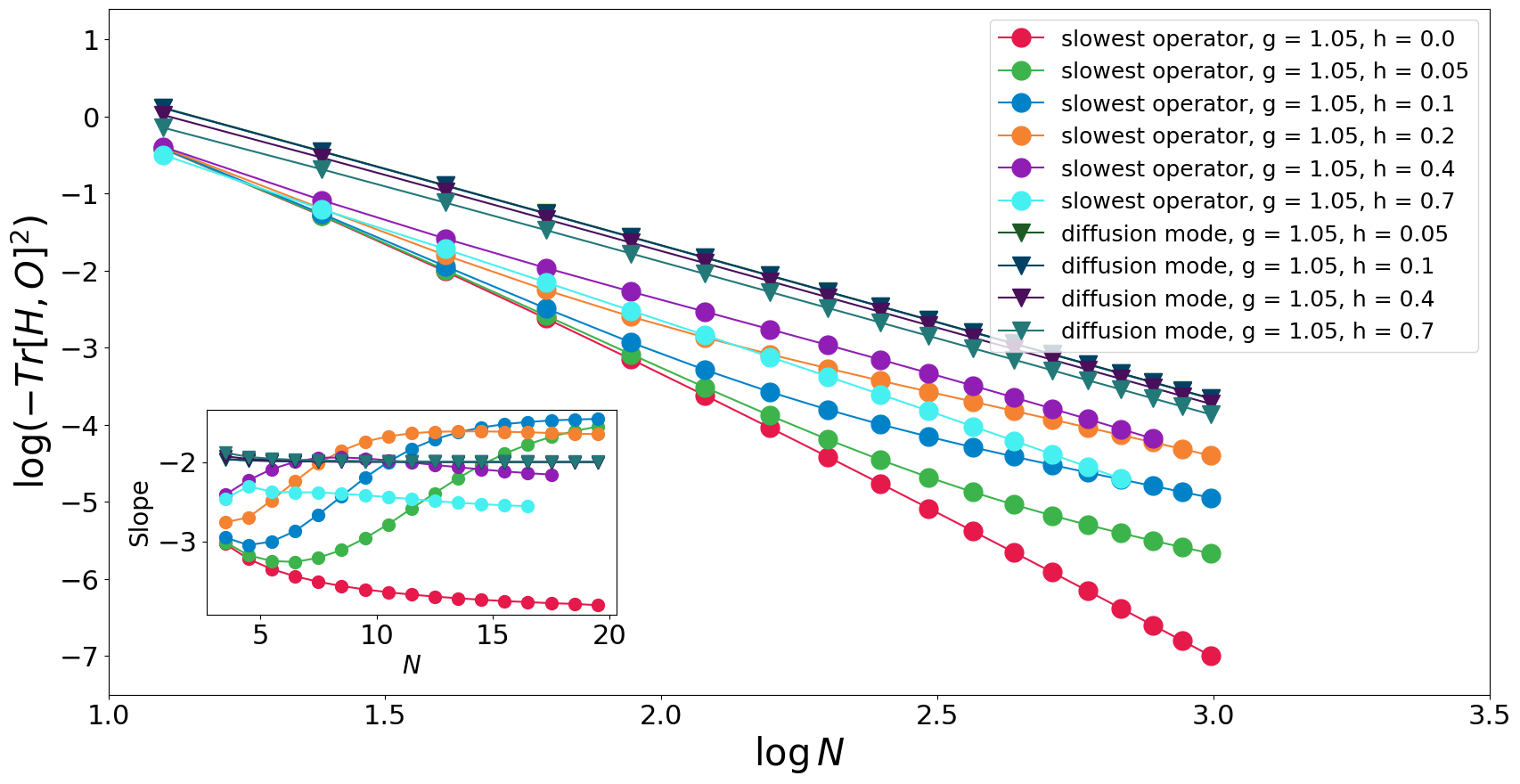}
    \caption{Local ($g=1.05$, various $h$)}
\end{subfigure}
\hfill
\begin{subfigure}{0.47\textwidth}
    \includegraphics[scale=0.22]{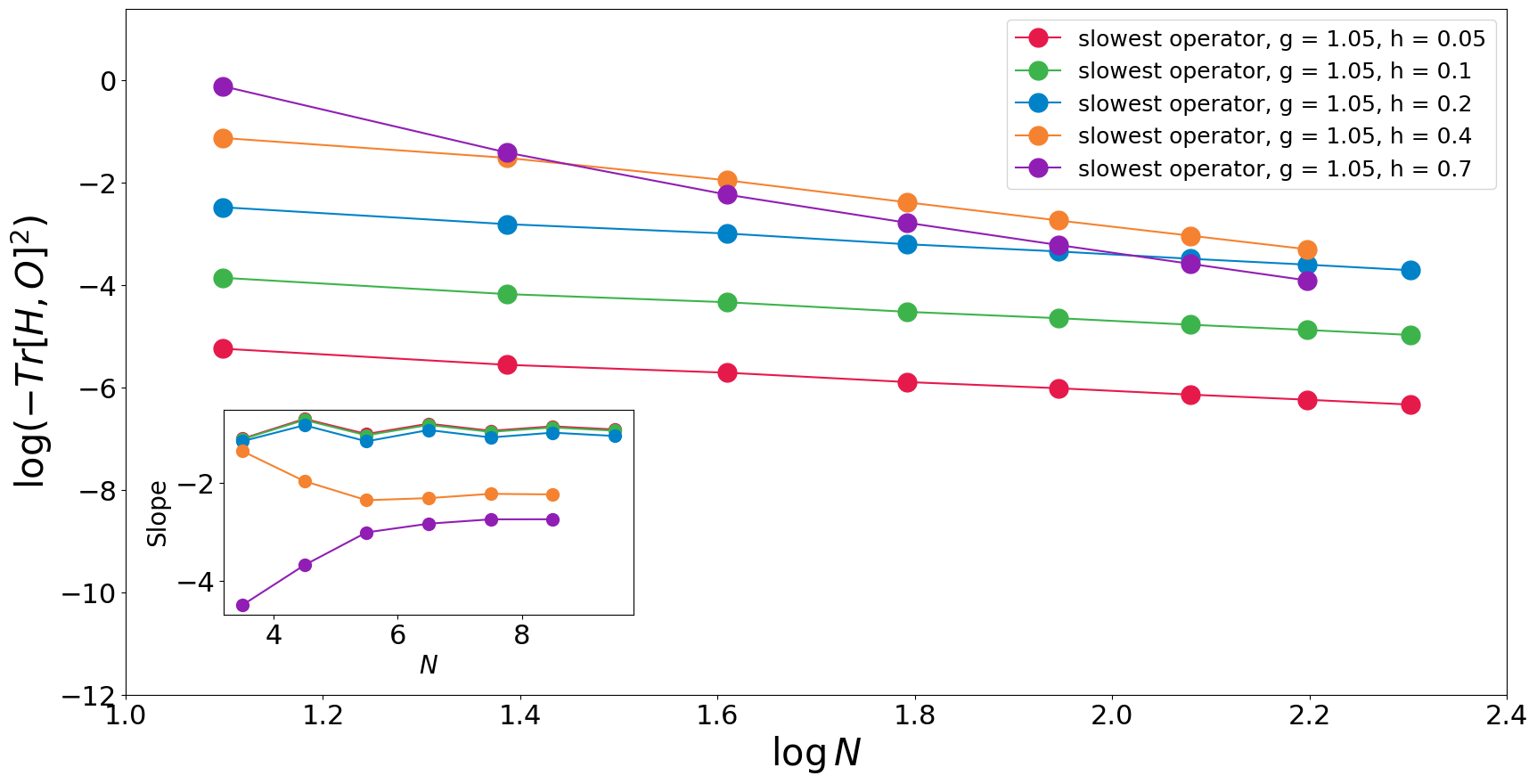}
    \caption{Translationally-invariant ($g=1.05$, various $h$)}
\end{subfigure}

\hfil

\begin{subfigure}{0.47\textwidth}
    \includegraphics[scale=0.22]{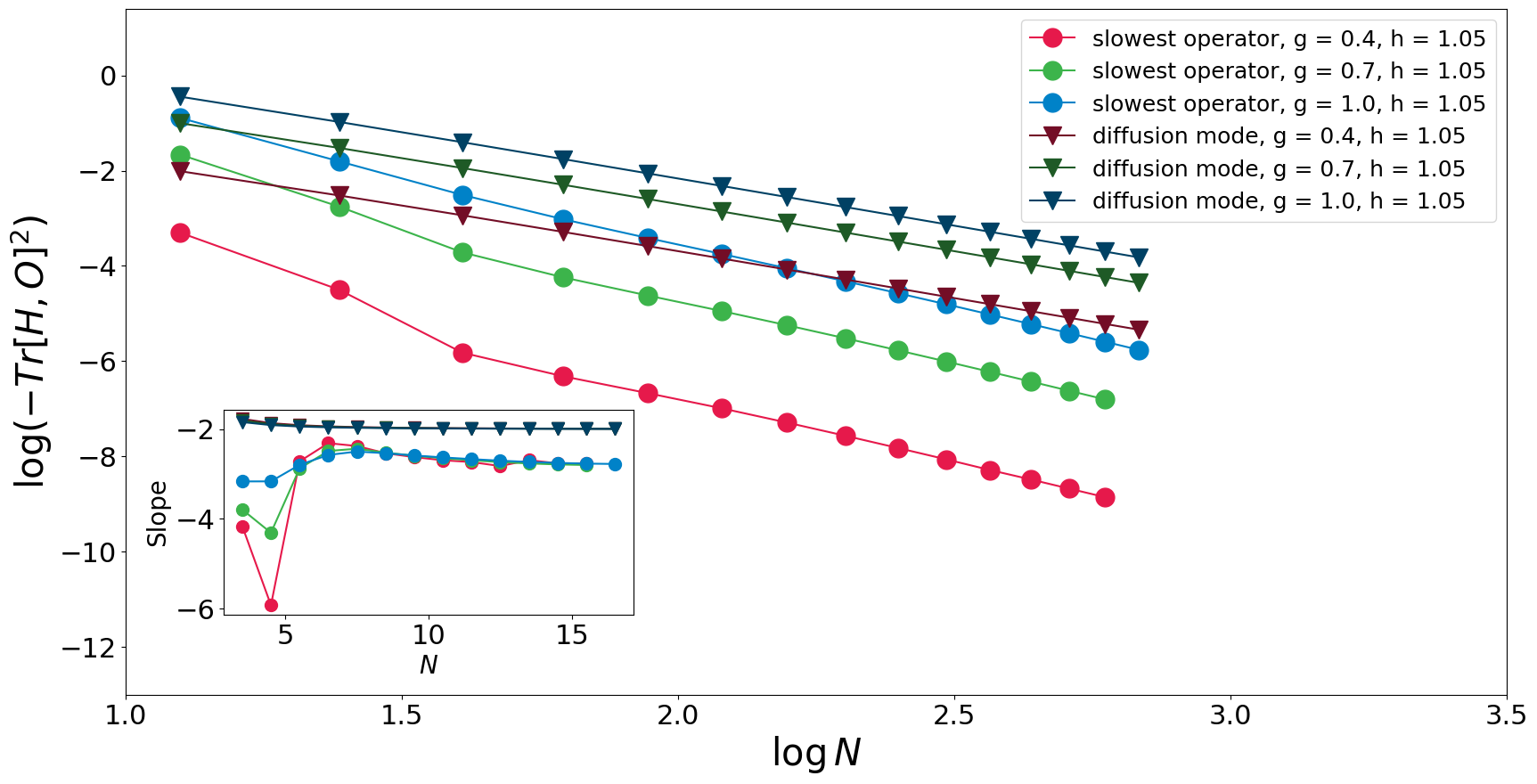}
    \caption{Local ($h=1.05$, various $g$)}
\end{subfigure}
\hfill
\begin{subfigure}{0.47\textwidth}
    \includegraphics[scale=0.22]{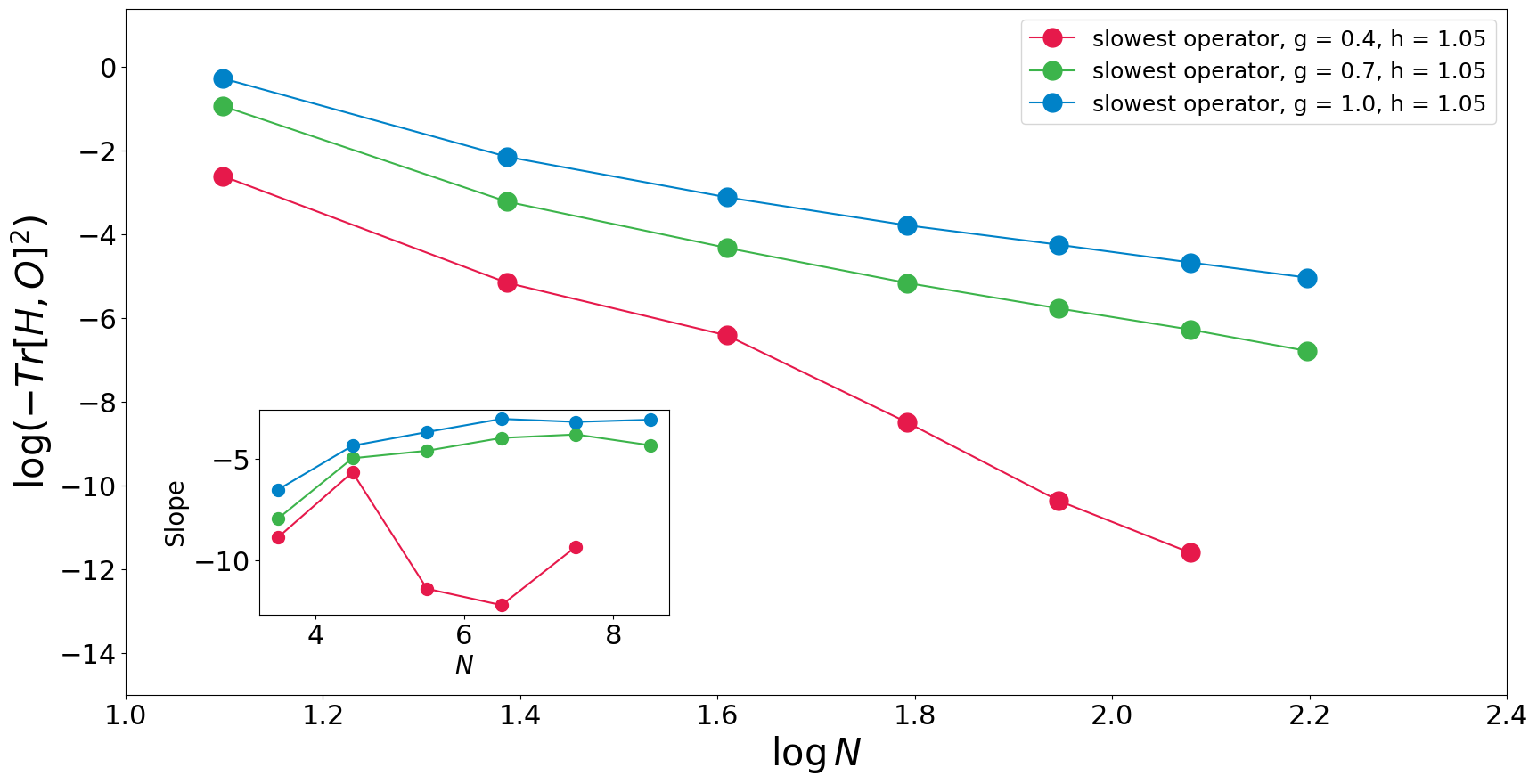}
    \caption{Translationally-invariant ($h=1.05$, various $g$)}
\end{subfigure}

\hfil

\begin{subfigure}{0.47\textwidth}
    \includegraphics[scale=0.22]{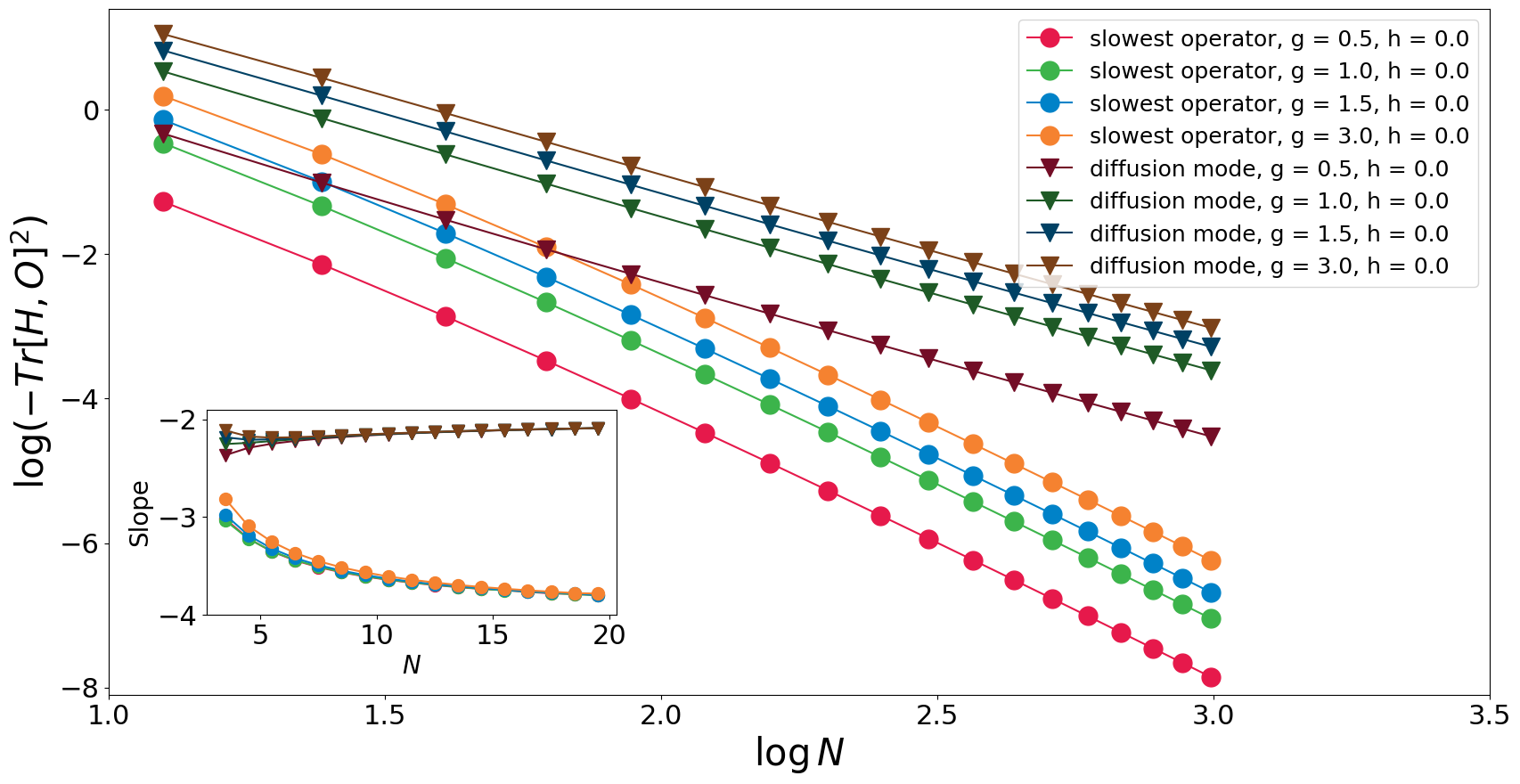}
    \caption{Local ($h=0$, various $g$)}
\end{subfigure}
\hfill
\begin{subfigure}{0.47\textwidth}
    \includegraphics[scale=0.22]{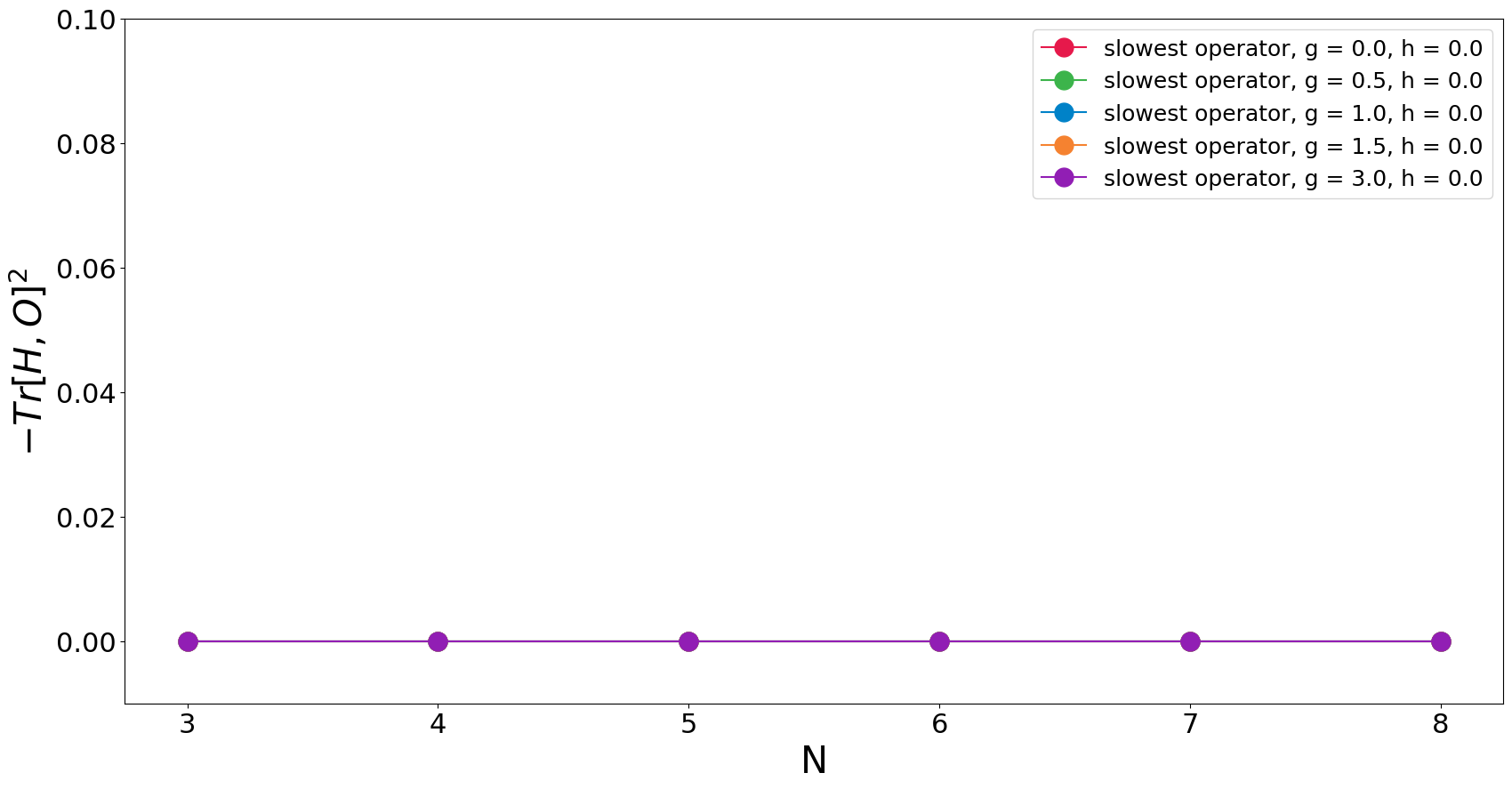}
    \caption{Translationally-invariant ($h=0$, various $g$)}
\end{subfigure}

\caption{Scaling of  $\log (-\Tr [H,O]^2)$ with $\log N$ for local and translationally-invariant slowest operators for various $g$ and $h$. Inset: "instant" slope of this graph (calculated from 2 nearby points) as a function of $N$. For local slowest operator, we compare the value of  $-\Tr [H,O]^2$ with that of diffusion mode (dark triangular points). [(f): we plot $-\Tr [H,O]^2$ as a function of $N$ to show that, for $h=0$, $-\Tr [H,O]^2$ is near $0$ for all $g$ and all $N$.] }
\label{scalingN}
\end{figure*}

In Fig. \ref{scalingN}, we show $\log -\Tr([H,O]^2$ as a function of $\log N$. The instant slope of this graph (for two nearby values $N$ and $N+1$) is depicted in the inset, as a function of $N$. Thus, we can see how the rate of expansion changes with the support size $N$ of the slowest operator.

In the case of the local slowest operator, we also plot the function $\log -\Tr([H,\widetilde{E^{(0)}}]^2$ for diffusion mode $\widetilde{E^{(0)}}$. It is defined as before, but with coefficients $\{a_i,b_i,c_i\}$:
\be E^{(0)} = \sum_{i=0}^{N-2} \cos \left(-\frac{\pi}{2} +\frac{i+\frac{1}{2}}{N}\pi \right) (- \boldsymbol{a_i} \sigma^{(i)}_z \sigma^{(i+1)}_z) + \nn\ee
\be + \sum_{i=0}^{N-1} \cos \left( -\frac{\pi}{2} + \frac{i}{N}\pi \right) (\boldsymbol{b_i} h \sigma^{(i)}_z + \boldsymbol{c_i} g \sigma^{(i)}_x ) \ee

We optimize the coefficients $\{a_i,b_i,c_i\}$, so that $-\Tr([H,\widetilde{E^{(0)}}]^2$ is minimal, provided the normalization is fixed: $\Tr \left( E^{(0)} \right)^2 = 1$.

We compare the slowest operator with the diffusion mode, because they have a big overlap (see above). We calculate their rate of expansion.

We do not plot the diffusion mode in the case of translationally-invariant operator, because they have different nature: the translationally-invariant slowest operator is orthogonal to Hamiltonian by definition ($\Tr H O = 0$).

\subsubsection{Overlap between $O$ and probe operators as a function of $N$}

Here, the probe operators are the same as in the previous section. We focus on the overlap of $O$ with the diffusion mode. If this overlap is significant, then the rate of expansion is close to that of diffusion.

%

\begin{figure*}
\centering
\begin{subfigure}{0.16\textwidth}
    \includegraphics[scale=0.17]{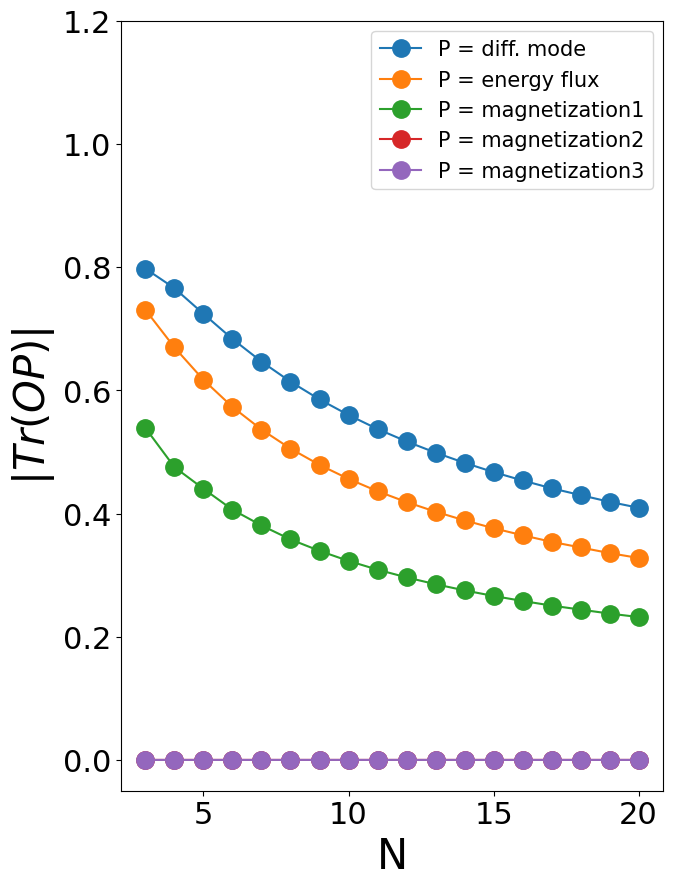}
    \caption{Loc. ($g=1.05, h = 0.0$)}
\end{subfigure}
\hfill
\begin{subfigure}{0.16\textwidth}
    \includegraphics[scale=0.17]{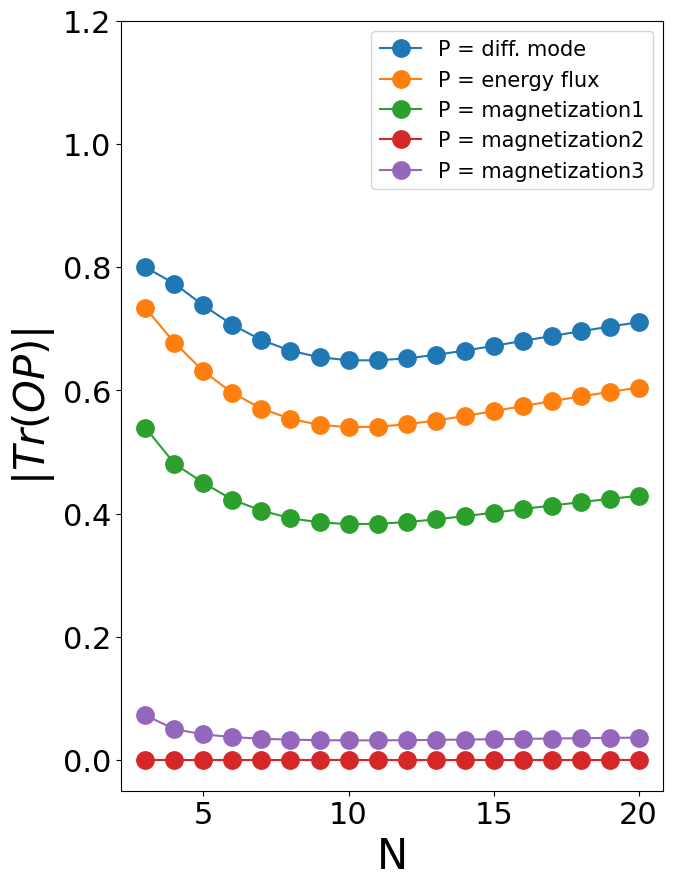}
    \caption{Loc. ($g=1.05, h = 0.1$)}
\end{subfigure}
\hfill
\begin{subfigure}{0.16\textwidth}
    \includegraphics[scale=0.17]{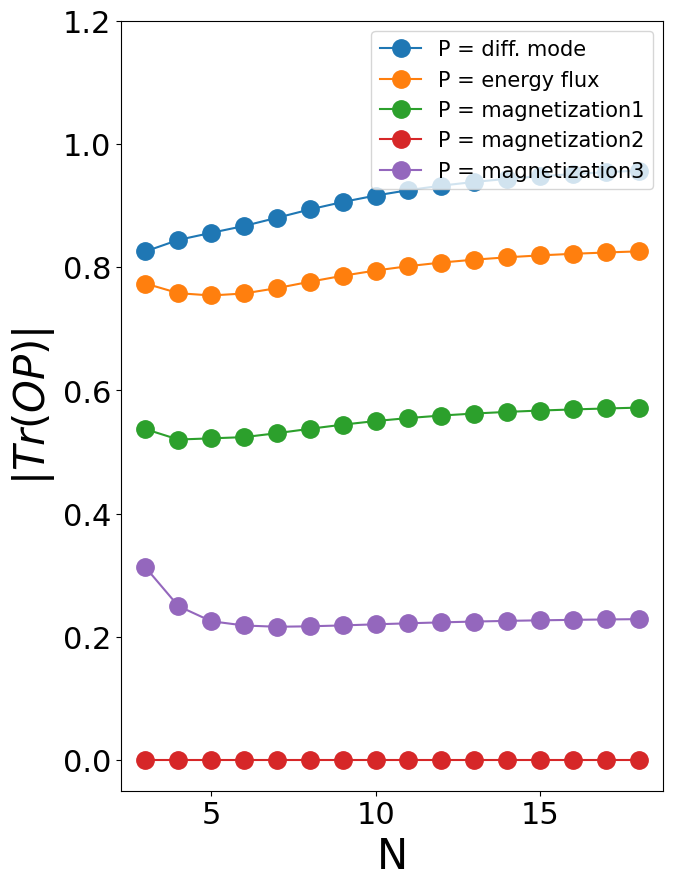}
    \caption{Loc. ($g=1.05, h = 0.4$)}
\end{subfigure}
\hfill
\begin{subfigure}{0.16\textwidth}
    \includegraphics[scale=0.17]{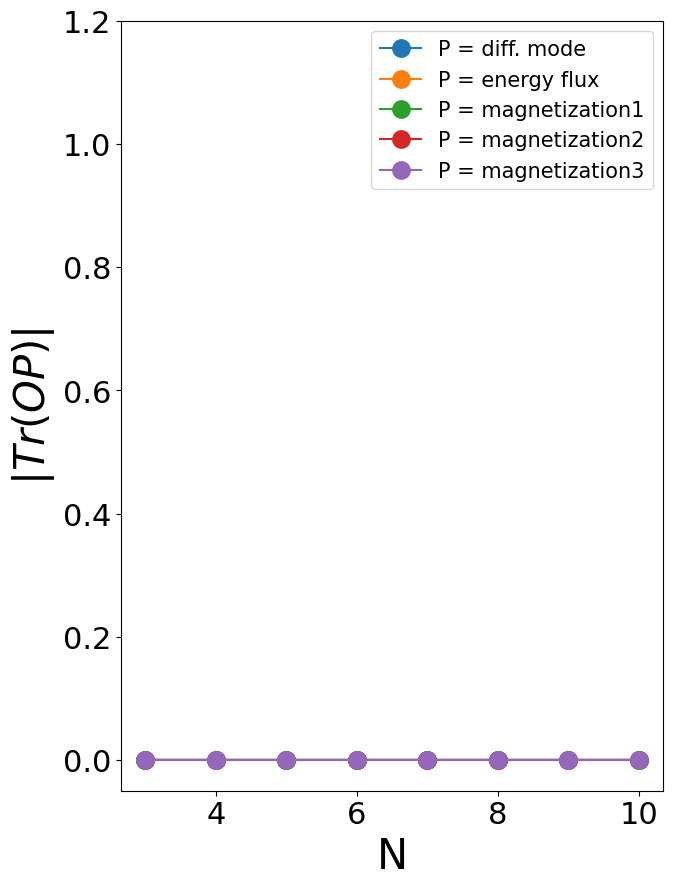}
    \caption{Tr.-inv. ($g=1.05, h = 0.1$)}
\end{subfigure}
\hfill
\begin{subfigure}{0.16\textwidth}
    \includegraphics[scale=0.17]{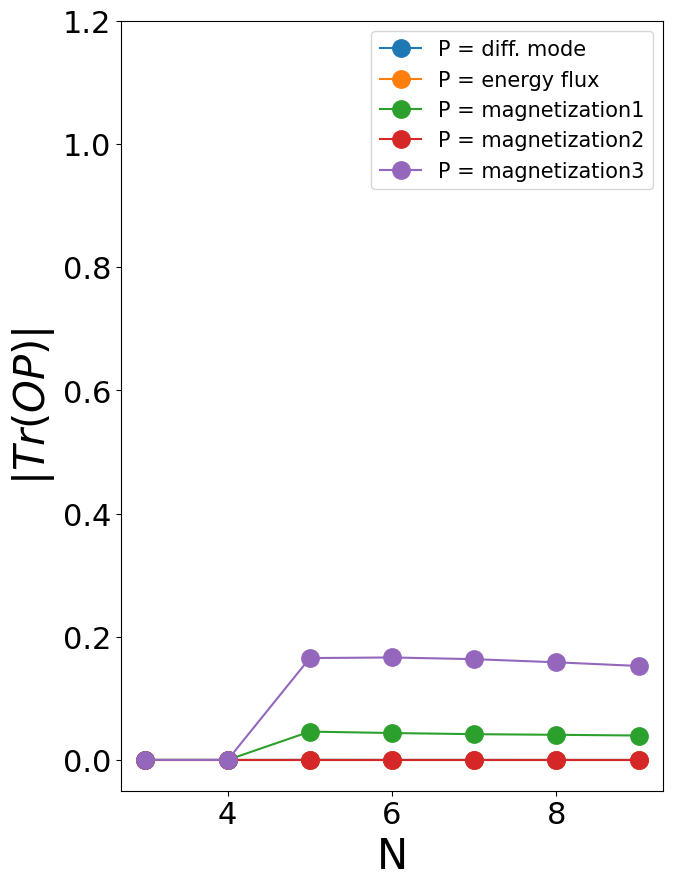}
    \caption{Tr.-inv. ($g=1.05, h = 0.4$)}
\end{subfigure}
\hfill
\begin{subfigure}{0.16\textwidth}
    \includegraphics[scale=0.17]{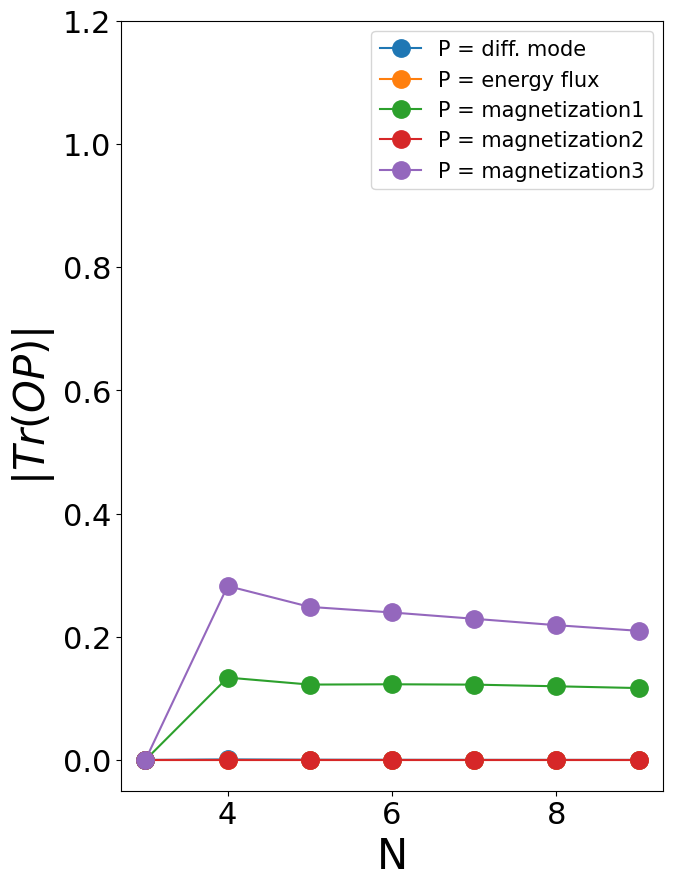}
    \caption{Tr.-inv. ($g=1.05, h = 0.7$)}
\end{subfigure}

\hfil

\begin{subfigure}{0.16\textwidth}
    \includegraphics[scale=0.17]{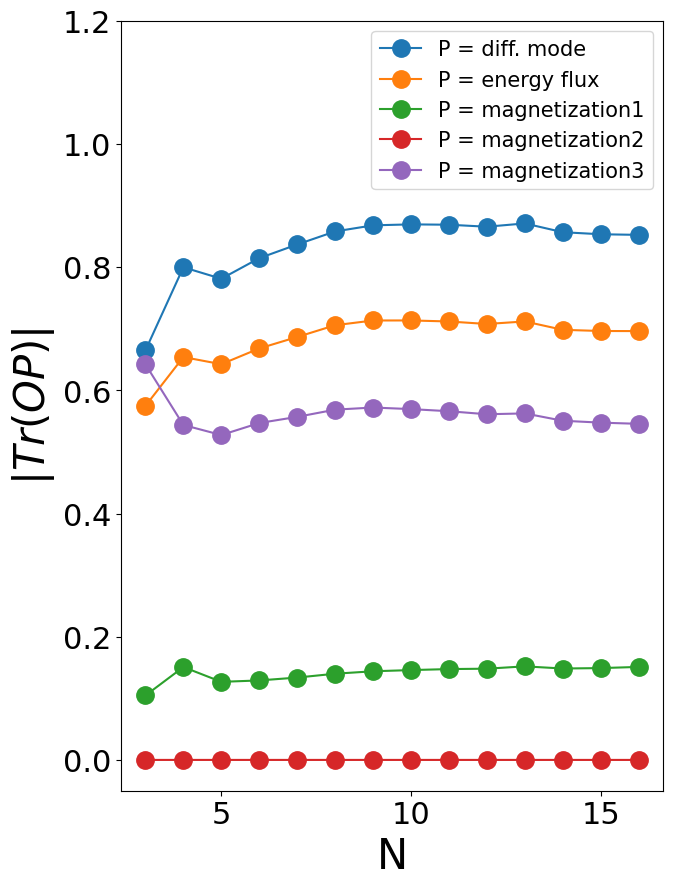}
    \caption{Loc. ($h = 1.05, g = 0.4$)}
\end{subfigure}
\hfill
\begin{subfigure}{0.16\textwidth}
    \includegraphics[scale=0.17]{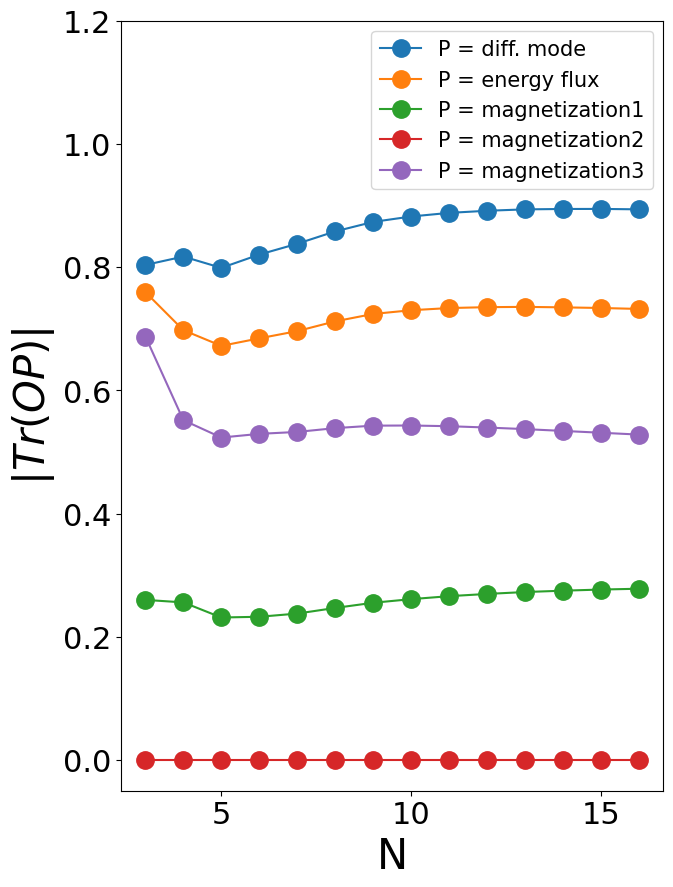}
    \caption{Loc. ($h = 1.05, g = 0.7$)}
\end{subfigure}
\hfill
\begin{subfigure}{0.16\textwidth}
    \includegraphics[scale=0.17]{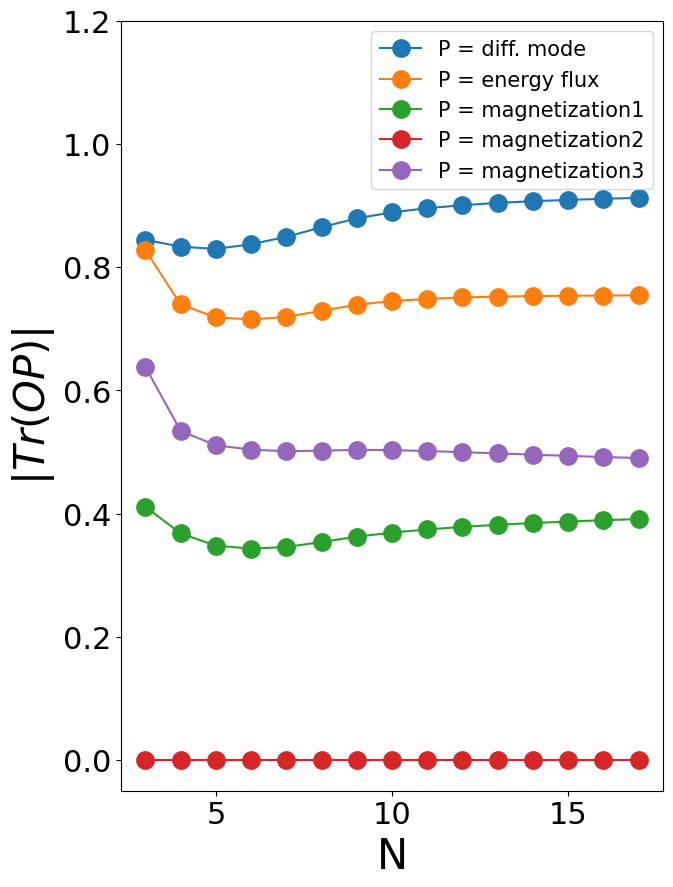}
    \caption{Loc. ($h = 1.05, g = 1.0$)}
\end{subfigure}
\hfill
\begin{subfigure}{0.16\textwidth}
    \includegraphics[scale=0.17]{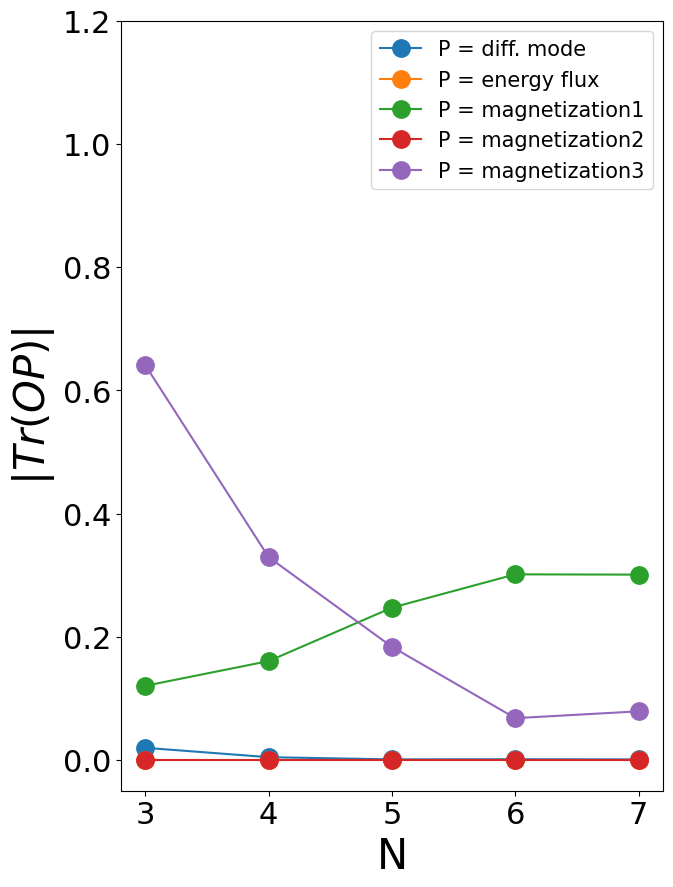}
    \caption{Tr.-inv. ($h = 1.05, g = 0.4$)}
\end{subfigure}
\hfill
\begin{subfigure}{0.16\textwidth}
    \includegraphics[scale=0.17]{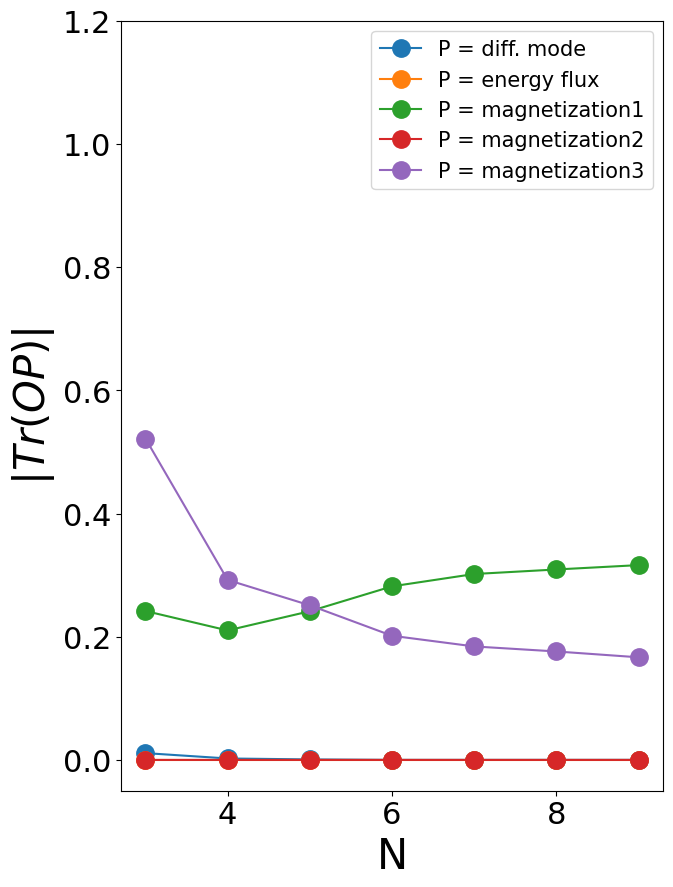}
    \caption{Tr.-inv. ($h = 1.05, g = 0.7$)}
\end{subfigure}
\hfill
\begin{subfigure}{0.16\textwidth}
    \includegraphics[scale=0.17]{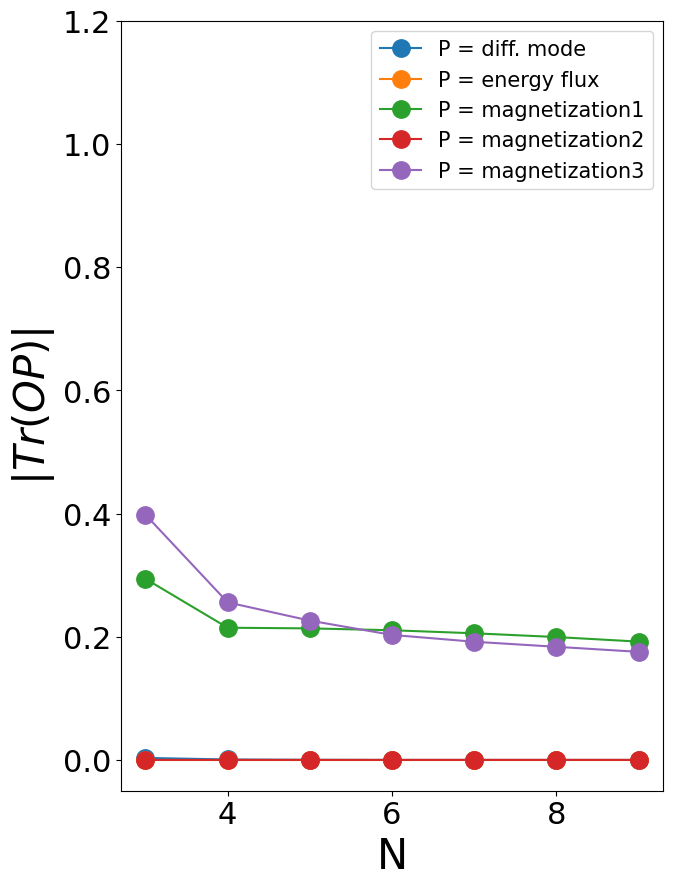}
    \caption{Tr.-inv. ($h = 1.05, g = 1.0$)}
\end{subfigure}

\hfil

\begin{subfigure}{0.16\textwidth}
    \includegraphics[scale=0.17]{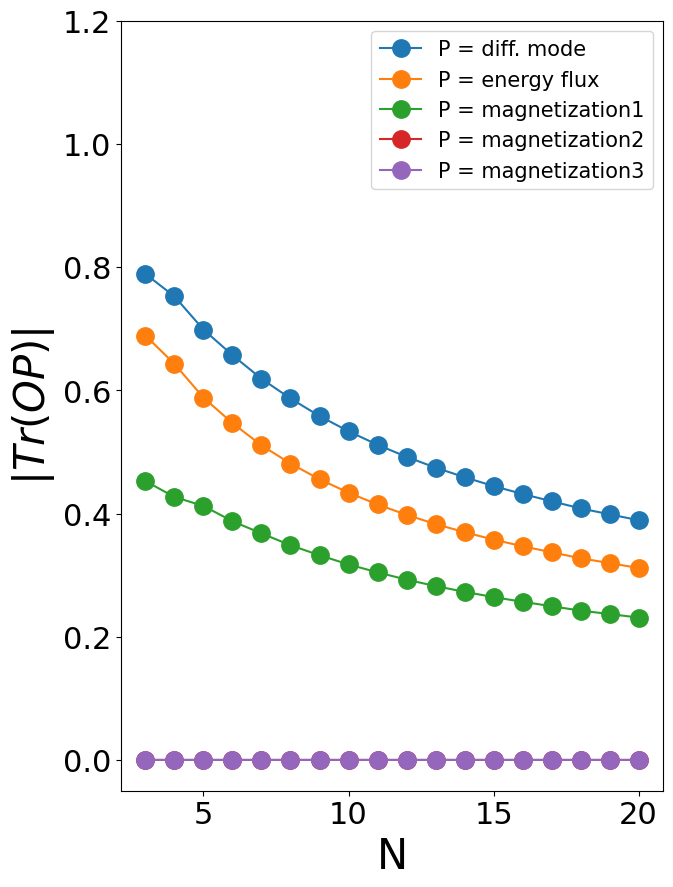}
    \caption{Loc. ($h = 0.0, g = 0.5$)}
\end{subfigure}
\hfill
\begin{subfigure}{0.16\textwidth}
    \includegraphics[scale=0.17]{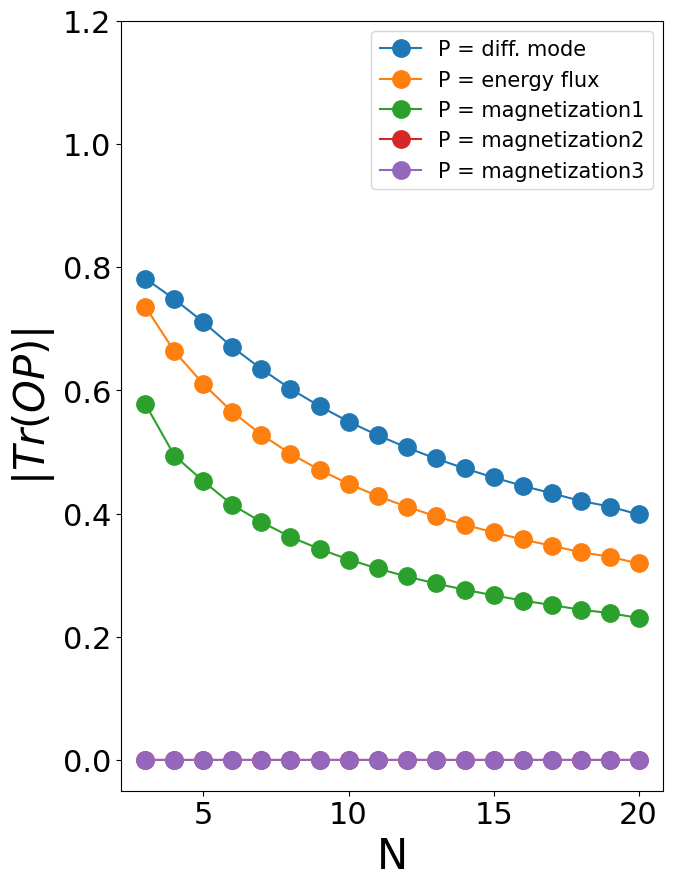}
    \caption{Loc. ($h = 0.0, g = 1.5$)}
\end{subfigure}
\hfill
\begin{subfigure}{0.16\textwidth}
    \includegraphics[scale=0.17]{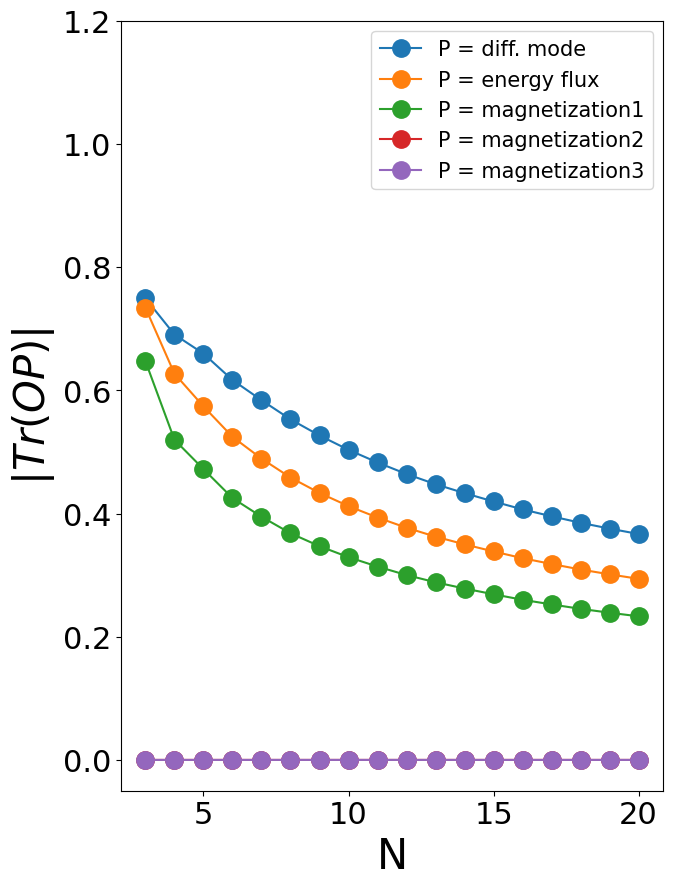}
    \caption{Loc. ($h = 0.0, g = 3.0$)}
\end{subfigure}
\hfill
\begin{subfigure}{0.16\textwidth}
    \includegraphics[scale=0.17]{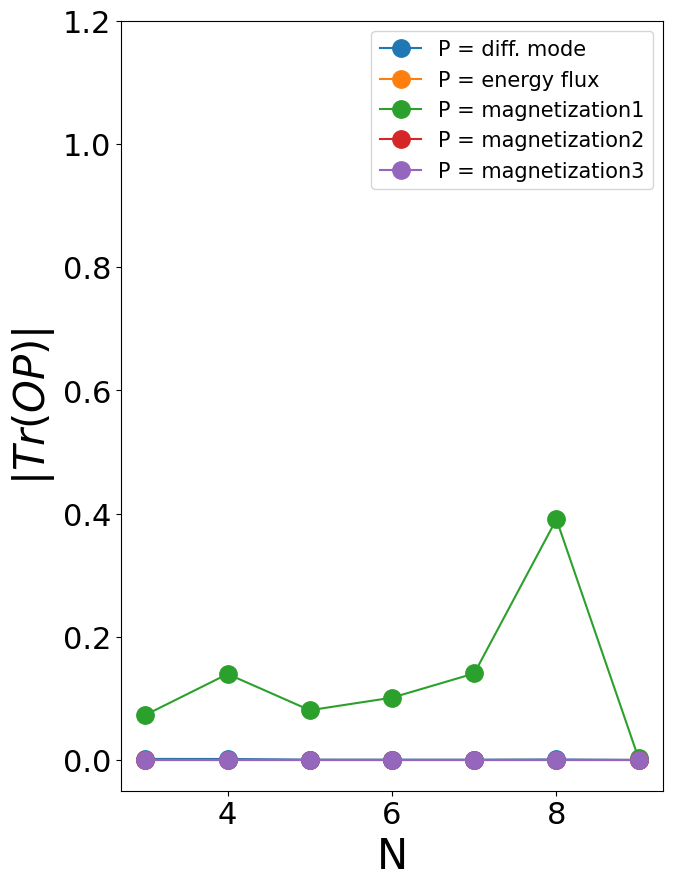}
    \caption{Tr.-inv. ($h = 0.0, g = 0.5$)}
\end{subfigure}
\hfill
\begin{subfigure}{0.16\textwidth}
    \includegraphics[scale=0.17]{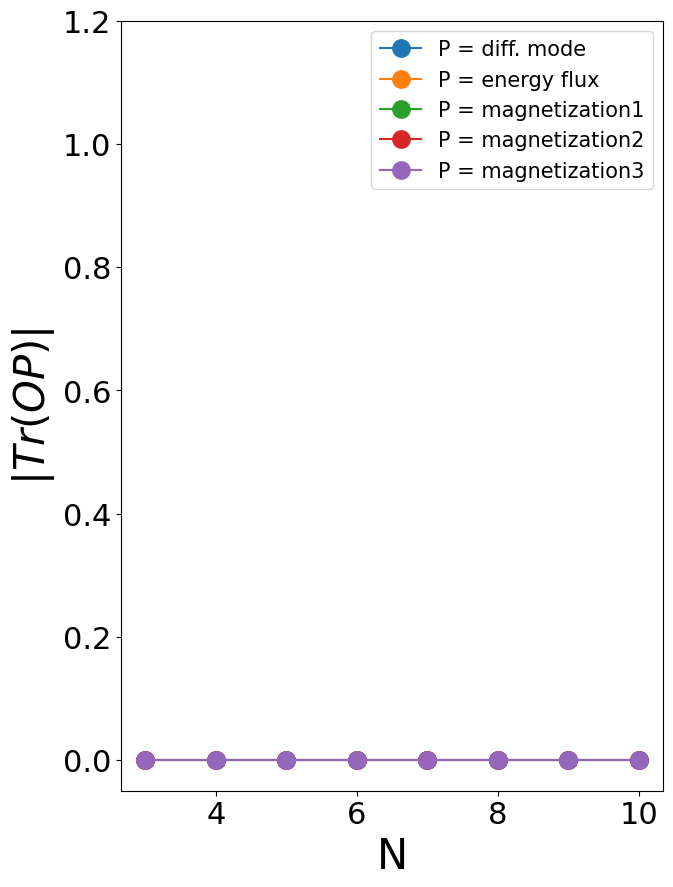}
    \caption{Tr.-inv. ($h = 0.0, g = 1.0$)}
\end{subfigure}
\hfill
\begin{subfigure}{0.16\textwidth}
    \includegraphics[scale=0.17]{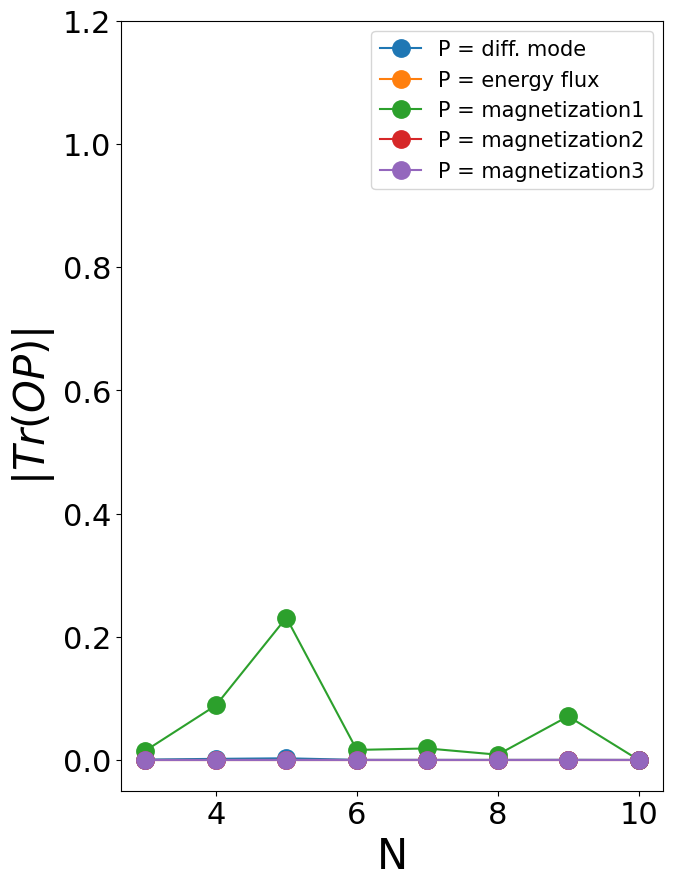}
    \caption{Tr.-inv. ($h = 0.0, g = 1.5$)}
\end{subfigure}

\caption{Overlap $\Tr (OP)$ of the slowest operator $O$ and a probe operator $P$ as a function of support size $N$ of the slowest operator. We take $P$ as diffusion mode, energy flux, magnetization1,2,3 (all operators are defined in the text).}
\label{overlapN}
\end{figure*}

\subsection{The results}

Here we discuss Fig. \ref{scalingN} and Fig. \ref{overlapN}. 

\subsubsection{The local diffusion mode has scaling $-\Tr [H,O]^2 \sim \frac{1}{N^2}$}

It can be seen in Fig. \ref{scalingN} (a), (c), (e). The diffusion mode is depicted with the dark points, they approach the value $-2$ in the inset. (It agrees with theoretical arguments, see \cite{kim2015slowest}).

\subsubsection{For the integrable case of $h=0$, the local slowest operator expands much slower, than in the non-integrable system. The slowest operator \textbf{does not} correspond to the ballistic transport of quasi-particles.}

As one can see in  Fig. \ref{scalingN} (a), (e), for integrable case of $h=0$ and any value of $g$, the scaling of $-\Tr [H,O]^2$ approaches approximately $\frac{1}{N^4}$. It corresponds to the rate of expansion much slower than diffusion $\left( \frac{1}{N^2} \right)$. Therefore, the slowest operator cannot corresond to the known ballistic transport of quasi-particles \cite{banchi2013ballistic, bastianello2022fragmentation, castro2016emergent}.

We also see in Fig. \ref{overlapN} (m), (n), (o), (a), that the overlap with diffusion mode decreases, as we increase $N$. The rate of expansion stops being diffusive (and becomes slower).

\subsubsection{For the non-integrable case, the expansion of the local slowest operator is slower than diffusion, but faster than in the integrable case.}

It can be seen in Fig. \ref{scalingN} (a), (c). The final slope (for big values of $N$) for all $h>0$ is less than $-2$, but greater than $-4$ (as in the integrable case). In Fig. (a), for small $h$, such as $0.05-0.2$, the maximum $N=20$ is not enough to see this. But, clearly, for all $h>0$ the slope curve goes up and then goes down, to the value less than $-2$ (this value is first found in \cite{kim2015slowest}).

On the other hand, Fig. \ref{overlapN} (b), (c), (h), (i) show that, as one increases $N$, the slowest operator increases its overlap with diffusion mode.

\subsubsection{As one increases $h$, there is an intermediate behavior of the local slowest operator between integrable and non-integrable ones.}



As one can see in  Fig. \ref{scalingN} (a), when one increases $h$, the curve $\log (-\Tr [H,O]^2)$ gradually changes: it does not reach $-4$, but instead goes up and then goes down to the value $<-2$. This process goes faster for bigger values of $h$, i.e. further away from the integrable point.

The similar behavior is observed in Fig. \ref{overlapN} (b). The overlap with diffusion mode decreases, as it is in the integrable case (Fig. \ref{overlapN} (a)), but then increases, as it is in highly non-integrable case (Fig. \ref{overlapN} (c)).

%
%
%

\subsubsection{The translationally-invariant slowest operator has overlap with magnetization1 and magnetization3, but with no other probe operators, for any value of $N$.}

See Fig. \ref{overlapN} (d), (e), (f), (j), (k), (l), (p), (q), (r).

\subsubsection{For a translationally-invariant slowest operator, for every $h$, there is a threshold value $N^*$, such that for $N>N^*$, the operator has non-zero overlap with magnetization1 and magnetization3.}

It can be seen in Fig. \ref{overlapN} (d), (e), (f). (Recall Fig. \ref{scaling_h_g} (b) and Fig. \ref{overlap} (b).)
%
%

\subsubsection{For a translationally-invariant slowest operator, the rate of expansion before the transition is faster than diffusion, but after the transition - slower than diffusion.}

As can be seen in \ref{scalingN} (b), for $h=0.05, 0.1, 0.2$, the slope is approximately equal to $-1$.

On the other hand, for $h=0.4, 0.7$, the slope is less than $-2$. Fig. \ref{scalingN} (d) also shows that, for $h=1.05$ after the transition, the slope is less than $-2$.


%

\subsubsection{In the integrable case, the translationally-invariant slowest operator corresponds to an integral of motion, for any value of $N$.}

As obvious from Fig. \ref{scalingN} (f), the translationally-invariant operator $O$ obeys $[H,O] = 0$. Therefore, it is an integral of motion.

\section{Time evolution \label{sec_timeevol}}

In this section we calculate time evolution of the slowest operator $O$. In particular, we observe how the slowest operator expands over the chain and how it thermalizes.

\subsection{The physical quantities}

We calculate the following physical quantities.

\subsubsection{Two-point correlation function $\Tr O(t) O(0)$}

As described in Section \ref{periods}, in the period of final thermalization, the average of the slowest operator changes from $\Tr (\rho_{\widetilde{GGE}} O(t))$ to $\Tr (\rho_{th} O(t))$. $\Tr (\rho_{\widetilde{GGE}} O(t))$ can be transformed as follows:
\be \Tr (e^{-\beta H + \mu O(0)} O(t)) \sim \Tr (e^{\mu O(0)} O(t)) \sim \Tr (O(t) O(0)) \label{twopoint}\ee

So that we suppose very high temperature ($T = \frac{1}{\beta} \rightarrow \infty$), and claim that two-point correlation function $\Tr (O(t) O(0))$ essentially describes the late-time dynamics of $O$.

The average of $O$ is expected to reach $\Tr (\rho_{th} O(t))$ in the late-time limit. The latter is equal to $\Tr (e^{-\beta H} O(t)) \sim \Tr O(t) = \Tr O(0) = 0$. Therefore, we believe that $\Tr (O(t) O(0))$ aims at $0$ at late times.


$\Tr (O(t) O(0))$ can be calculated using exact diagonalization of Hamiltonian. One can find a complete set of eigenvectors of Hamiltonian $\{ \ket{E_i} \}$: $H \ket{E_i} = E_i \ket{E_i}$; then, replace the trace with $\sum_i \bra{E_i} \dots \ket{E_i}$ and insert identity operator $\sum_j \ket{E_j} \bra{E_j}$:
\be \frac{1}{2^L} \Tr O(t) O(0) = \frac{1}{2^L} \sum_{ij} e^{i(E_i - E_j)t} |\bra{E_i} O(0) \ket{E_j}|^2 \label{exactdiag}\ee

But exact diagonalization can only be implemented for small dimension of the Hilbert space ($=2^L$): up to $L \sim 11$ for a reasonable time.

Therefore, instead, we use random vector approximation \cite{iitaka2004random,goldstein2006canonical,dymarsky2017canonical, elsayed2013regression}. We substitute the trace as: $\frac{1}{2^L} \sum_i \bra{E_i} \dots \ket{E_i} \rightarrow \frac{1}{K} \sum^{K}_{k=1} \bra{\psi_k} \dots \ket{\psi_k}$,where $\ket{\psi_k}$ is a vector with real and imaginary parts given by Gaussian random variables with zero mean and unit variance and normalized as $\braket{\psi_k | \psi_k} = 1$; in the following, we take $K = 50$. Then, $\Tr O(t) O(0)$ reads:
\be \frac{1}{2^L} \Tr O(t) O(0) = \frac{1}{2^L} \Tr \left( O(0) e^{-iHt} O(0) e^{iHt}\right) \sim \nn\ee
\be \sim  \frac{1}{K} \sum_{k=1}^K \bra{\psi_k} O(0) e^{-iHt} O(0) e^{iHt}\ket{\psi_k} = \nn \ee
\be = \frac{1}{K} \sum_{k=1}^K \bra{\phi_k (-t)} O(0) \ket{\chi_k (-t)} \label{2pointcorr}\ee
where $\ket{\chi_k (-t)} \equiv e^{iHt}\ket{\psi_k} $ and \\ $\ket{\phi_k (-t)} \equiv e^{iHt} O(0) \ket{\psi_k}$.

Time evolution of any vector $\ket{\psi}$ ($\ket{\chi_k}$ or $\ket{\phi_k}$) can be calculated using the expansion in Chebyshev polynomials \cite{tchebychev1853theorie,fehske2007computational,khlebnikov2013thermalization}:
\be \ket{\psi(t)} = e^{-iHt} \ket{\psi} = \nn\ee
\be = J_0(2\bar{E}t)\ket{\psi} + 2 \sum_{n=1}^\infty (-i)^n J_n (2\bar{E}t) T_n\left(\frac{\bar{H}}{2}\right)\ket{\psi} \label{Chebyshev}\ee
where $T_n$ are Chebyshev polynomials of the first kind, $J_n$ are Bessel functions of the first kind, $\bar{H} = \frac{H}{\bar{E}}$, where $\bar{E}$ is chosen such that eigenvalues of $\bar{H}$ get inside the interval $[-1,1]$ (we take $\bar{E} = 1000$).

This row quickly converges \cite{khlebnikov2013thermalization}. For a fixed $t$, we calculate the terms in (\ref{Chebyshev}) one by one and put them inside $\ket{\psi_{reduced}}$ until $|\braket{\psi_{reduced}|\psi_{reduced}} - 1|<10^{-13}$. We use the resulting $\ket{\psi_{reduced}}$ as a final answer for this time $t$.

We do not use standard tensor network method - time-evolution block-decimation (TEBD) \cite{suzuki1976generalized} - because it only allows for calculation for small time $t$ (one needs to divide a time evolution into very small intervals $dt$). Here we do not have such limitation and, thus, can reach later times $t$.

%


To conclude, we calculate $\ket{\chi_k (-t)}$ and $\ket{\phi_k (-t)}$ according to the described procedure, and then obtain $\Tr O(t) O(0)$ from (\ref{2pointcorr}).

Below we calculate $\Tr O(t) O(0)$:
\begin{enumerate}
\item For different full size of the system $L$.
\item For various parameters $g$ and $h$.
\end{enumerate}


\subsubsection{Out-of-time-ordered commutator (OTOC)}

We calculate the following quantity, out-of-time-ordered commutator (OTOC) \cite{rozenbaum2017lyapunov,lin2018out,maldacena2016bound,fine2014absence}:
\be \Tr ([O(t),\sigma_{x,y,z}^{(i)}(0)]^\dag [O(t),\sigma_{x,y,z}^{(i)}(0)]) = - \Tr [O(t),\sigma_{x,y,z}^{(i)}(0)]^2\label{OTOC_}\ee
where $i$ is the site on the chain ($i = 0 \dots L-1$), where $\sigma_{x,y,z}^{(i)}$ is located. (The same Pauli matrix is in both places.)

We calculate it to observe how the slowest operator delocalizes over the chain (see \cite{swingle2018quantum} for details). Initially, it has non-zero OTOC only with those Pauli matrices, which are located inside the support of the slowest operator. But, as it expands over the chain, it starts to have non-zero OTOC with Pauli matrices at other locations. Finally, we expect OTOC to be equal for any location of the Pauli matrix.

One can estimate the contributions of $\{ \sigma_x, \sigma_y, \sigma_z\}$ to the slowest operator $O$ by the final value of OTOC: it is smaller for that of $\{ \sigma_x, \sigma_y, \sigma_z\}$, which has a bigger contribution to $O$ (it best commutes with $O$).

Here we do not use random vector approximation, because we have $O(t)$ twice in the expression (\ref{OTOC_}). One would need two random vectors to calculate it. Therefore, the  error of this approximation would significantly increase. Instead, we use exact diagonalization of Hamiltonian (see (\ref{exactdiag})). (\ref{OTOC_}) is further transformed as
\be 2 - 2 \Tr (O(t) \sigma_{x,y,z}^{(i)}(0) O(t) \sigma_{x,y,z}^{(i)}(0)) = \nn\ee
\be = 2 - \sum_{ijmn} e^{i(E_i-E_j+E_m-E_n)t} W_{ij} S_{jm} W_{mn} S_{ni} \ee
where $W_{ij} = \braket{E_i|O(0)|E_j}$, $S_{jm} = \braket{E_j|\sigma_{x,y,z}^{(i)}(0)|E_m}$.

\begin{figure*}
\centering
\begin{subfigure}{0.57\textwidth}
    \includegraphics[scale=0.22]{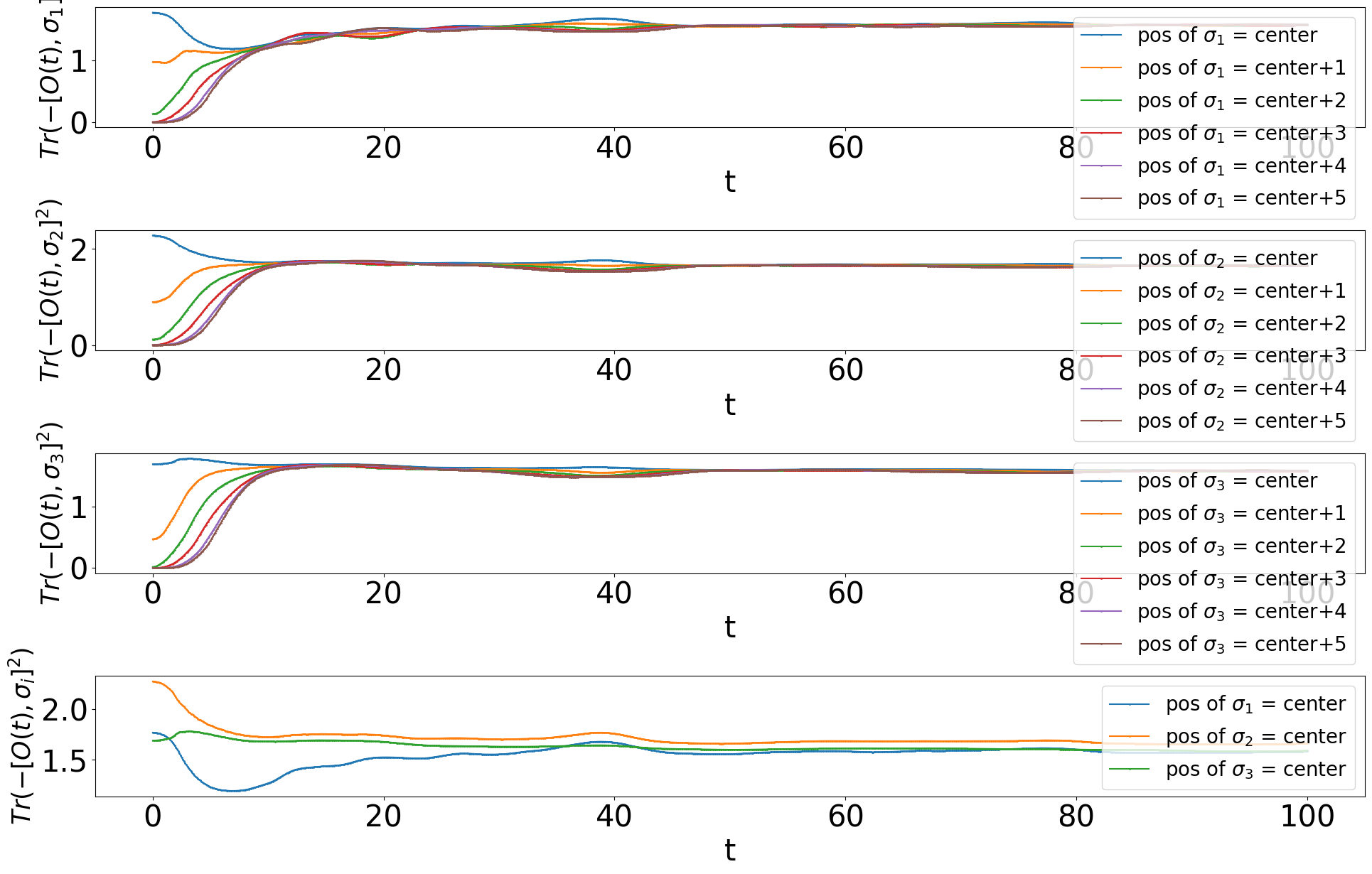}
    \caption{Local, $g=1.05, h = 0.1$}
\end{subfigure}
\hfill
\begin{subfigure}{0.38\textwidth}
    \includegraphics[scale=0.16]{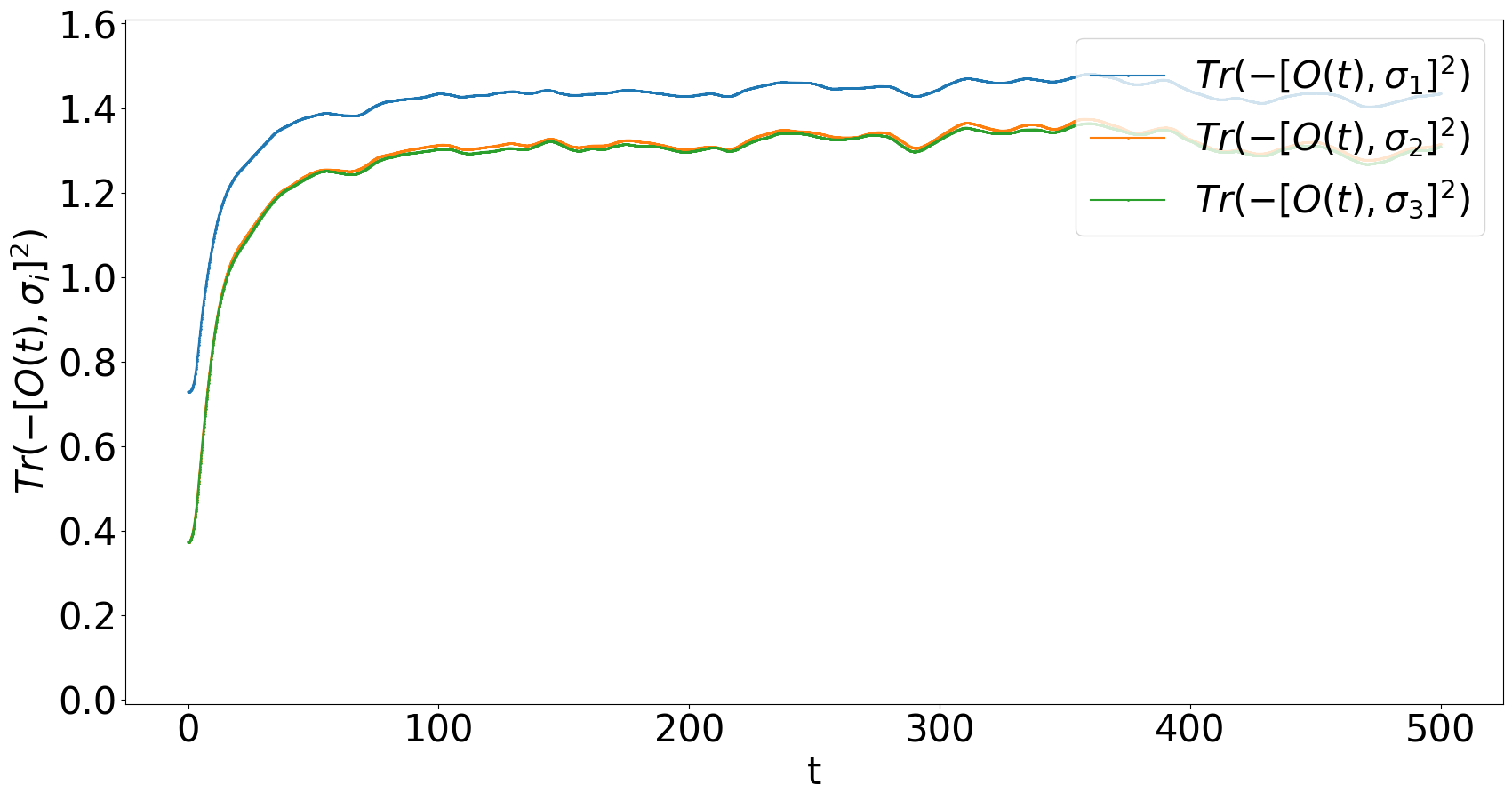}
    \caption{Translationally-invariant, $g=1.05, h = 0.1$}
\end{subfigure}
\hfill
\begin{subfigure}{0.57\textwidth}
    \includegraphics[scale=0.22]{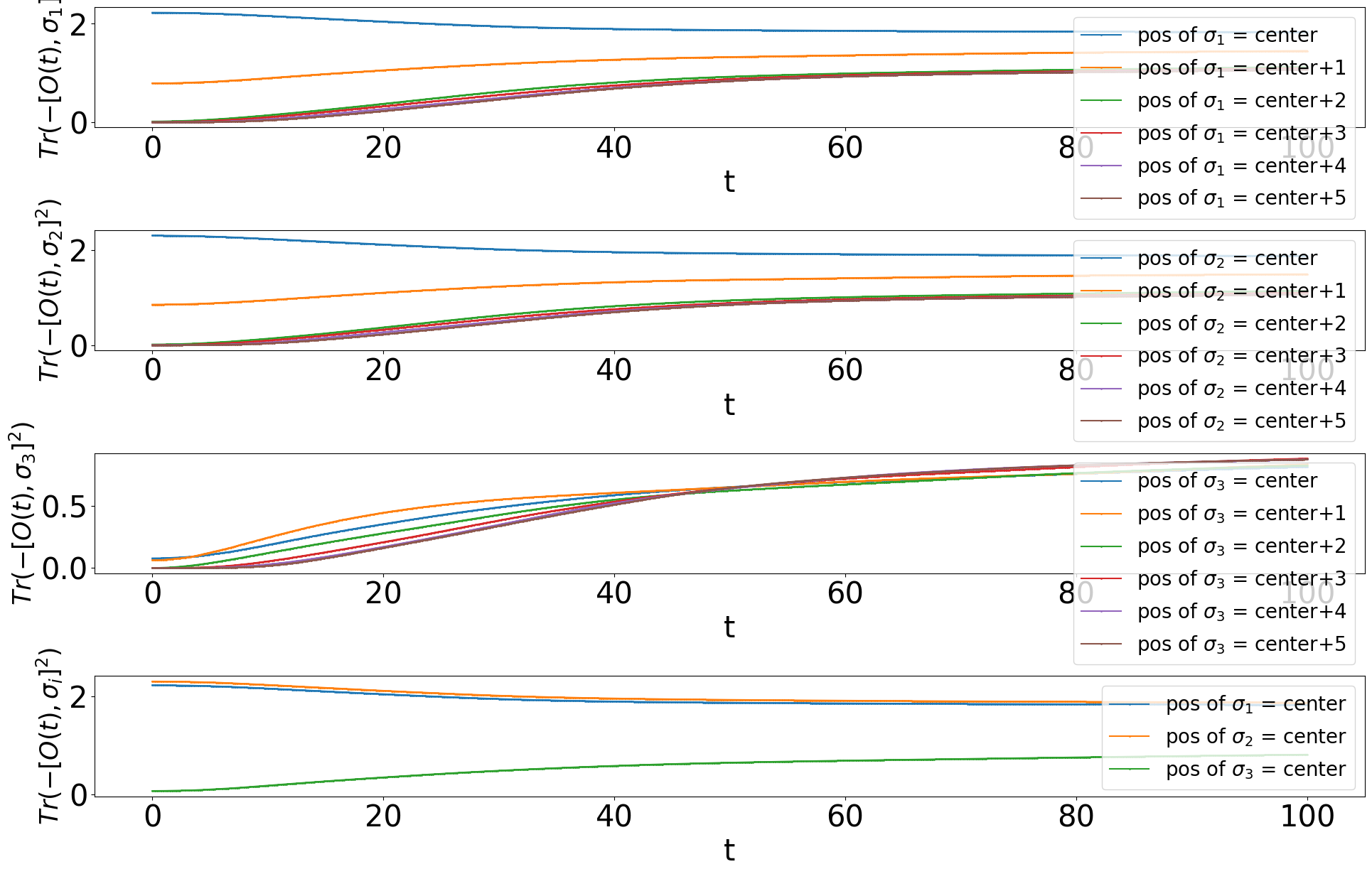}
    \caption{Local, $g=0.4, h = 1.05$}
\end{subfigure}
\hfill
\begin{subfigure}{0.38\textwidth}
    \includegraphics[scale=0.16]{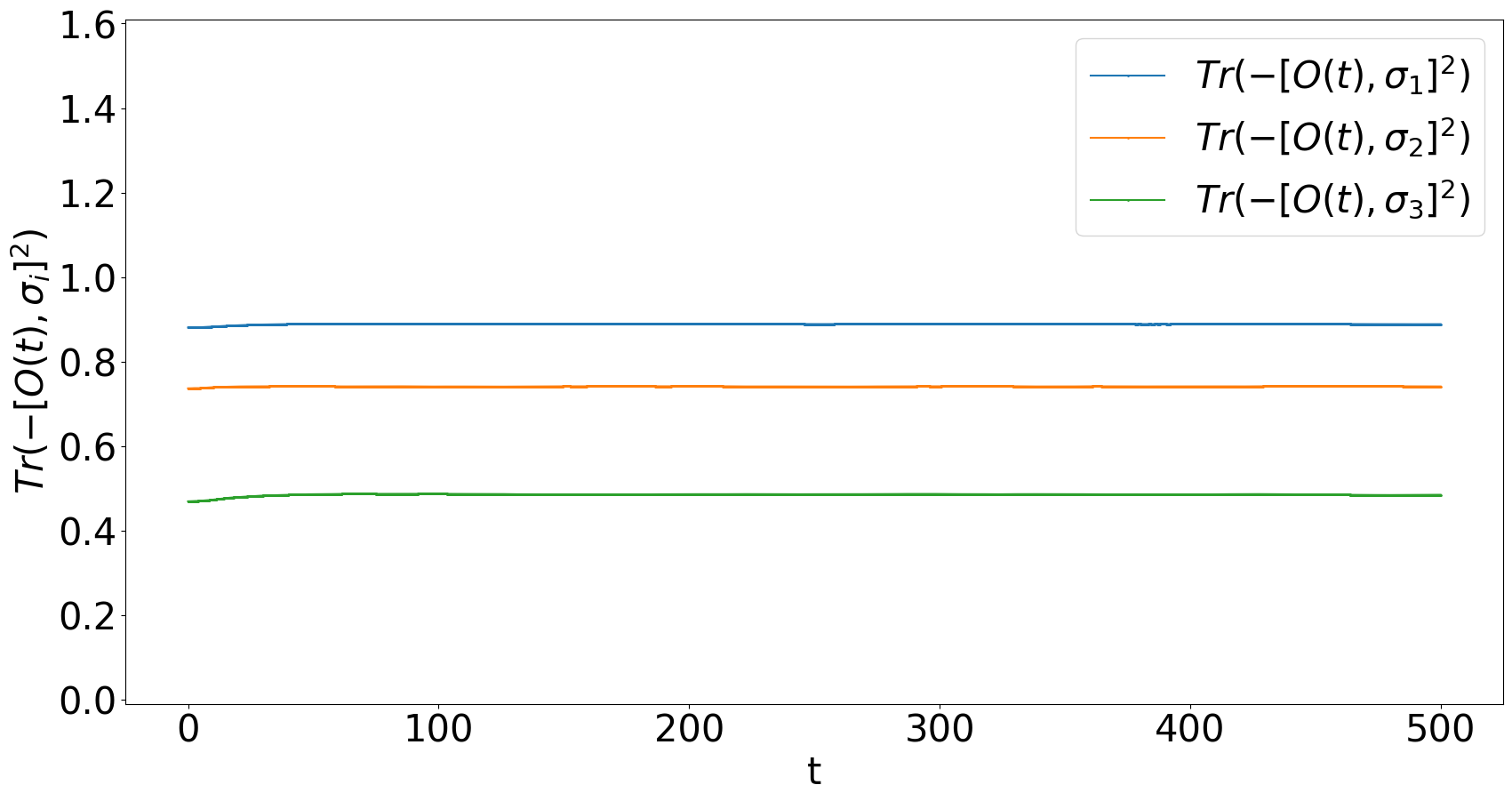}
    \caption{Translationally-invariant, $g=0.4, h = 1.05$}
\end{subfigure}
\hfill
\begin{subfigure}{0.57\textwidth}
    \includegraphics[scale=0.22]{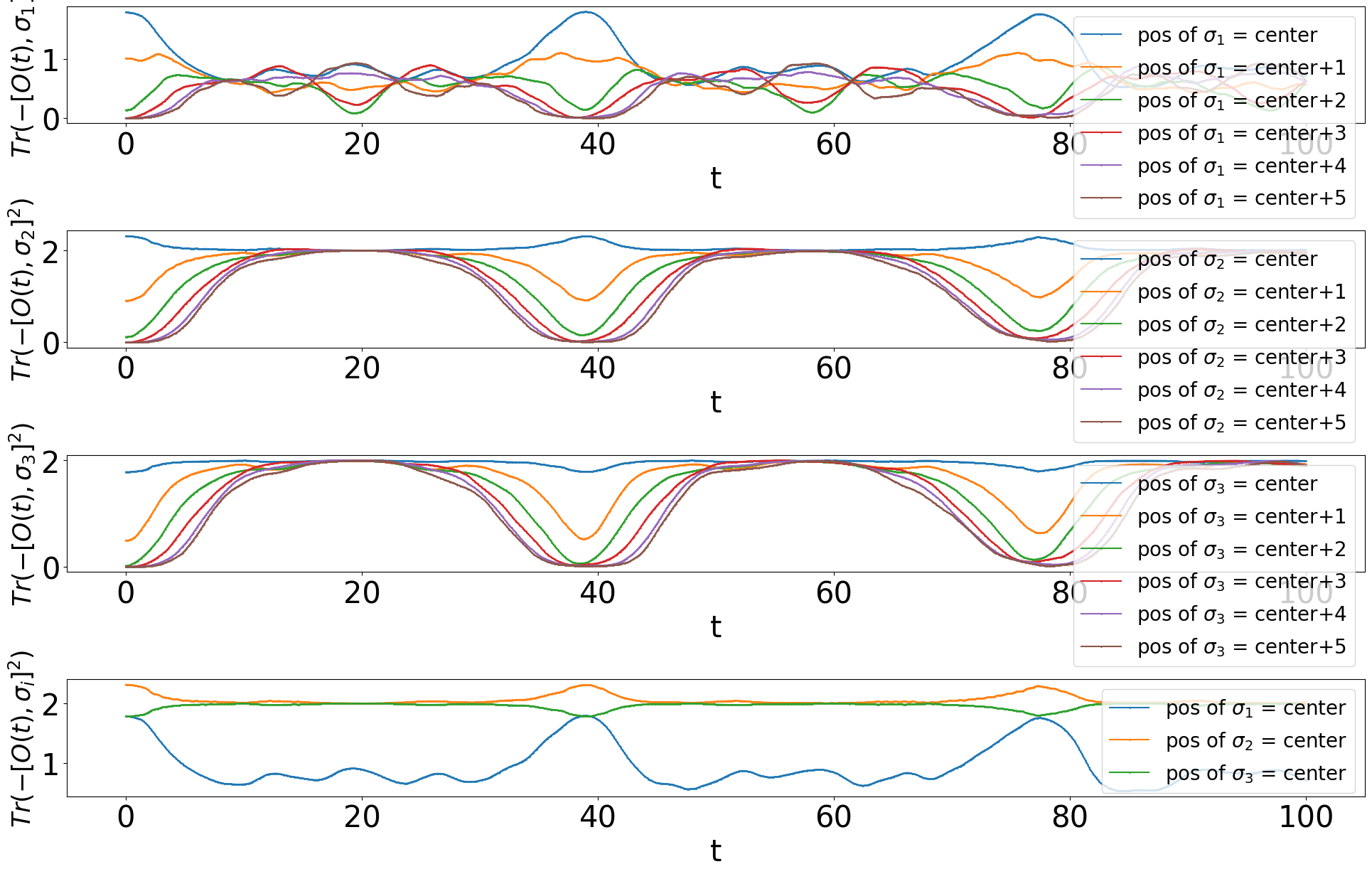}
    \caption{Local, $g=1.05, h = 0.0$}
\end{subfigure}
\hfill
\begin{subfigure}{0.38\textwidth}
    \includegraphics[scale=0.16]{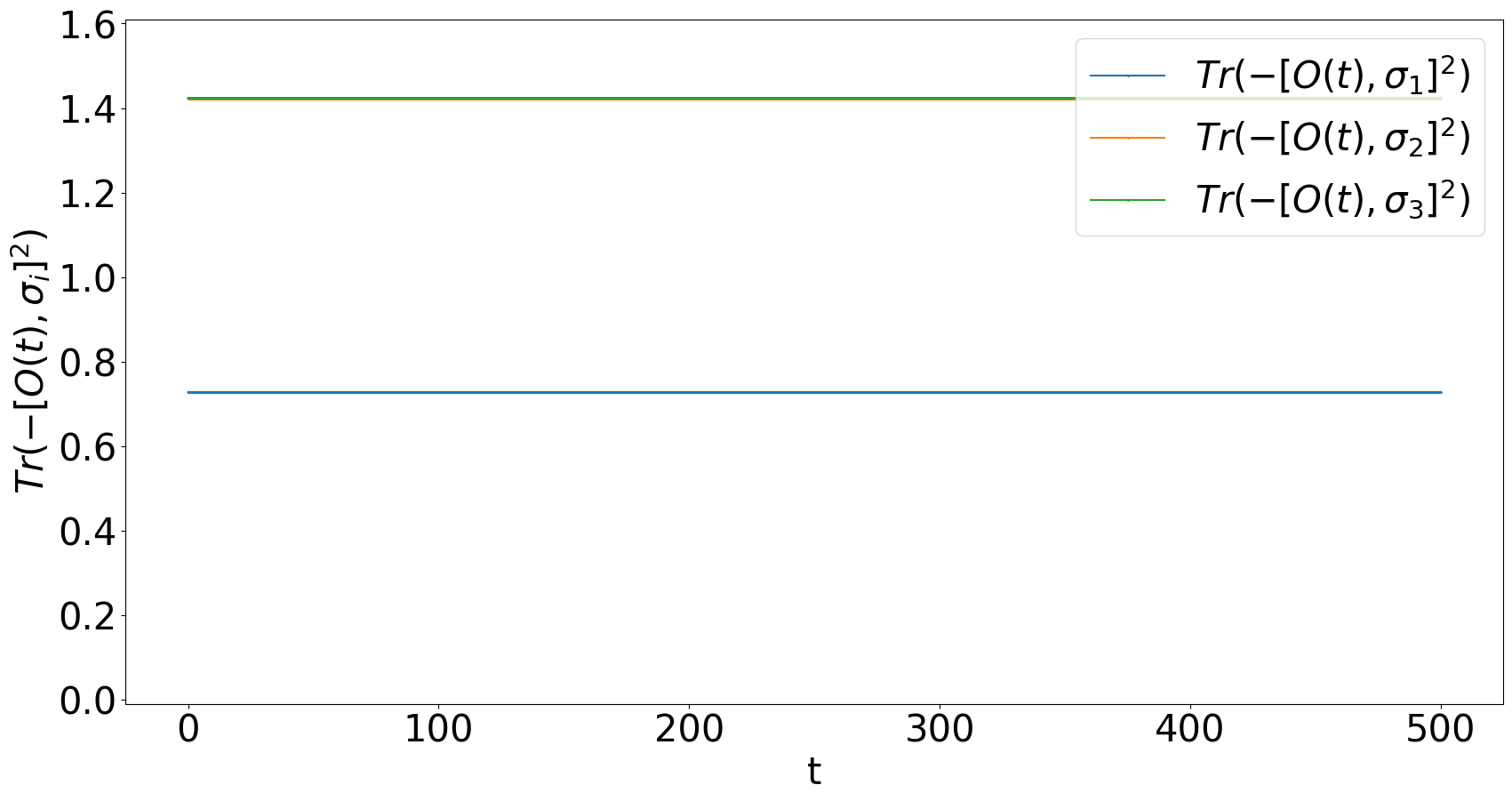}
    \caption{Translationally-invariant, $g=1.05, h = 0.0$}
\end{subfigure}

\caption{OTOC of the slowest operator with the Pauli matrix at a particular site: $Tr(-[O(t),\sigma_i(0)]^2)$ ($L=11$, $N=6$).}
\label{OTOC}
\end{figure*}

We plot OTOC in Fig. (\ref{OTOC}).

On the left, we calculate OTOC for the local slowest operator. The location $i$ of $\sigma_{x,y,z}^{(i)}(0)$ is "center", "center+1", etc. By "center" we mean the central site of the slowest operator. For instance, if it has support $5$ and is located at sites $0 \dots 4$, then there is one such site $2$. Then, "center+1" is sites $1$ and $3$, "center+2" is sites $0$ and $4$. "center+3" is not inside the support of $O$ and it corresponds to sites $-1$ and $5$, etc. For the considered in Fig. \ref{OTOC} case of $N=6$ ($O$ has support at sites $0 \dots 5$), there are 2 central sites  - $2$ and $3$, "center+1" corresponds to sites $1$ and $4$, etc.

The first 3 graphs for each set of $(g,h)$ correspond to OTOC with $\sigma_{x}$, $\sigma_{y}$, $\sigma_{z}$ respectively. Each graph compares OTOC for different locations of the Pauli matrix. In the fourth graph, we fix the position of the Pauli matrix at the "center" and compare OTOCs with $\sigma_x$, $\sigma_y$ and $\sigma_z$.




For the translationally-invariant operator ($O = \sum^{L-1}_{k=0} O_k$), there is no difference, where to put a Pauli matrix. Therefore, we fix its location at $i=0$ and plot only the fourth graph (Fig. \ref{OTOC} (b), (d), (f)).

\begin{figure*}
\centering
\begin{subfigure}{0.96\textwidth}
    \includegraphics[scale=0.35]{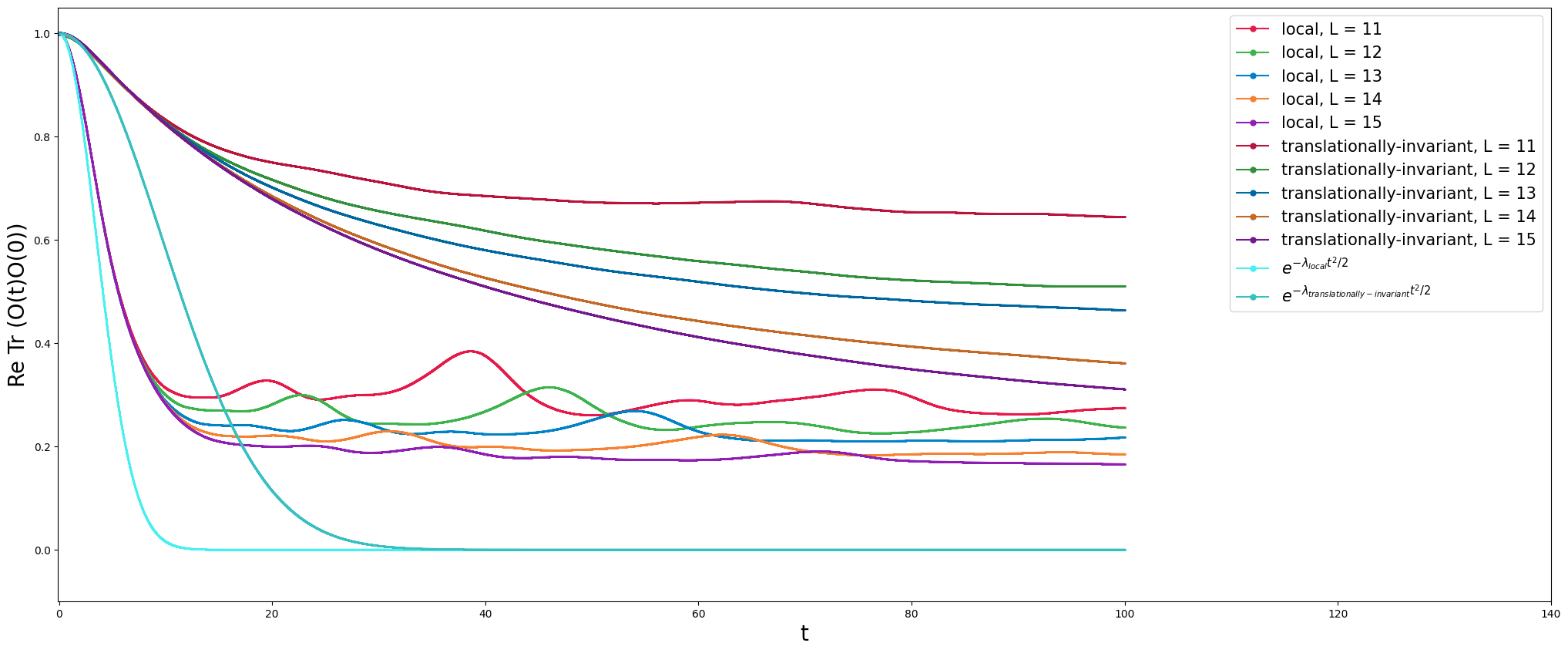}
    \caption{$g=1.05$, $h = 0.1$}
\end{subfigure}

\hfil

\begin{subfigure}{0.96\textwidth}
    \includegraphics[scale=0.35]{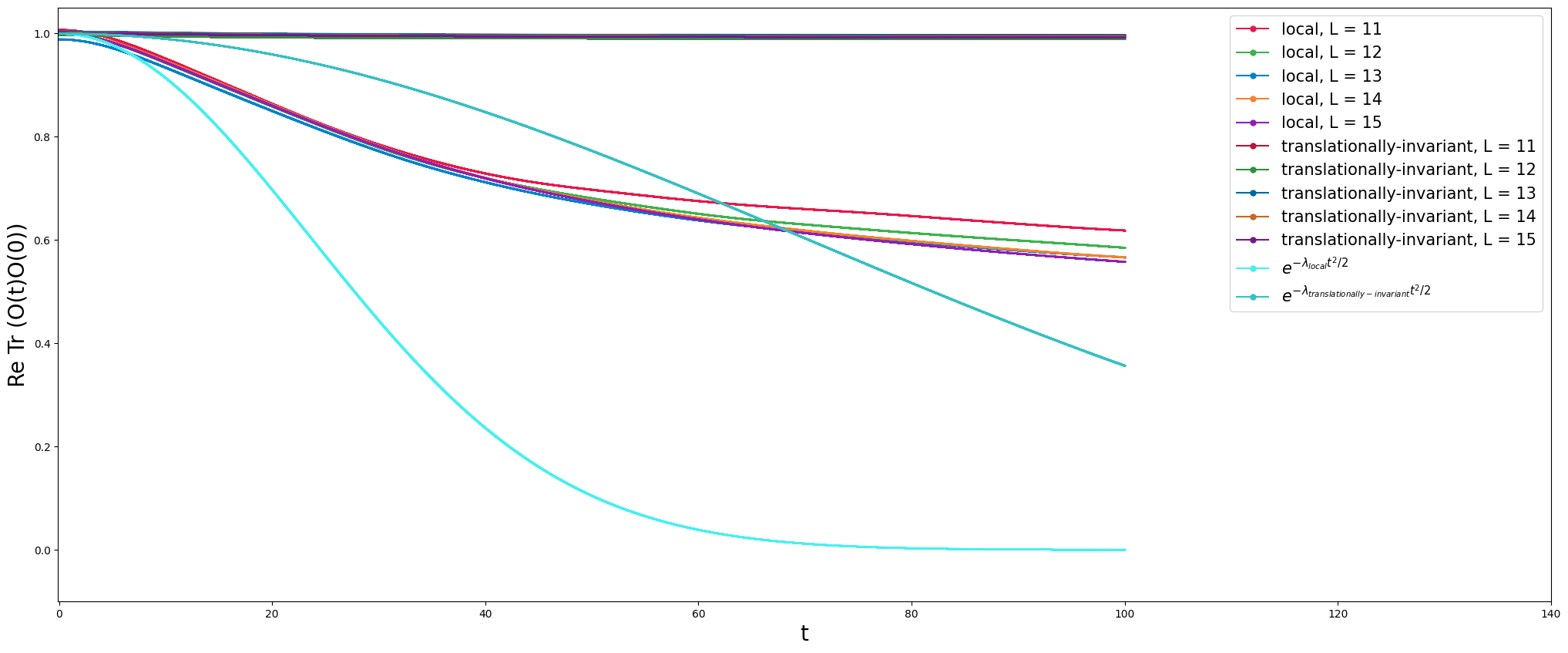}
    \caption{$g=0.4$, $h=1.05$}
\end{subfigure}

\hfil

\begin{subfigure}{0.96\textwidth}
    \includegraphics[scale=0.35]{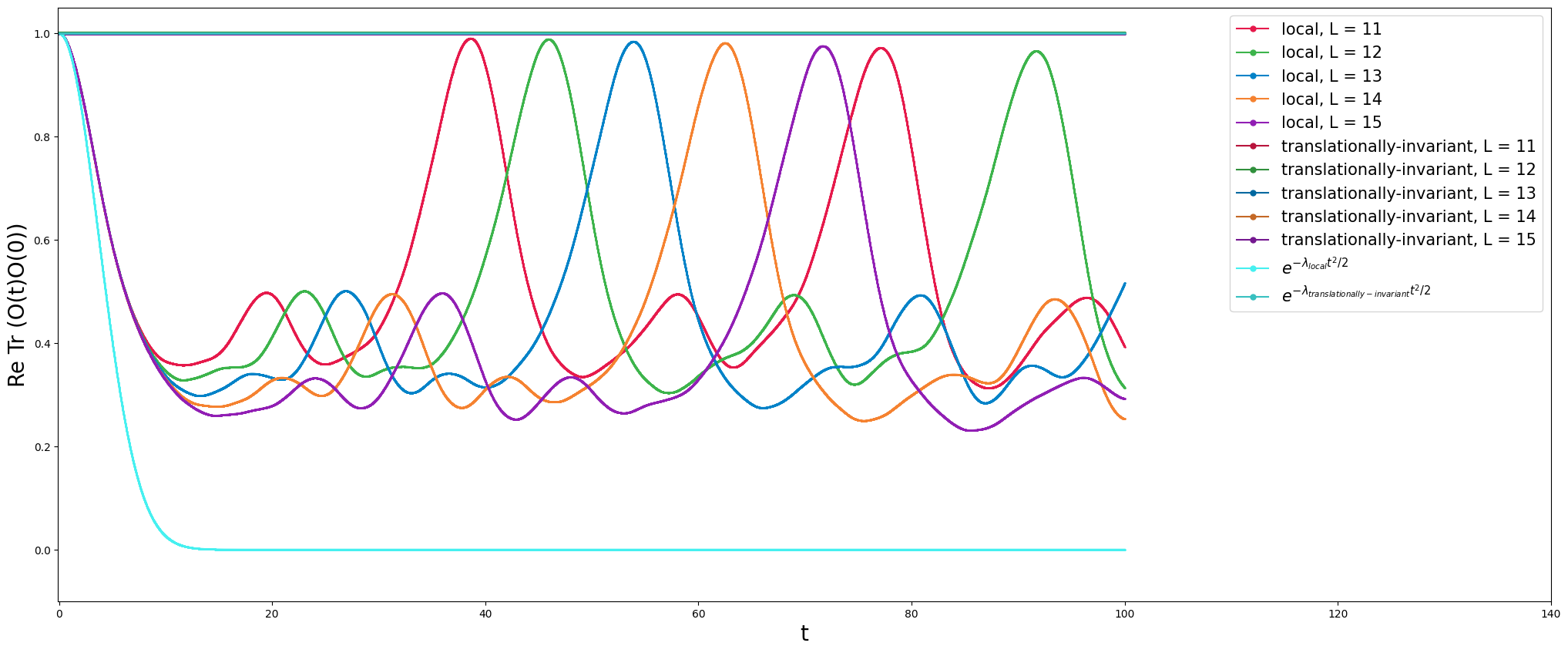}
    \caption{$g=1.05$, $h=0.0$}
\end{subfigure}

\caption{Time evolution as a function of the full size of the system $L$. $N=6$; $\lambda$ is defined as $-\Tr [H,O]^2$.}
\label{timeevol_L}
\end{figure*}

\begin{figure*}
\centering
\begin{subfigure}{0.47\textwidth}
    \includegraphics[scale=0.22]{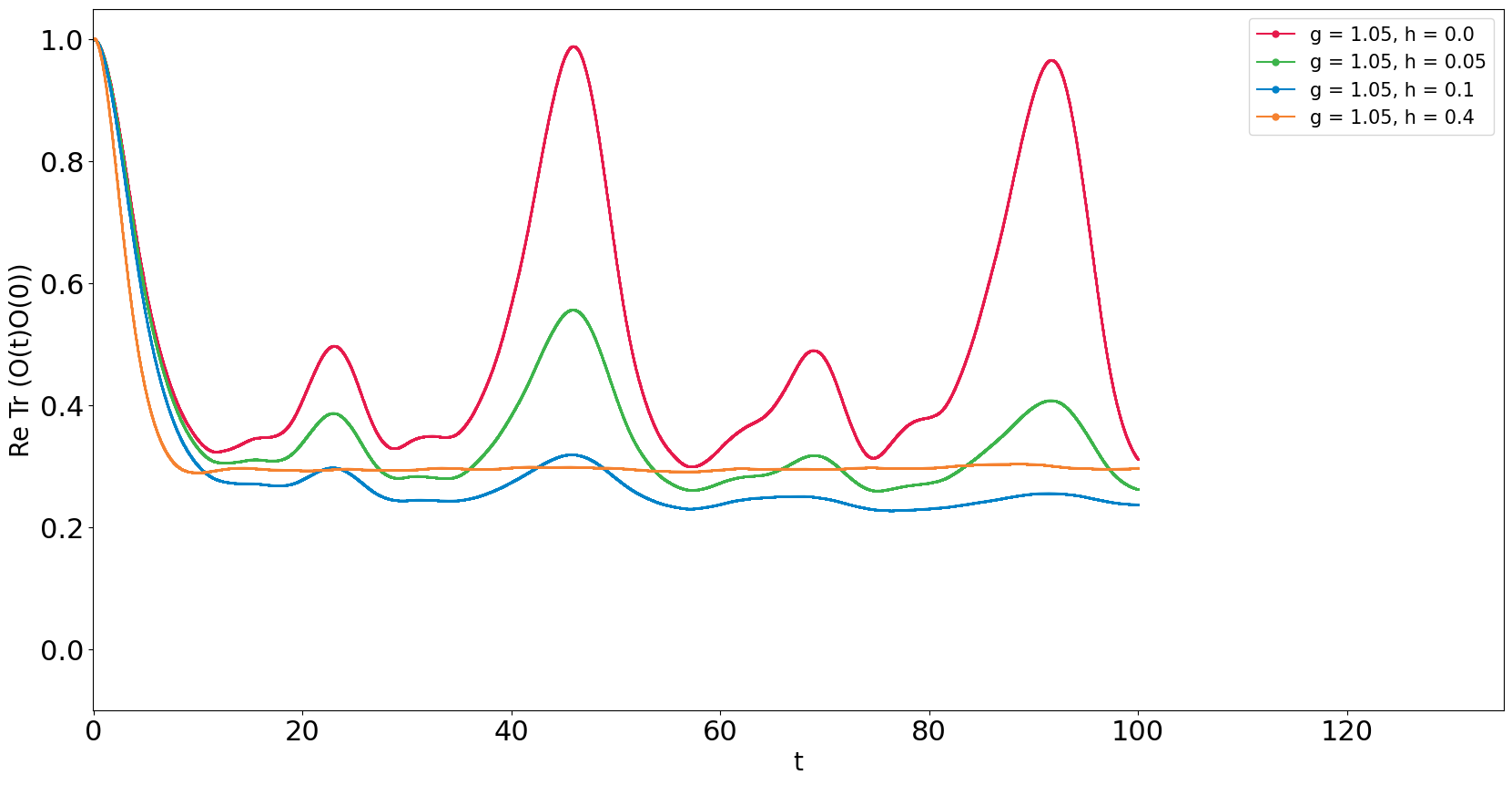}
    \caption{Local, $g=1.05$, different $h$}
\end{subfigure}
\hfill
\begin{subfigure}{0.47\textwidth}
    \includegraphics[scale=0.22]{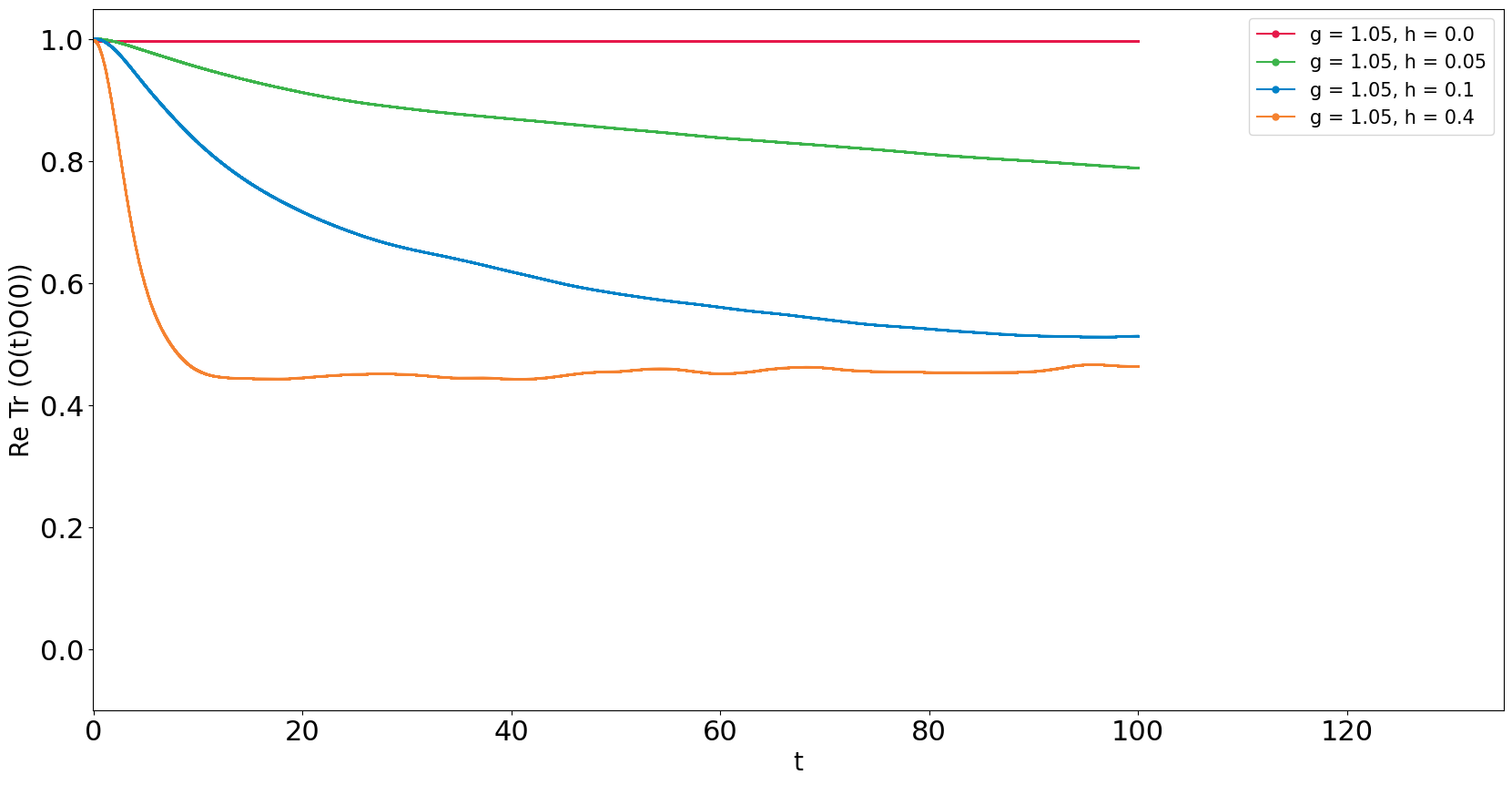}
    \caption{Translationally-invariant, $g=1.05$, different $h$}
\end{subfigure}
\hfill
\begin{subfigure}{0.47\textwidth}
    \includegraphics[scale=0.22]{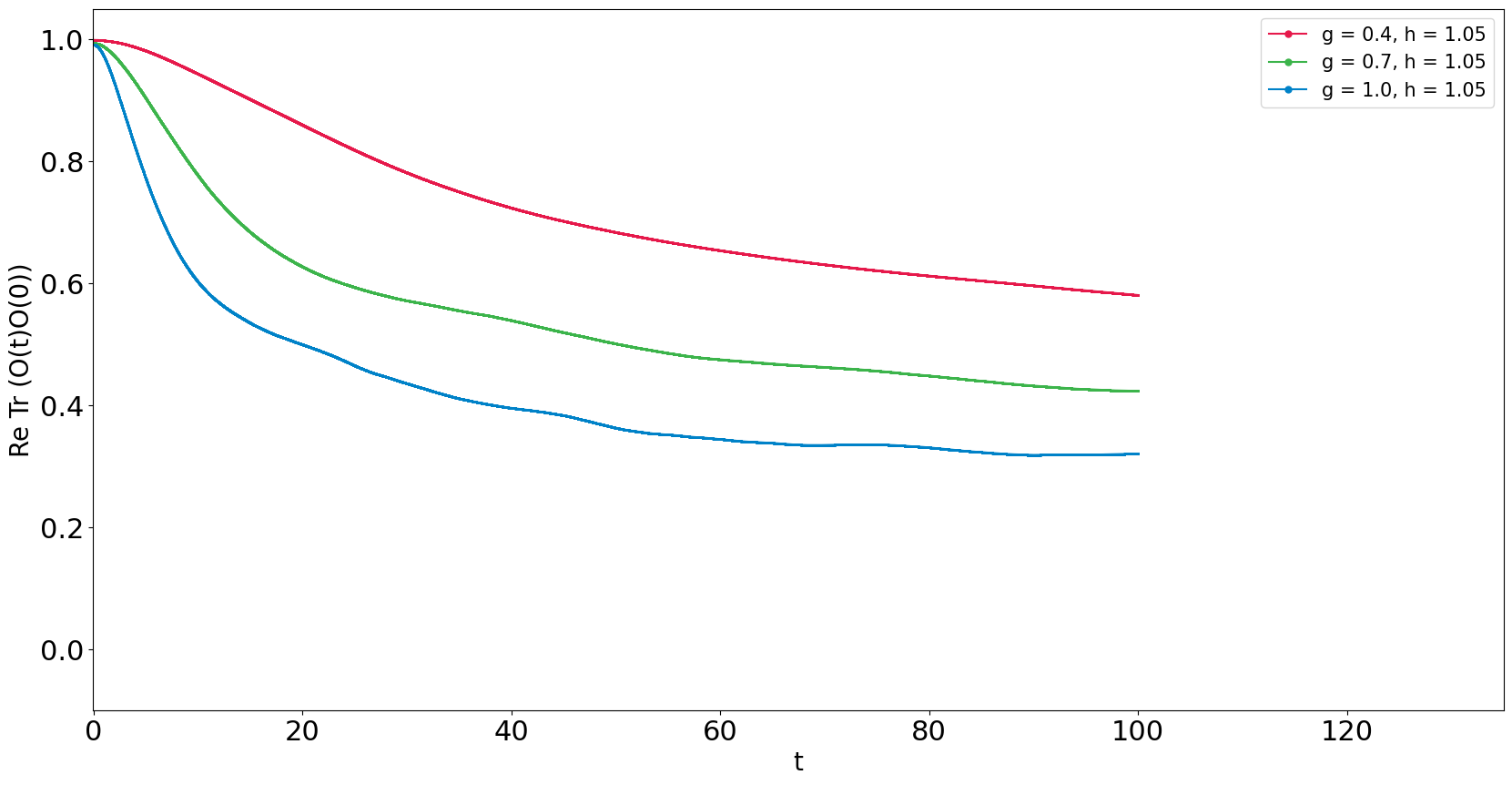}
    \caption{Local, $h=1.05$, different $g$}
\end{subfigure}
\hfill
\begin{subfigure}{0.47\textwidth}
    \includegraphics[scale=0.22]{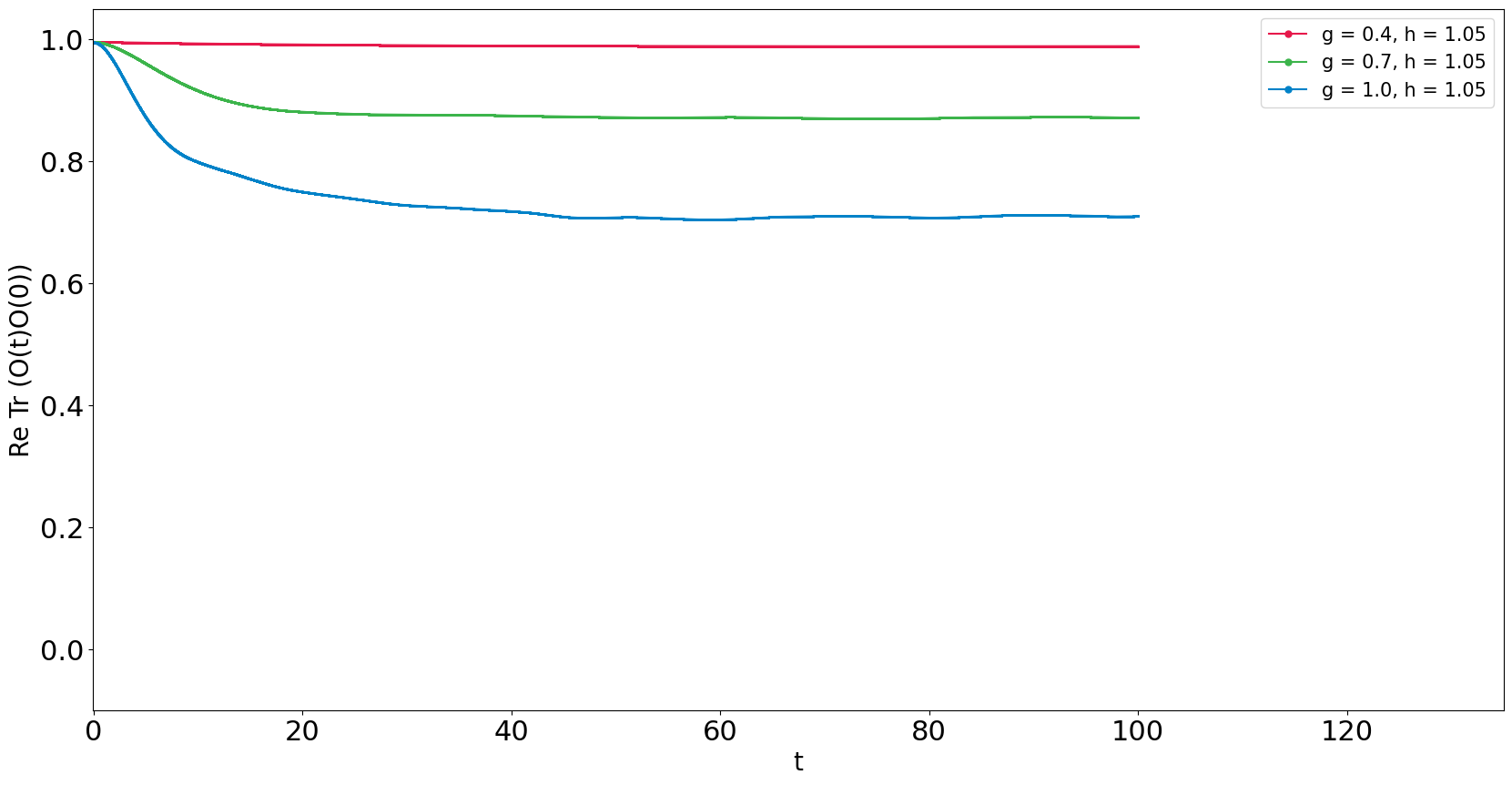}
    \caption{Translationally-invariant, $h=1.05$, different $g$}
\end{subfigure}
\hfill
\begin{subfigure}{0.47\textwidth}
    \includegraphics[scale=0.22]{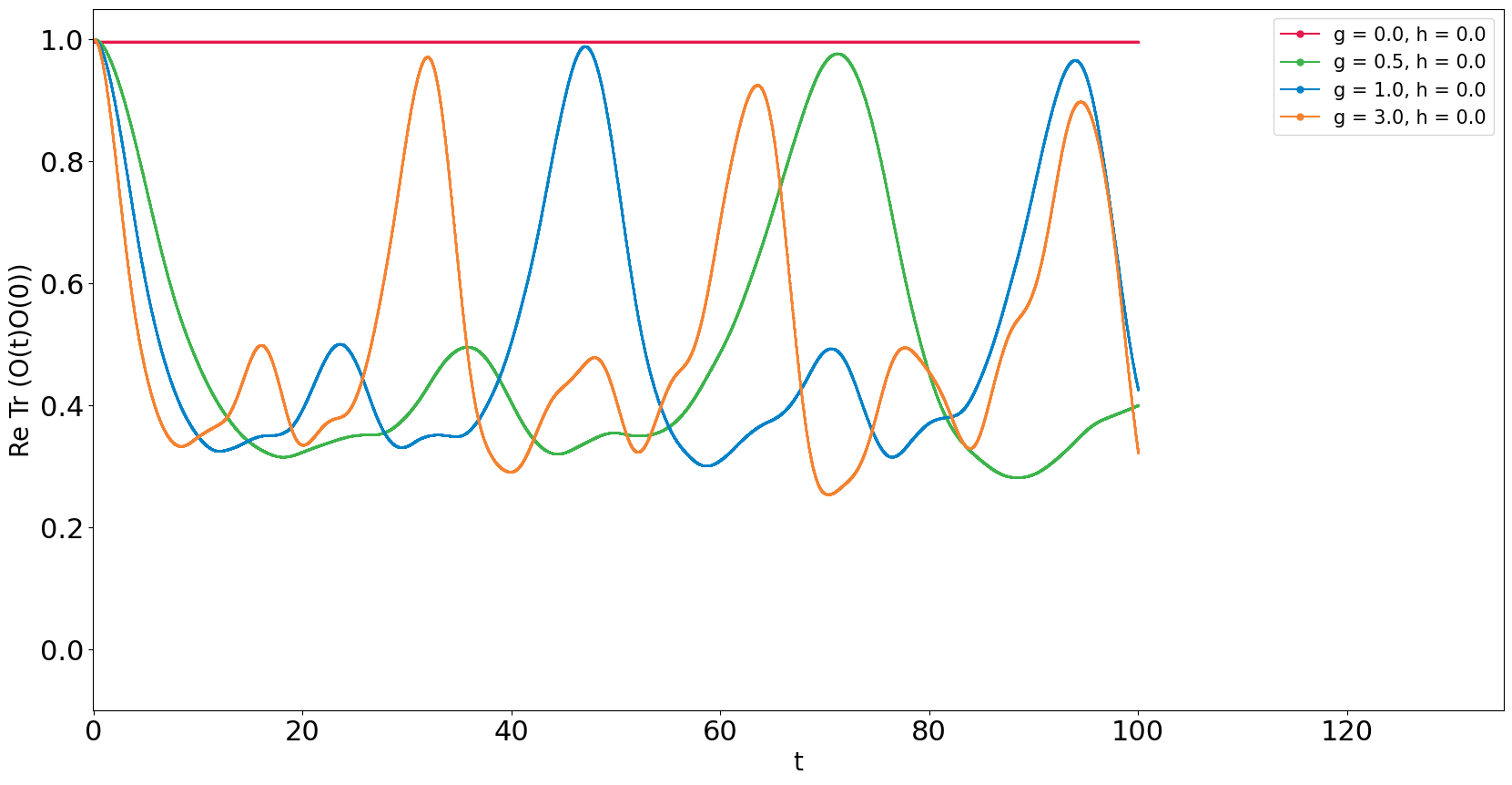}
    \caption{Local, $h =0.0$, different $g$}
\end{subfigure}
\hfill
\begin{subfigure}{0.47\textwidth}
    \includegraphics[scale=0.22]{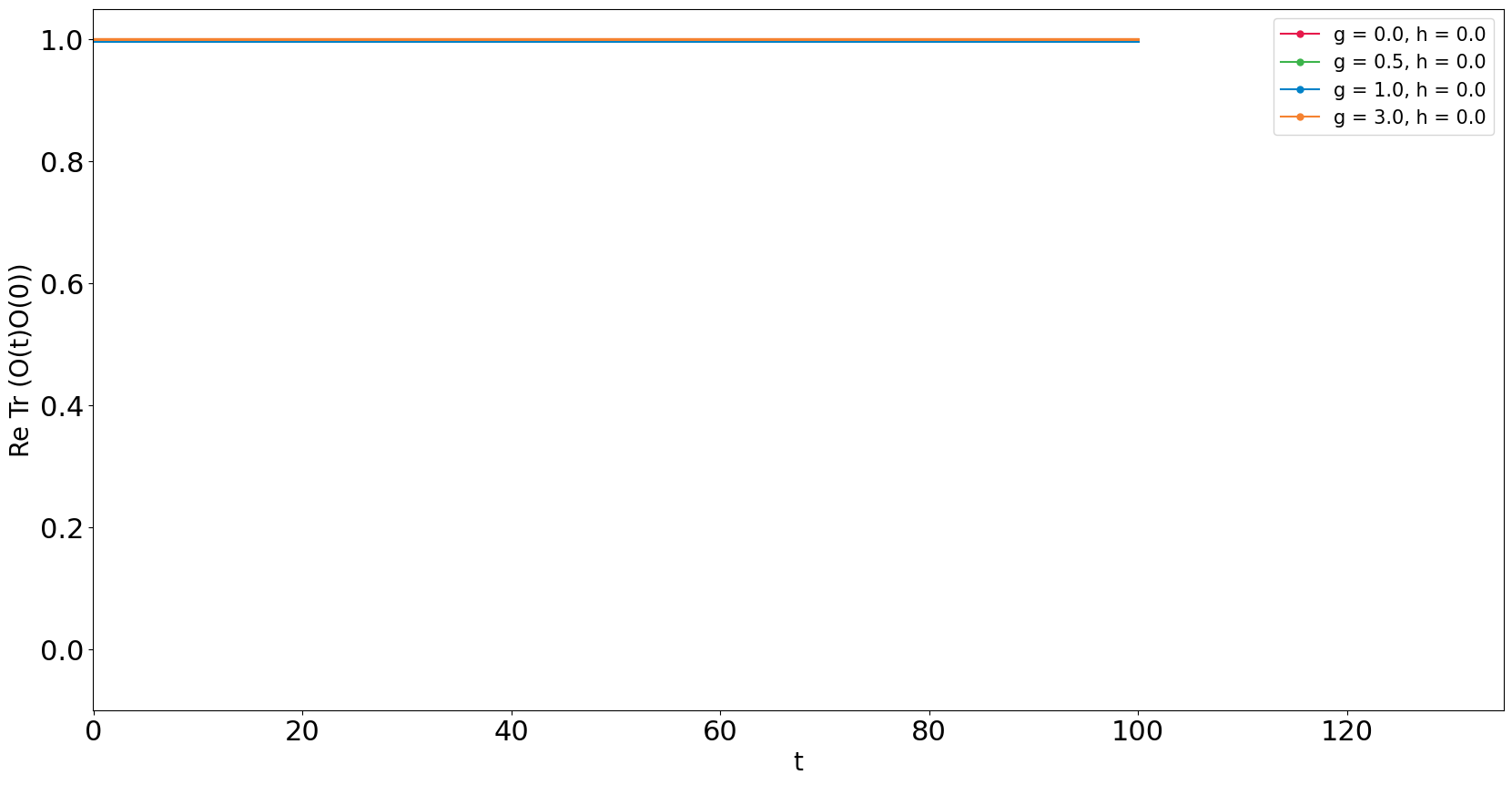}
    \caption{Translationally-invariant, $h =0.0$, different $g$}
\end{subfigure}

\caption{Time evolution as a function of parameters $g$ and $h$ ($N=6$, $L=12$).}
\label{timeevol_h_g}
\end{figure*}

\subsection{The results}

Here we outline the main results of the calculation of
\begin{itemize}
\item $\Tr O(t) O(0)$ as a function of the full size of the system $L$ 

See Fig. \ref{timeevol_L}.

\item $\Tr O(t) O(0)$ as a function of $g$ and $h$

See Fig. \ref{timeevol_h_g}.

\item Out-of-time-ordered commutator (OTOC) (see (\ref{OTOC_}))

See Fig. \ref{OTOC}.

\end{itemize}


\subsubsection{In the integrable case $h=0$, there are revivals of the local slowest operator, and no dynamics of the translationally-invariant slowest operator.}

For the local slowest operator, one can see that there are revivals \cite{michailidis2020slow,franco2012revival,ermakov2021almost} in the function $\Tr O(t) O(0)$ (Fig. \ref{timeevol_L} (c), Fig. \ref{timeevol_h_g} (e)). They occur, when the slowest operator have explored the full chain and passed through the boundaries (periodic boundary conditions). As we increase $L$, the operator takes more time to explore the chain, and the revival shows up later. After each revival the evolution repeats itself. The same behavior is observed in Fig. \ref{OTOC} (e). There is no thermalization, OTOC for Pauli matrices located at different sites does not get equal in the long-time limit, only for some limited period of time.

We also see the revivals of half the amplitude. As clear from Fig. \ref{OTOC} (e), they come from $\sigma_1$ (probably, because of high $g$ component in Hamiltonian and in the slowest operator).


On the other hand, translatonally-invariant operator experiences no dynamics (Fig. \ref{timeevol_L} (c), Fig. \ref{timeevol_h_g} (f), Fig. \ref{OTOC} (f)). It is a consequence of the fact that in the integrable case it is one of the integrals of motion (see Fig. \ref{scaling_h_g} (f) and Fig. \ref{scalingN} (f)).

\subsubsection{As one goes away from the integrable point $h=0$, the revivals of the local slowest operator get suppressed.}

As one can see in Fig. \ref{timeevol_h_g} (a), when one increases $h$, the revivals get smaller, until they are gone. The same phenomenon is visible, if one compares Fig. \ref{timeevol_L} (c) and (a), Fig. \ref{OTOC} (e) and (a).

This behavior reflects the fact that the system becomes less integrable and more thermalizing.


\subsubsection{During thermalization, the local slowest operator experiences fluctuations, while the translationally-invariant slowest operator does not.}

See Fig. \ref{timeevol_L} (a), and compare Fig. \ref{timeevol_h_g} (a) and (b). The fluctuations of the local slowest operator come from the revivals of the integrable system. The fluctuations do not appear for the translationally-invariant slowest operator, because it has no dynamics in the integrable case ($\rightarrow$ no revivals).

\subsubsection{In non-integrable case, there are 3 distinct periods of dynamics of the slowest operator: the initial dependence on one parameter $-\Tr [H,O]^2$, then approaching the boundaries and final thermalization.}

You can observe them in Fig. \ref{timeevol_L} (a).

Initially, the dynamics is well-fit with the function $e^{-\lambda t^2/2}$, where $\lambda = -\Tr [H,O]^2$. The reason is that $e^{-\lambda t^2/2}$ and $\Tr O(t)O(0)$ have identical small $t$ behavior (see (\ref{lambdacorrfuncconnection})). In other words, the early time dynamics depends on one parameter $\lambda=-\Tr [H,O]^2$. 


Then, the slowest operator expands over the chain, and it does not feel any boundaries (the curves coincide). But, after some time, it has explored the full chain and reached the boundaries: it is expressed via separation of curves for different $L$ (recall that we have periodic boundary conditions).

After this, the slowest operator ultimately thermalizes, and during this process $\Tr O(t) O(0)$ reaches the final value. As discussed above, we believe this value to be $0$.

In Fig. \ref{timeevol_L} (b), the dynamics has the same pattern, but the process runs slower (see below).

We note that there is no initial common dynamics for different $h$ (or $g$) in Fig. \ref{timeevol_h_g}. For every pair $(h,g)$, there is a unique value of $- \Tr [H,O]^2$ (see Fig. \ref{scaling_h_g}), that governs the dynamics at early times.

In the integrable case (Fig. \ref{timeevol_L} (c)), we see the first two periods of dynamics, but no thermalization in the end.


\subsubsection{The dynamics of the slowest operator (of any kind) depends on $-\Tr[H,O]^2$. In particular, as one increases $g$ or $h$, the dynamics becomes faster.}

As one increases $g$ or $h$, the quantity $-\Tr[H,O]^2$ increases (see Fig. \ref{scaling_h_g}), and the dynamics becomes faster. We observe this in Fig. \ref{timeevol_h_g} (a), (b), (c), (d), (e): the curves move to the left. In Fig. \ref{timeevol_h_g} (e), aside from that, the revivals become more narrow.

One can also compare the dynamics of the local slowest operator for the parameters $g=1.05, h=0.1$ and $g=0.4, h=1.05$. In the latter case, $-\Tr[H,O]^2$ is much smaller (see Fig. \ref{scaling_h_g} (a), (c)). As a consequence, the dynamics is much slower (compare Fig. \ref{timeevol_L} (b) and (a)). The delocalization is also slower: in Fig. \ref{OTOC} (c) the OTOCs with Pauli matrices at different locations quickly become equal, but in Fig. \ref{OTOC} (c) this process runs much slower.



We note that for near-zero $h$ or $g$, the dynamics is extremely slow (almost none) for translationally-invariant operator, as it becomes an integral of motion ($-\Tr[H,O]^2 \rightarrow 0$). See Fig. \ref{scaling_h_g} (b), (d), (f); Fig. \ref{timeevol_h_g} (b), (d), (f); Fig. \ref{timeevol_L} (b), (c).







\subsubsection{Delocalization of the slowest operator is expressed as follows: for the local slowest operator, OTOCs for different locations of the Pauli matrix become equal, while, for translationally-invariant slowest operator, OTOC rapidly increases at early times.}


See Fig. \ref{OTOC} (a) for the local slowest operator. Initially, OTOC is equal to $0$ for $\sigma_{x,y,z}$ located at sites "center+3", "center+4", "center+5", because the slowest operator is not there just yet and commutes with those Pauli matrices. But it has non-zero OTOC with $\sigma_{x,y,z}$ located at "center", "center+1", "center+2". OTOC has the biggest value for "center", smaller value for "center+1", etc. We conclude that the slowest operator is gradually vanishing to its boundaries. As it expands over the chain, OTOC becomes equal for different locations $i$ of $\sigma^{(i)}_{x,y,z}$. In Fig. \ref{OTOC} (b), the dynamics is similar, but much slower.



For the translationally-invariant operator $O = \sum_k O_k$, we see a period of growth at early times (see Fig. \ref{OTOC} (b)). The reason is that, initially, the Pauli matrix fixed at some site has non-zero commutator only with nearby $O_i$, but then each $O_k$ delocalizes and starts to contribute to OTOC. As a result, OTOC quickly grows. For $g=0.4, h=1.05$, the dynamics is much slower, and the initial growth is hardly visible (see Fig. \ref{OTOC} (d)). (Compare also the rate of dynamics in Fig. \ref{timeevol_L} (b) and (a).)


In the integrable case, there is no delocalization, and, therefore, no initial period of rapid growth (Fig. \ref{OTOC} (f)).

\subsubsection{For the local slowest operator, the OTOCs for $\sigma_x, \sigma_y, \sigma_z$ reflect the contributions of different magnetizations into the slowest operator. In non-integrable case, these OTOCs become equal in the long-time limit.}

As one can see in the fourth subplot of Fig. \ref{OTOC} (a), (c), (e), OTOC is smaller for that Pauli matrix, which corresponds to magnetization with the greatest contribution (see Fig. \ref{overlap} (a), (c), (e)). More generally, the hierarchy of OTOCs follows that of magnetizations. For instance, for $g=1.05, h=0.1$, magnetization1 has biggest contribution, then follows magnetization3, and there is no contribution from magnetization2 (Fig. \ref{overlap} (a)). And the values of corresponding OTOCs increase (fourth subplot of Fig. \ref{OTOC} (a)).

In non-integrable case, OTOCs for $\sigma_x, \sigma_y, \sigma_z$ become equal in the long-time limit (fourth subplot of Fig. \ref{OTOC} (a) and (b)). But this does not happen in the integrable case (Fig. \ref{OTOC} (c)).


For the translationally-invariant operator ($O = \sum_k O_k$), there is no connection between OTOCs and magnetizations. Probably, the reason is the interplay between different $O_k$ in the OTOC.

\section{Conclusion \label{sec_conclusion}}


In this work we considered the quantum Ising model in external magnetic field (see (\ref{Ham})) close to an integrable point. We studied the slowest operator, as it plays an important role in the final period of dynamics. We introduced local and translationally-invariant definitions of the slowest operator. We showed that both operators have low entanglement, and, therefore, we were able to construct them using tensor networks.



Throughout the paper, we extensively compared their properties. Here we emphasize their main characteristic features.






The local slowest operator is not an integral of motion of the integrable system ($h=0$). As one increases $h$, there is a transition from integrable to thermalizing behavior. In the integrable system, there are revivals (of full and half amplitude). As one increases $h$, the revivals get suppressed, but the fluctuations remain. The rate of delocalization changes from extremely slow to slower than diffusion. The operator has a significant overlap with diffusion mode/energy flux.



The translationally-invariant slowest operator corresponds to an integral of motion of an integrable system ($h=0$). It changes its nature at a specific value $h^*$: before the transition ($h<h^*$) it does not  have an overlap with any magnetization and expands over the chain faster than diffusion; after the transition ($h>h^*$) it has non-zero overlap with magnetization1 and magnetization3 and expands slower than diffusion. The time evolution shows no fluctuations.



The two definitions have common features in the dynamics (consider non-integrable system): the initial period of dependence on one parameter $-\Tr [H,O]^2$, then delocalization and approaching the boundaries and final thermalization.






\section{Discussion \label{discussion}}

We found the distinct features of the local and translationally-invariant slowest operators. But several questions remain, regarding their dynamics and overall thermalization.

To start with, the integrable Ising model is known to have the ballistic transport of quasi-particles in the final period of its dynamics. But we obtain the local slowest operator with extremely slow dynamics, which has a significant overlap with energy flux. It is important to understand what role this operator plays in the ballistic transport picture.


One of the findings of this paper is that, as one goes away from the integrable point, the dynamics of the local slowest operator changes from extremely slow to slower than diffusion. But the concrete processes in the spin chain, leading to this behavior, are not understood.




From more technical side, we find that the parameter $h$ plays a special part in the dynamics of the local slowest operator - it suppresses the revivals. But $g$ only regulates the rate of dynamics. On the other hand, $h$ and $g$ both regulate the rate of dynamics of the translationally-invariant slowest operator. Thus, one needs to better understand the scopes of action of $h$ and $g$ in the final period of dynamics and in thermalization process as a whole.


Finally, there are several quantities describing the rate of dynamics. The first one is $-\Tr [H,O]^2$. But there are also others, describing the rate of delocalization:  the scaling $-\Tr [H,O]^2 \sim \frac{1}{N^k}$ and out-of-time-ordered commutator (OTOC). One needs to understand, which one plays a decisive role in delocalization, and what is the meaning of the residual dynamics, not leading to delocalization.

\section{Acknowledgements}

The author thanks Anatoly Dymarsky for introduction into this topic, formulation of the problem and useful discussions.



\bibliography{mybib}

\end{document}